\documentclass{aa}  

\usepackage{graphicx}

\usepackage{txfonts}
\usepackage[pdftex, colorlinks=true, linkcolor=blue, citecolor=blue, filecolor=blue, urlcolor=blue]{hyperref}
\newcommand{\nodata}{ ~$\cdots$~ }     
\graphicspath{{./}{figures/}}          

\begin{document}

\title{Streamlined lensed quasar identification in multiband images via ensemble networks}

\author{
	Irham Taufik Andika \inst{\ref{affil:tum}, \ref{affil:mpa}}
	\and
	Sherry H. Suyu \inst{\ref{affil:tum}, \ref{affil:mpa}, \ref{affil:asiaa}}
	\and
	Raoul Ca\~nameras \inst{\ref{affil:mpa}, \ref{affil:tum}}
	\and
	Alejandra Melo \inst{\ref{affil:mpa}, \ref{affil:tum}}
	\and
	Stefan Schuldt \inst{\ref{affil:unimi}}
	\and
	Yiping Shu \inst{\ref{affil:pmo}}
	\and
	Anna-Christina Eilers \inst{\ref{affil:mit}}\fnmsep\thanks{Pappalardo Fellow}
	\and
	Anton Timur Jaelani \inst{\ref{affil:itb}, \ref{affil:ucoe}}
	\and
	Minghao Yue	\inst{\ref{affil:mit}}
}

\institute{
	Technical University of Munich, TUM School of Natural Sciences, Department of Physics, James-Franck-Str. 1, D-85748 Garching, Germany \\
	\email{irham.andika@tum.de}
	\label{affil:tum}
	\and
	Max-Planck-Institut f\"{u}r Astrophysik, Karl-Schwarzschild-Str. 1, D-85748 Garching, Germany
	\label{affil:mpa}
	\and
	Academia Sinica Institute of Astronomy and Astrophysics (ASIAA), 11F of ASMAB, No.1, Section 4, Roosevelt Road, Taipei 10617, Taiwan
	\label{affil:asiaa}
	\and
	Dipartimento di Fisica, Universit\`a degli Studi di Milano, via Celoria 16, I-20133 Milano, Italy
	\label{affil:unimi}
	\and
	Purple Mountain Observatory, No. 10 Yuanhua Road, Nanjing, Jiangsu, 210033, People's Republic of China
	\label{affil:pmo}
	\and
	MIT Kavli Institute for Astrophysics and Space Research, 77 Massachusetts Ave., Cambridge, MA 02139, USA
	\label{affil:mit}
	\and
	Astronomy Research Group and Bosscha Observatory, FMIPA, Institut Teknologi Bandung, Jl. Ganesha 10, Bandung 40132, Indonesia
	\label{affil:itb}
	\and
	U-CoE AI-VLB, Institut Teknologi Bandung, Jl. Ganesha 10, Bandung 40132, Indonesia
	\label{affil:ucoe}
}


\date{}

 
\abstract{
	Quasars experiencing strong lensing offer unique viewpoints on subjects related to the cosmic expansion rate, the dark matter profile within the foreground deflectors, and the quasar host galaxies.
	Unfortunately, identifying them in astronomical images is challenging since they are overwhelmed by the abundance of non-lenses.
	To address this, we have developed a novel approach by ensembling cutting-edge convolutional networks (CNNs) -- for instance, ResNet, Inception, NASNet, MobileNet, EfficientNet, and RegNet -- along with vision transformers (ViTs) trained on realistic galaxy-quasar lens simulations based on the Hyper Suprime-Cam (HSC) multiband images.
	While the individual model exhibits remarkable performance when evaluated against the test dataset, achieving an area under the receiver operating characteristic curve of $>$97.3\% and a median false positive rate of 3.6\%, it struggles to generalize in real data, indicated by numerous spurious sources picked by each classifier.
	A significant improvement is achieved by averaging these CNNs and ViTs, resulting in the impurities being downsized by factors up to 50.
	Subsequently, combining the HSC images with the UKIRT, VISTA, and unWISE data, we retrieve approximately 60 million sources as parent samples and reduce this to 892,609 after employing a photometry preselection to discover $z>1.5$ lensed quasars with Einstein radii of $\theta_\mathrm{E}<5\arcsec$.
	Afterward, the ensemble classifier indicates 3080 sources with a high probability of being lenses, for which we visually inspect, yielding 210 prevailing candidates awaiting spectroscopic confirmation.
	These outcomes suggest that automated deep learning pipelines hold great potential in effectively detecting strong lenses in vast datasets with minimal manual visual inspection involved.
}


\titlerunning{Streamlined Lensed Quasar Identification via Ensemble Networks}
\authorrunning{Andika et al.} 

\maketitle

%

\nolinenumbers 

\section{Introduction} \label{sec:intro}

Quasars are fueled by matter accretion onto supermassive black holes (SMBHs) and are among the most luminous objects in the universe, emitting enormous amounts of energy. This attribute makes them ideal probes for studying the distant universe and the physical processes that govern the emergence of the SMBHs and their host galaxies across cosmic time \citep[e.g.,][]{2020ARA&A..58...27I,2022arXiv221206907F}.

In rare occurrences, the presence of a nearby galaxy in the observer's line of sight can distort the light originating from a distant quasar in the background, resulting in so-called gravitational lensing \citep{2015eaci.book.....S}. In the event of strong lensing, where highly magnified and multiple images of the quasar are produced, the mass distribution of the deflectors can be examined by analyzing the observed lens configuration, providing insights into the dark matter profile and the processes that drive the mass assembly of galaxies and clusters \citep[e.g.,][and references therein]{2022arXiv221010790S}.

Lensed quasars also serve as crucial tracers for understanding the fundamental physics of our universe. 
For example, the cosmic expansion rate, age, and critical density are related to the Hubble constant ($H_0$), which can be inferred via lens mass distribution modeling and time delay analysis of lensed quasar images \citep[e.g.,][]{1964MNRAS.128..307R,2022A&ARv..30....8T}.
Considering the current tension of $H_0$ values inferred from the different late and early universe probes, independent methods using the lensed quasars analysis are critically important \citep[see e.g.,][]{2020MNRAS.498.1420W}, and more precise measurement is expected with higher number statistics.
In addition, lensing could provide flux magnification and increase the effective spatial resolution of the target of interest that would otherwise be too faint (compact) to be detected (resolved).
This effect enables us to study quasars with intrinsically lower luminosity along with their host galaxies in unprecedented detail \citep{2021ApJ...917...99Y,2022MNRAS.517.3377S,2023ApJ...943...25G}.

At the time of writing, around 300 lensed quasars have been discovered through various observational techniques, including locating multiple point sources with quasar-like colors or selecting objects with unusual shapes consistent with lensing configurations.
In the early days, for example, the Cosmic Lens All-Sky Survey identified many multiply-imaged flat-spectrum radio sources, which are then confirmed as radio-loud lensed quasars \citep{2003MNRAS.341....1M,2003MNRAS.341...13B}.
Shortly after, in the optical wavelength, the Sloan Digital Sky Survey Quasar Lens Search confirmed a few tens of lenses based on the morphological analysis and color selection of spectroscopically classified quasars \citep{2006AJ....132..999O,2008AJ....135..496I,2010AJ....140..403I,2012AJ....143..119I}.

Over time, modern wide-field sky surveys can reach deeper limiting magnitudes and deliver great data quality, making it feasible to select more lensed quasar candidates via imaging data alone without the need for spectroscopic preselection.
For instance, data mining on photometric catalogs to detect multiple quasar sources allows for the discovery of lensed quasars in numerous projects, such as the Dark Energy Survey \citep{2018MNRAS.481L.136S,2018MNRAS.480.5017A,2019MNRAS.489.2525A,2020MNRAS.494.3491L}, the Kilo-Degree Survey \citep{2018MNRAS.480.1163S,2019A&A...632A..56K}, and the Dark Energy Spectroscopic Instrument Legacy Imaging Surveys \citep{2022arXiv220806356D,2023A&A...672A.123H}.
In addition, complementing optical data with infrared photometry and adding astrometric measurements could further reduce the number of false detections \citep{2018A&A...616L..11K,2018A&A...618A..56D,2018MNRAS.479.5060L,2019MNRAS.489.4741S,2019MNRAS.483.4242L,2022MNRAS.509..738D,2023MNRAS.520.3305L}.
We note that despite the high success rates of the previous lensed-finding approaches, they still face a substantial challenge: the inevitability of human involvement in time-consuming and exhaustive visual inspection stages to acquire a final list of candidates with high purity \citep[e.g.,][]{2023arXiv230405425C,2023AJ....165..191Y}.

Over the past few years, automated lens-finding methods, either using conventional point sources and lens arcs finder or state-of-the-art machine learning algorithms -- for example, convolutional neural network (CNN), variational autoencoder (VAE), and vision transformer (ViT) -- are being explored to reduce human interventions further \citep{2020MNRAS.494.3750C,2022MNRAS.517.1156R,2022MNRAS.513.2407A}.
However, more optimizations are still required since applying these classifiers to real survey data frequently yields samples dominated by false detections.
Often, manual visual inspection still needs to be done at the final stage on more than ten thousand candidates recommended by automated classifiers to select high-grade lenses and remove the contaminants \citep[e.g.,][]{2022A&A...659A.140C}.
This hassle might be caused by the need for more realistic training datasets to improve the classifier performance, coupled with complications caused by the very low fraction ($\lesssim$10$^{-3}$) of strong lenses in all galaxies per sky area \citep[see, for example,][]{2010MNRAS.405.2579O}.

Although machine learnings are now widely used for finding galaxy-galaxy strong lenses and performing lens modeling to those systems \citep[e.g.,][]{2017Natur.548..555H,2022AandA...662A...4S,2023A&A...671A.147S}, their use case for lenses containing galaxy-quasar pairs is still limited and has not been explored much.
They might have worked well for lensed galaxies since these sources display extended lens arcs that can be distinguished from other astronomical sources. 
On the other hand, lensed quasars often show only two or more point sources that outshine the light from the lensing galaxy.
They are also overwhelmed by visually identical impurities such as binary stars or quasar-star projections.

As mentioned earlier, previous lensed quasar searches are proven to be effective, but they might not be efficient enough and not scalable in larger datasets. 
Specifically, current estimations for the upcoming data from the next-generation surveys such as Euclid \citep{2011arXiv1110.3193L,2022A&A...662A.112E} and Vera C. Rubin Observatory's Legacy Survey of Space and Time \citep[LSST;][]{2019ApJ...873..111I} expect that these projects would expand the number of candidates for strong lenses by at least a few orders of magnitude \citep{2015ApJ...811...20C,2022AJ....163..139Y,2023arXiv230402784T}. 
Therefore, developing a highly efficient, automated lensed quasar selection algorithm is very much indispensable.

Here, we develop a novel lens finder using the ensemble of state-of-the-art convolutional and transformer-based neural networks \citep[e.g.,][]{2019arXiv190503288S,2020arXiv201011929D}.
Our classifier is particularly optimized to detect lensed quasars in multiband images of the Hyper Suprime-Cam Subaru Strategic Program \citep[HSC-SSP;][]{2022PASJ...74..247A}, extending the selection space to higher redshift ranges that might be missed by previous surveys.
Complementing the primary optical data with infrared photometry, we further apply spectral color modeling to obtain lensed quasar candidates with minimal contaminants.

This paper is presented as follows. Section~\ref{sec:data} begins with a description of data collection and target preselection using photometric color cuts. 
Section~\ref{sec:lens_simulation} explains the simulation for understanding the color and morphology of the galaxy--quasar lens systems. 
Section~\ref{sec:method} then goes through the specifics of lens detection using automated classifiers, including the datasets utilized for training and evaluating the neural networks. Section~\ref{sec:result} then discusses the classification outputs and the lensed quasar candidates. 
Section~\ref{sec:conclusion} completes with a summary and our conclusions.

Throughout this paper, we employ the $\Lambda$CDM cosmological model where $\Omega_\Lambda = 1 - \Omega_\mathrm{m} = 0.685$ to simulate lensed images in the training set \citep{2020A&A...641A...6P}.
It is worth noting that the resulting lensed images do not depend on the exact value of $H_0$.
In addition, the written magnitudes are reported using the AB system.

\section{Dataset and preselection} \label{sec:data}

Our lensed quasar hunt comprises two steps: (1) selecting candidates based on their photometric color using multiband data, and (2) calculating the relative likelihoods of the candidates being a lens or contaminant utilizing a machine learning classifier.
In this first step, we want to increase the purity of the candidates by restricting the search to objects that we suspect, based on catalog-level photometry, are more likely to be lenses. 
This approach offers a strategy for efficiently distinguishing the candidates from the majority of the contaminants while requiring the least amount of computer resources. 
The following section will describe the first part of our search method in more detail.

\subsection{Primary optical photometric data}
As the primary catalog in the optical regime, we make use of the wide-layer data of HSC-SSP Public Data Release 3 \citep[shortened to HSC;][]{2022PASJ...74..247A}.
The observations are conducted by utilizing the Hyper Suprime-Cam mounted on the Subaru 8.2m telescope \citep{2018PASJ...70S...8A,2019PASJ...71..114A}, capturing the sky image of $670\deg^2$ wide in five bands ($grizy$) at the full depth of $\sim$26~mag ($5\sigma$ for point sources), a pixel scale of 0\farcs168, and seeing of 0\farcs6--0\farcs8.
It should be noted that if we also account for the partially observed areas, the current data release covers up to approximately 1300~$\deg^2$ instead.
This larger HSC footprint will be used to construct the parent sample of our lensed quasar candidates selection.

\begin{table*}[htb!]
	\caption{Summary of the selection criteria applied to find the lensed quasar candidates.}
	\label{tab:preselection}
	\centering
	\begin{tabular}{clccccc}
		\hline\hline
		Step & Selection & Candidates & Lenses & Galaxies & Quasars & Stars \\
		\hline
		1 & Initial flag and S/N limits: & 57,464,157 & 22 & 246,918 & 77,650 & 65,666 \\
		& ~~\textbullet~~ S/N($i$, $z$, $y) > (3, 5, 8)$ \\
		& ~~\textbullet~~ \texttt{[grizy]\_apertureflux\_20\_flag is False} \\
		2 & Presence of neighboring sources within a 2\arcsec\ radius & 4,854,831 & 22 & 244,080 & 77,603 & 65,656 \\
		& or objects with available spectroscopic data \\
		3 & Detection in the unWISE MIR catalog: & 911,263 & 22 & 184,282 & 56,971 & 5629 \\
		& ~~\textbullet~~ S/N(W1, W2) $>$ (5, 3) \\
		& ~~\textbullet~~ $0.1 < y-\mathrm{W1} < 3.6$ \\
		& ~~\textbullet~~ $-0.7 < \mathrm{W1-W2} < 0.7$ \\
		4 & Detection in the UKIRT or VISTA NIR data: & 621,713 & 22 & 177,900 & 48,588 & 5138 \\
		& ~~\textbullet~~ S/N($J$) $> 3$ \\
		5 & Optical and NIR color cuts: & 601,277 & 22 & 177,262 & 44,531 & 4726 \\
		& ~~\textbullet~~ $-0.8 < z-y < 3.9$ \\
		& ~~\textbullet~~ $-0.2 < y-J < 2.8$ \\
		6 & Sources with at least one neighbor within a 2\arcsec\ radius & 389,263 & 15 & 14,706 & 1792 & 2406 \\
		\hline\hline
		7 & Sources including their companions & 892,609 & 31 & 17,263 & 1990 & 3748 \\
		8 & Ensemble network classification & 3080 & 31 & 439 & 40 & 8 \\
		9 & Astrometric information & 2604 & 31 & 380 & 38 & 5 \\
		& ~~\textbullet~~ Astrometric excess noise $<10$~mas \\
		& ~~\textbullet~~ Proper motion significance $<10\sigma$ \\
		10 & Visual inspection & 210 & \nodata & \nodata & \nodata & \nodata \\
		\hline
	\end{tabular}
	\tablefoot{
		From Step~7 and beyond, the reported numbers are composed of the total primary sources added with their corresponding neighbors.
		For example, in Step~6 we recover 15 unique known lenses, and in Step~7 this number becomes 31 (see ``Lenses'').
		This addition is because the 14 lenses consist of 2 components while the other 1 system has 3 detected sources in the HSC catalog.
		Concerning the numbers shown in the columns named ``Galaxies'', ``Quasars'', and ``Stars'', only primary targets and companions with spectroscopic data are counted.
	}
\end{table*}

To begin the initial selection, we pick all sources detected in the $i$, $z$, and $y$ bandpasses with signal-to-noise ratio (S/N) values of more than 3, 5, and 8, respectively.
These sources should also have $g$ and $r$ images in the HSC data, but we do not impose any S/N cuts for these bands. 
As a note, these S/N cut values are derived based on our lens simulation, which will be discussed in the later section.
Also, in this case, we adopt the flux and magnitude measurements within the 2$\arcsec$ ($\approx$12~pixels) aperture diameter from the HSC \texttt{pdr3\_wide.summary}\footnote{\url{https://hsc-release.mtk.nao.ac.jp/schema/\#pdr3.pdr3\_wide.summary}} catalog entries.
This table incorporates forced photometry on stacked images containing frequently-selected columns of primary objects.
Furthermore, we apply the flags: (1) \texttt{[grizy]\_apertureflux\_20\_flag is False}, and (2) \texttt{[grizy]\_is\_clean\_centerpixels is True}, to retrieve only the sources with reliable photometry.
The science images with a size of $72\times72$ pixels and their corresponding point spread function (PSF) cutouts are subsequently downloaded using the HSC data access tools\footnote{\url{https://hsc-gitlab.mtk.nao.ac.jp/ssp-software/data-access-tools/}}.
At this point, 57,464,157 unique sources, defined as our parent sample, pass our preliminary S/N cuts and flag criteria, implying that a lot of computing power is required to process them all.
As additional information, the summary of our selection criteria will be reported in Table~\ref{tab:preselection}.

After that, since most kpc-scale quasar pairs and lensed quasars have separations of $\lesssim3\arcsec$ \citep[e.g.,][]{2023AJ....165..191Y}, we try to narrow the selection to sources that show the presence of nearby companions within a 2$\arcsec$ radius.
We are aware that this choice might be too strict.
As an illustration, out of 22 optically bright lensed quasars with spectroscopic confirmation in the HSC catalog \citep{2023arXiv230405425C}, we only recover 15 of them via the above preselection --  that is, a recovery rate of 68\%.
Seven lensed quasars are missed due to the absence of neighboring sources in their vicinity, probably because they are too faint or some failures in the HSC object deblending process.
Nevertheless, this method managed to reduce the number of selected objects significantly while keeping many known lenses to be recovered with minimal contaminations and required computational power later.
We then crossmatch our parent sample with catalogs of known quasars \citep{2021arXiv210512985F,2022arXiv221206907F}, galaxies/stars \citep{2023arXiv230107688A}, strong lenses\footnote{
	The list of previously published lens systems is compiled from the Master Lens Database (hereafter MLD; \url{https://test.masterlens.org/}) and the Gravitationally Lensed Quasar Database (dubbed as GLQD; \url{https://research.ast.cam.ac.uk/lensedquasars/})
}, 
and brown dwarfs\footnote{The brown dwarf catalogs consist of late-M stars plus L and T dwarfs.} \citep{2018ApJS..234....1B,2019MNRAS.489.5301C} to identify the spectroscopic classification of these sources when available.
In the end, after selecting only sources that: (1) have at least one neighboring source within a 2$\arcsec$ radius or (2) have spectroscopic classifications, we managed to reduce the number of objects to only 4,854,831.

\subsection{Infrared photometry from public surveys}
The near-infrared (NIR) data is then acquired from the catalogs of the UKIRT Infrared Deep Sky Survey \citep[UKIDSS;][]{2007MNRAS.379.1599L}, the UKIRT Hemisphere Survey \citep[UHS;][]{2018MNRAS.473.5113D}, the VISTA Hemisphere Survey \citep[VHS;][]{2013Msngr.154...35M}, and the VISTA Kilo-degree Infrared Galaxy (VIKING) Survey \citep{2013Msngr.154...32E}.
We use here the photometry in the $J$, $H$, and $K$ (or $K_\mathrm{s}$) bands, when available.
As a note, most of the southern hemisphere is covered by VHS and VIKING, while UKIDSS and UHS capture a large sky area in the north.
When we began the candidate selection, UKIDSS and VIKING had completed their observations.
However, UHS had only published its $J$-band photometry, while VHS had only provided its $J$ and $K_\mathrm{s}$-band photometry for most sky regions.
As a consequence, the precise photometry accessible for each source is determined by its location in the sky.
We also exploit mid-infrared (MIR) observations from the unWISE catalog \citep{2019ApJS..240...30S}, which contains about two billion objects identified by the Wide-field Infrared Survey Explorer \citep[WISE;][]{2010AJ....140.1868W} throughout the whole sky. 
With its $\approx$0.7 magnitudes of deeper imaging data and better source extraction in crowded sky areas, unWISE data surpasses the quality of the predecessor WISE catalog.

To combine the HSC data with the compiled infrared catalogs, we use a crossmatching radius of 2$\arcsec$ between the sources.
Together with the NIR photometry, the MIR W1~(3.4\,$\mu$m) and W2~(4.6\,$\mu$m) bands from unWISE are highly valuable for determining if the sources are quasars, stars, or brown dwarfs \citep[e.g.,][]{2020ApJ...903...34A,2022AJ....163..251A}.
This crossmatching technique also works for removing unwanted sources, such as cosmic rays or moving sources that are present in one survey but not in others \citep[e.g.,][]{2022PhDT.........1A}.
Subsequently, we retrieve the candidates with fluxes of respectively at least 5$\sigma$ and 3$\sigma$ in the W1 and W2 bands, as well as having the color of $0.1 < y-\mathrm{W1} < 3.6$ and $-0.7 < \mathrm{W1-W2} < 0.7$, resulting in a remaining 911,263 sources.
The NIR color cut is then conducted by keeping the sources with $J$-band S/N~$>$~3, 
leaving us only 621,713 candidates.
Further cut is made by taking only sources with $-0.8 < z-y < 3.9$ and $-0.2 < y-J < 2.8$, which yields 601,277 objects.
We note that the criteria we employed so far are derived empirically and managed to preserve 68\% previously discovered lensed quasars and 57\% known unlensed quasars within the HSC footprint while removing 93\% and 28\% of contaminating stars and galaxies.

Next, we will focus on sources that present the existence of nearby companions within a 2$\arcsec$ radius, reducing the number of candidates further to 389,263.
Since a lens candidate could have one, two, or more detected companions, we also need to take account of the neighbors around the primary targets, so the total number of sources that will be analyzed at the next stage is 892,609.
At the end of this preselection, we are still able to recover all of the 15 known lensed quasars mentioned before.
As a reminder, Table \ref{tab:preselection} contains an overview of all the selection steps employed up to this point.
It is also worth mentioning that all photometric measurements have been corrected from Galactic reddening employing the dust map from \cite{1998ApJ...500..525S} and with the updated bandpass corrections from \cite{2011ApJ...737..103S} and \cite{1999PASP..111...63F} extinction relation, implemented via the \texttt{dustmaps} library of \cite{2018JOSS....3..695G}.

\section{Simulating the lenses} \label{sec:lens_simulation}

\begin{figure*}[htb!]
	\centering
	\resizebox{\hsize}{!}{\includegraphics{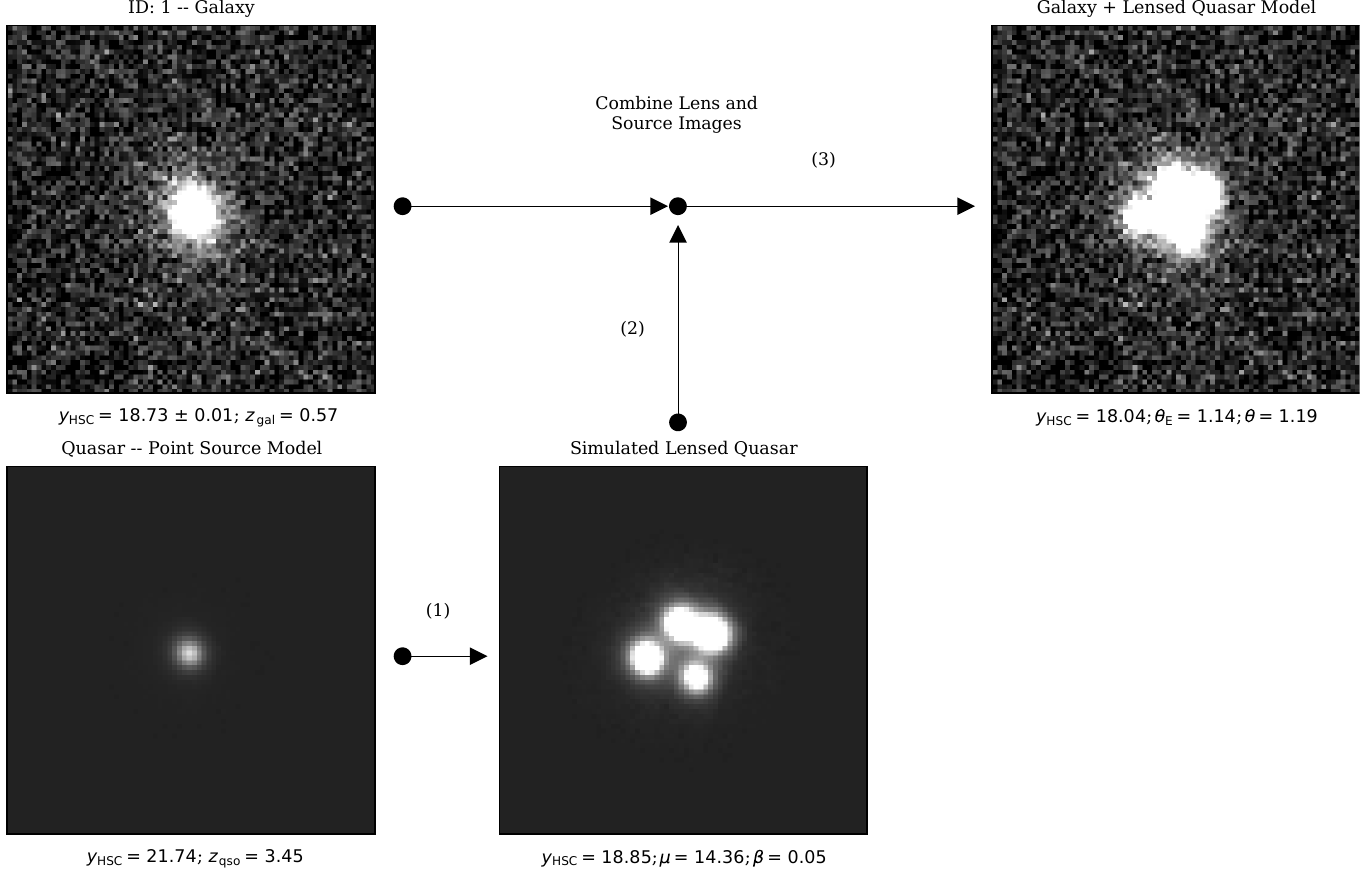}}
	\caption{
		Simulation workflow that we adopt for generating the mock lensed quasar images.
		The cutouts are based on the HSC images with $72\times72$~pixels wide ($\approx$12\arcsec\ on a side).
		The real galaxy, acting as a lens, is shown in the upper left panel.
		On the other hand, the quasar, which serves as a background source, is depicted in the lower left panel as a Gaussian light profile.
		The deflected source's light convolved with the respective HSC PSF model is shown in the lower middle panel, which is determined based on the associated lens arrangement.
		Ultimately, we paint the multiply-imaged sources on top of the galaxy image and display it in the upper right panel.
		Below each panel, the lens parameters (in arcseconds, $\beta$ and $\theta_\mathrm{E}$) and the photometry is reported.
		In this case, we construct mock images for all HSC bandpasses -- namely, $grizy$ bands.
	}
	\label{fig:lens_simflow}
\end{figure*}

To find lensed quasar candidates based on their multiband images, we need to understand their spectral energy distribution (SED) and morphology, composed by the addition of lights between the galaxy in the foreground and the quasar in the background.
Following some lensing configurations, we can ray trace the lights and produce highly-realistic mock lens images by overlaying the lights of deflected point sources, representing the background quasar emission, over the real deflector images.
These mock images will serve as input for the training dataset for building our neural networks model at the later selection step.
As a brief illustration, the outline of our simulation workflow is presented in Figure~\ref{fig:lens_simflow}.
More details on (1) the deflector galaxies data retrieval, (2) the mock quasar spectra generation, and (3) the galaxy-quasar lens image production are explained in this part of the paper.

\subsection{Assembling the deflector galaxy samples}

We first need to look for a sample of spectroscopically verified galaxies in the Sloan Digital Sky Survey Data Release 18 catalog \citep[hereafter SDSS;][]{2023arXiv230107688A}, accessible through the \texttt{CasJobs}\footnote{\url{https://skyserver.sdss.org/CasJobs/}} website, to assemble the deflectors for our lens simulation.
Since the velocity dispersion ($\sigma_v$) is a critical metric for computing the lensing effect later, we pick all pipeline-classified ``GALAXY'' sources and narrow our search to those with the ratio of velocity dispersion to its error of $\sigma_v/\sigma_{v,\,\mathrm{err}} > 5$ to retrieve samples with accurate measurements.
We also exclude galaxies with $\sigma_v \leq 50$~km~s$^{-1}$ to discard lenses with too small or potentially inaccurate mass.

\begin{figure*}[htb!]
	\centering
	\resizebox{\hsize}{!}{\includegraphics{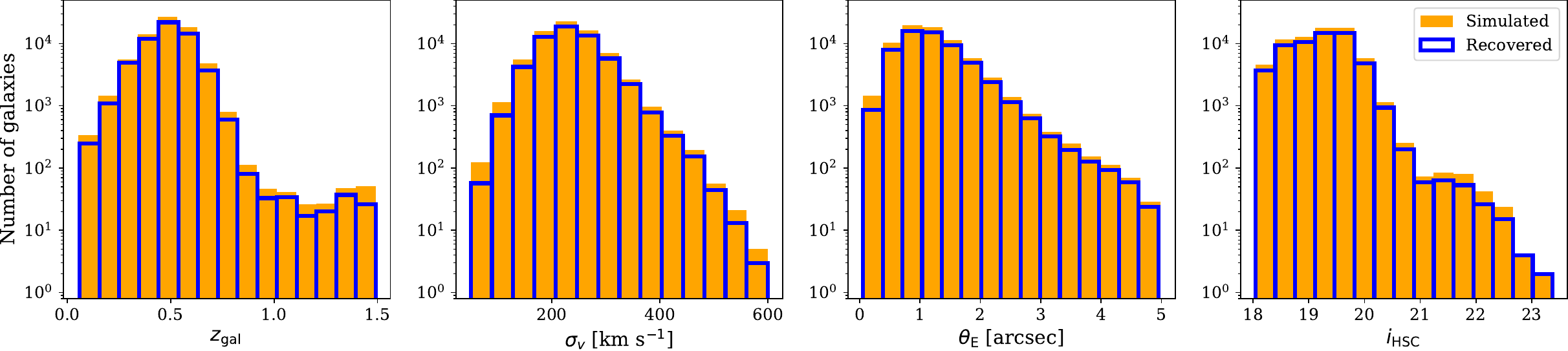}}
	\caption{
		Redshifts ($z_\mathrm{gal}$), stellar velocity dispersions ($\sigma_v$), Einstein radii ($\theta_\mathrm{E}$), and HSC $i$-band magnitudes ($i_\mathrm{HSC}$) distributions of the galaxies used for making the mock lenses. 
		These configurations are utilized to create mock lenses and are shown as orange histograms.
		The distribution of recovered lenses predicted by our ensemble network classifier is presented with blue lines.
	}
	\label{fig:lensgal_dist}
\end{figure*}

Furthermore, because the bulk of the lensing optical depth for high-$z$ sources originates from the early-type lens galaxies at a redshift of $z\sim1$ \citep{2015ApJ...805...79M,2019ApJ...870L..12P}, we limit our selection to deflectors at $z=0.05$ to 4 (see Figure~\ref{fig:lensgal_dist} as a reference).
The resulting samples are then matched to the HSC catalog, with a radius of search of 1\arcsec, to get their associated magnitudes and image cutouts when present.
As a result, we acquire a sample of 78,619 deflectors dominated by the luminous red galaxies (LRGs) population, peaked at $z\sim0.5$ and $\sigma_v\sim250$~km~s$^{-1}$, extending out to $z\lesssim1.5$.

\subsection{Generating the quasar spectral colors}

We proceed now to create simulated quasar emissions by generating a thousand quasar spectra, distributed uniformly at redshifts of $1.5 \leq z \leq 7.2$ and absolute magnitudes of $-30 \leq M_{1450} \leq -20$ at the rest-frame wavelength of 1450~\AA.
The simulation is done using the \texttt{SIMQSO}\footnote{\url{https://simqso.readthedocs.io/en/latest/}} module \citep{2013ApJ...768..105M}, following the prescription of \citet[][see their Section 2.2 for details]{2023ApJ...943..150A}.
This kind of simulation has been proven to mimic the SDSS quasar colors in high accuracy while also frequently used to assess the completeness of various quasar surveys \citep[e.g.,][]{2016ApJ...829...33Y,2018AJ....155..131M}.

As a quick summary, the foundation of our quasar spectral model consists of continuum emission, represented by a broken power-law function.
The slopes of the continuum ($\alpha_\nu$) follow a Gaussian distribution with mean values of $-1.5$ and $-0.5$ for the wavelengths at $\leq$1215~\AA\ and $>$1215~\AA, respectively, while each of their dispersions is fixed to 0.3.
Afterward, the series of iron emissions at the rest wavelengths of $<$2200~\AA, 2200--3500~\AA, and 3500--7500~\AA\ are consecutively appended to the model following the templates from \cite{2001ApJS..134....1V}, \cite{1992ApJS...80..109B}, \cite{1992ApJS...80..109B}.
The broad and narrow lines are then added to the spectra, complying with the ratio and width distributions of SDSS quasars \citep[e.g.,][]{2013AJ....145...10D,2016AJ....151...44D,2017AJ....154...28B}. 
Furthermore, the mock spectra incorporate the intergalactic medium (IGM) absorption by the Ly$\alpha$ forest in the sightline \citep{2010ApJ...721.1448S,2011ApJ...728...23W}.
On top of that, for $z\gtrsim5.5$ quasars, we apply the Ly$\alpha$ damping wing effect based on the theoretical approximation proposed by \cite{1998ApJ...501...15M}, with a fixed proximity zone size of 3~Mpc and a randomly assigned neutral hydrogen fractions of 0--10\% \citep[e.g.,][]{2019A&A...631A..85E,2020ApJ...903...34A}.
Finally, using the \cite{2000ApJ...533..682C} model and randomly picked $E(B-V)$ values of $-0.02$ to 0.14, the internal reddening effect from the dust is applied to the spectra. 
The negative reddening parameters are for creating quasar models with bluer continua than the original templates can accommodate.
The photometry is then estimated from the mock spectra, and the associated errors are calculated using the magnitude--error relations of each survey \citep[e.g.,][]{2016ApJ...829...33Y}.

\subsection{Producing the multiband images of lensed quasars}

As the next step, we adopt a singular isothermal ellipsoid \citep[SIE;][]{1998ApJ...502..531B} model to characterize the lens mass profile, which is specified by the Einstein radius ($\theta_\mathrm{E}$), axis ratio translated into a complex ellipticity, position angle, and image centroid \citep[e.g.,][]{2022AandA...668A..73R}.
Subsequently, the Einstein radius can be calculated from $\sigma_v$ enclosed by the gravitational potential using:
\begin{equation}
	\label{eq:sie_lens}
	\theta_\mathrm{E} = 4\pi \frac{\sigma_v^2}{c^2} \frac{D_\mathrm{ds}}{D_\mathrm{s}},
\end{equation}
where the light speed is $c$, while the angular diameter distances of the lens to source is $D_\mathrm{ds}$ and the observer to source is $D_\mathrm{s}$.
Given the ratio of distances in Equation~\ref{eq:sie_lens}, $\theta_\mathrm{E}$ is independent of $H_0$.
Nonetheless, for computing each of the distances, a value of $H_0 = 67.4$~km~s$^{-1}$~Mpc$^{-1}$ is used \citep[e.g.,][]{2020A&A...641A...6P}.
The SIE axis ratio, centroid, and position angle are then estimated directly by fitting the light distribution of each deflector on its HSC $i$-band image.
Here, we perform the light profile fitting using the combination of elliptical S\'{e}rsic and exponential functions implemented in the \texttt{PyAutoGalaxy}\footnote{\url{https://pyautogalaxy.readthedocs.io/en/latest/}}, an open-source library for investigating the galaxy morphologies and structures in multiwavelength data \citep{2018MNRAS.478.4738N,2023JOSS....8.4475N}.
The external shears are then added at random following a Gaussian distribution with a mean strength of 0 and a standard deviation of 0.058 \citep[e.g.,][]{2022AandA...662A...4S}, while the corresponding position angles are selected randomly in the range of 0 to 180~$\deg$.

Next, the simulated lens images are generated by coupling each real galaxy with a mock quasar taken at random.
The quasar is then randomly positioned behind the lens within $0\farcs01 \leq \beta \leq \theta_\mathrm{E}$, where $\beta$ is the true angular position of the respective source. 
After that, the source image is projected onto the lens plane, while the magnification and deflection angle are traced based on the lensing structure using the \texttt{PyAutoLens}\footnote{\url{https://pyautolens.readthedocs.io/en/latest/}} code \citep{2021JOSS....6.2825N}.
We also convolve the deflected quasar lights with the associated HSC PSF model before overlaying them to the original HSC galaxy images. 

The quasar pairing and placement can be repeated up to 500 times to find a suitable lens configuration that satisfies the following criteria: 
(1) the mock image has a strong lensing effect with a magnification factor of $\mu > 5$, 
(2) the lensed quasar $y$-band peak flux is detected at $\geq5\sigma$ against the mean background noise, 
and (3) its $y$-band magnitude is $>$15~mag to exclude unusually bright objects or saturated images.
Otherwise, we drop the current deflector and move to the next one.
Throughout this simulation, we also exclude systems with $\theta_\mathrm{E}\geq5\arcsec$ since the largest Einstein radius detected so far in the cases of galaxy-scale lensing corresponds to that limit \citep{2007ApJ...671L...9B,2021A&A...646A.126S}.
At last, we acquire 72,626 surviving lens configurations that fit our criteria from the initial 78,619 deflectors and 1000 mock quasars.

\begin{figure}[htb!]
	\centering
	\resizebox{\hsize}{!}{\includegraphics{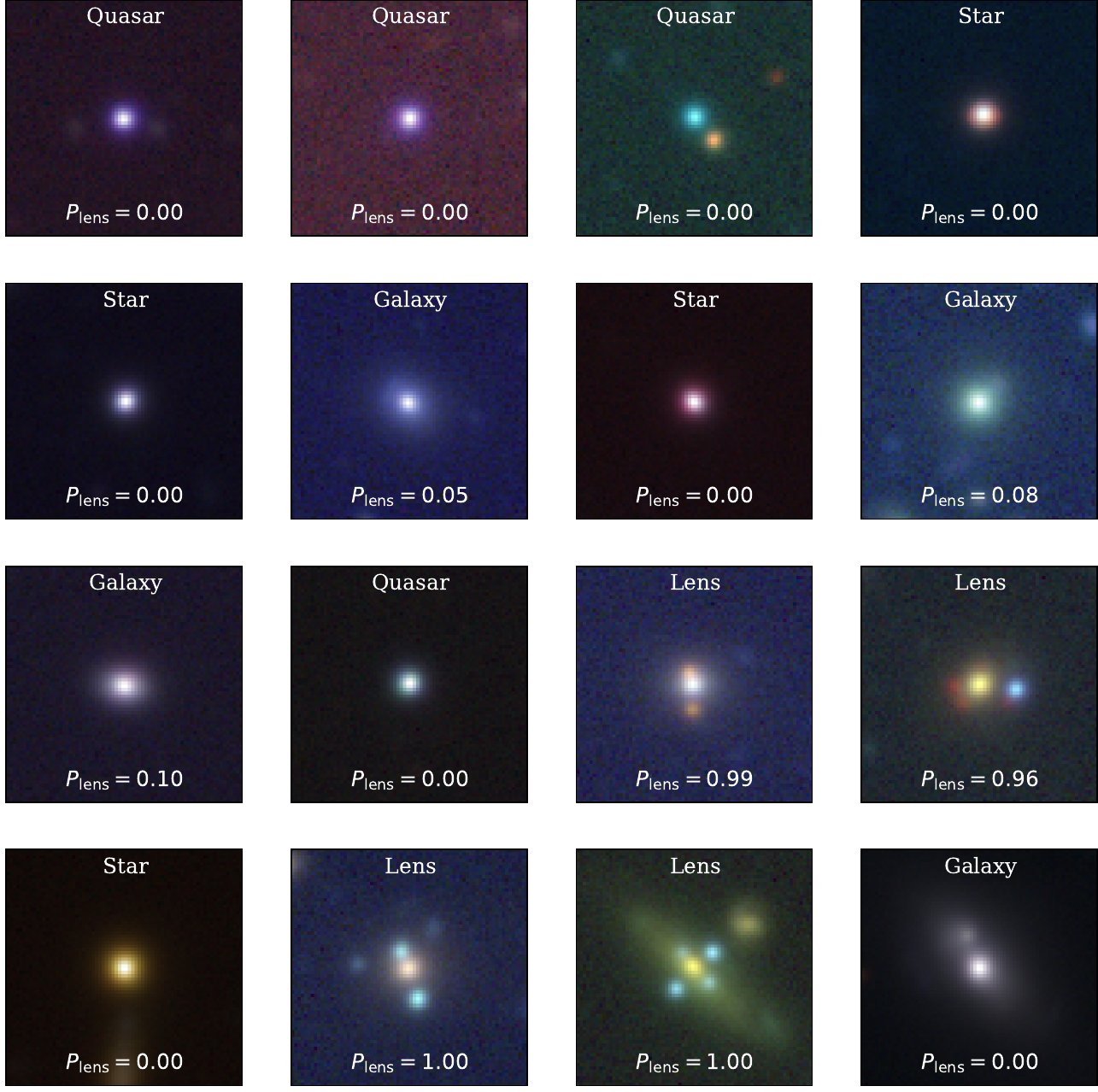}}
	\caption{
		HSC $grz$-band images of the unlensed galaxies, quasars, and stars as the negatives, along with the mock lenses as the positive examples.
		The square-root stretching is applied to the cutouts with a size of $72\times72$ pixels ($\approx$12\arcsec\ on a side) to emphasize features with low fluxes and improve the visual appearance.
		We also present the ground-truth labels and the estimated lens probability for each source in the figure.
	}
	\label{fig:lens_train}
\end{figure}

Figure~\ref{fig:lensgal_dist} depicts the distribution of the lens galaxy redshifts, velocity dispersions, Einstein radii, and $i$-band magnitudes adopted in our simulation. 
We also refer to Figure~\ref{fig:lens_train} for the resulting $grz$-band color images of the previously created mock lens systems.
It is apparent that the redshifts of our deflectors peak at $z\approx0.5$ and extend out to $z\lesssim1.5$, as mentioned before. 
Concerning the $i$-band magnitudes, we witness a spike in the deflector galaxy numbers up to $i_\mathrm{HSC}\approx19.5$, followed by an abrupt decrease near the faint end.
As a result, our training dataset is weighted toward brighter and larger lens galaxies. 
This occurrence is mostly produced by how SDSS picks its target galaxies for spectroscopy, which fulfills the guidelines outlined by \cite{2013AJ....145...10D,2016AJ....151...44D} and \cite{2016ApJS..224...34P} to investigate the universe's large-scale structure.
The majority of the targets are luminous elliptical galaxies at $z<1$, which are ideal tracers for studying the baryon acoustic oscillation signal and, hence, the expansion of our cosmos \citep[e.g.,][]{2022MNRAS.516.4307W,2022MNRAS.511.5492Z}. 
The magnitude boundaries for galaxies selected for spectroscopic surveys are $i = 19.9$ for the SDSS III and $i = 21.8$ for the SDSS IV projects \citep{2016ApJS..224...34P}.

\section{Lens finding with deep neural networks} \label{sec:method}

The second step of our lensed quasar search strategy involves supervised, deep neural network classification, requiring realistic training datasets as inputs to function. 
CNNs, for example, have been demonstrated to be successful in pattern recognition, such as discovering gravitational lenses in enormous sets of data \citep[e.g.,][]{2019A&A...625A.119M,2022MNRAS.515.5121B,2022MNRAS.510..500G,2022MNRAS.512.3464W,2022A&A...664A...4T}. 
While the exact design of the CNNs is generally determined by the challenge at hand, it typically consists of images as data inputs, which are subsequently processed by a sequence of convolutional, fully connected, and output layers.
In this part, we describe our automated classifier that has been trained to distinguish lensed quasars from non-lensed sources.
The section that follows will discuss the principles of our approach.

\subsection{Preparing the input data} \label{sec:training_input}

The inputs used to train our classifier will be divided into four categories:
(1) the mock lensed quasars created in the previous section,
(2) the real HSC galaxies that are not picked for the lensing simulation \citep{2023arXiv230107688A},
(3) previously discovered quasars from the local universe up to $z\sim7$ \citep{2021arXiv210512985F,2022arXiv221206907F},
and (4) a sample of stars and brown dwarfs \citep{2018ApJS..234....1B,2019MNRAS.489.5301C}.
Here, the distribution is balanced so that each class contains around 60,000 objects, and in sum, we use approximately 240,000 sources.

As later explained in Section~\ref{sec:final_lens}, we will build and train the classifier iteratively, where some false positives identified by our lens finder -- for instance, ring galaxies, spiral arms, irregular galaxies, CCD artifacts, and groups of multiple sources that imitate lensing arcs -- will be also included the in the training dataset.
About 90\% of them are galaxies, and the rest are labeled as either quasars or stars, depending on their spectroscopic classification when available. 
In total, there are 9562 of these additional negative examples.
It should be noted that the images utilized for the training inputs have been built based on the $grizy$ bands of HSC cutouts with a size of 72 pixels on a side, which is comparable to an angular dimension of $\approx$12\arcsec. 

Next, the images are min-max adjusted so that the fluxes vary from zero to one and are square-root stretched to boost features with low fluxes and enhance the visual appearance.
The relative pixel brightnesses across bandpasses are maintained, and therefore the colors of the associated sources are retained.
The images are subsequently augmented with random $\pm\,\pi/2$ rotations, 7-pixel translations, and horizontal or vertical flips on the fly each time they are called for training. 
This strategy will expand the quantity of training data while increasing the possibility that the network will properly categorize several perspectives of the same image.

\subsection{Network architectures} \label{sec:network_architecture}

Ensemble networks, which combine the predictions of multiple classifiers (e.g., CNNs or ViTs), have been shown to outperform individual models in various machine-learning tasks \citep{2021arXiv210402395G}. 
There are several reasons why this approach is often superior.
First, we can leverage the diversity of individual models. 
Each classifier in the ensemble is trained with a different initialization, architecture variation, or data augmentation scheme, leading to diverse learned representations. 
By combining these diverse models, the grouped networks can capture a broader range of patterns and variations in the data, improving overall generalization performance.
Second, ensemble networks can reduce overfitting, mitigate the impact of individual model biases or errors, and offer improved performance stability.
By averaging the predictions of multiple classifiers, they compensate for these biases and reduce the impact of individual errors, resulting in more reliable and robust predictions.
Therefore, several network models will be discussed in this section to assemble our ensemble network architecture.

\subsubsection{Baseline convolutional network}
We start with creating a simple CNN as a baseline, dubbed as \textbf{BaseNet}, following the same model presented by \cite{2023ApJ...943..150A} and motivated by other classical network designs \citep[e.g.,][]{726791,2014arXiv1409.1556S,10.1145/3065386}.
BaseNet has three convolutional layers containing kernels with sizes of $3\times3\times C$, with $C=32$,~64,~and~64 for the first, second, and third layers, respectively, along with a stride of $1\times1$ and the same padding\footnote{
	Same padding refers to the padding of additional rows and columns of zeros around the input image in such a way that the output feature map has the same spatial dimensions as the input.
}.
Each convolutional layer is then followed by a max pooling with the stride of $2\times2$, the size of $2\times2$, and the same padding.
At the end of this sequence, a fully connected layer of 128 neurons is appended, and the final output layer is attached to retrieve four outputs -- specifically, the chances of a target being a lensed quasar, a galaxy, an unlensed quasar, and a star.
The dropout regularizations are utilized everywhere, where the drop rates are set to 0.2 and 0.5 for convolutional and fully connected layers, respectively.
At the start, the learning rate is adjusted to $10^{-4}$ while the bias and weight of each neuron are set randomly and subsequently updated during the training.
The activation functions based on the Rectified Linear Unit (ReLU) are utilized throughout the networks, except for the output layer, which employs the softmax activation.
The \texttt{TensorFlow}\footnote{\url{https://www.tensorflow.org/}} deep learning platform is used to carry out all training procedures and CNN modeling \citep{2016arXiv160508695A,2022zndo...4724125D}.

After establishing the baseline network, additional classifiers will be added to the ensemble model, and an overview of each network architecture will be discussed. 
However, here we only provide a concise, high-level understanding and comparisons of the network architectures while acknowledging that in-depth information, technical specifications, and implementation details can be found in the respective cited references \citep[also see, e.g.,][for a review]{2021RemS...13.4712C}.

\subsubsection{Residual learning network}
Residual network, which is often abbreviated as \textbf{ResNet}, is a deep CNN architecture familiarized by \cite{2015arXiv151203385H}.
The core idea of ResNet is based on the observation that deeper networks could suffer from diminishing performance or even degradation due to vanishing/exploding gradients. 
The introduction of skip connections bypasses certain layers, allowing the network to learn residual functions or the difference between the input and the desired output. 
This strategy enables the model to focus on learning the residual information, which is often easier to optimize. 

Further progress is made with the introduction of ResNetV2, which follows the original ResNet designs but integrates the bottleneck residual blocks \citep{2016arXiv160305027H}. 
It also presents the concept of identity shortcuts to handle the skip connections, enhancing the overall network performance.
Another variant of of this family is ResNetRS, which incorporates the Squeeze-and-Excitation (SE) modules into the residual blocks, aiming to capture channel-wise dependencies and recalibrate the feature maps adaptively \citep{2021arXiv210307579B}.
By selectively amplifying informative features, these modules enhance the representation capacity of the network and improve its discriminative ability.
Here, ResNet50V2 and ResNetRS50, which have 50 layers, will be picked as our choices for building the ensemble model components.
This starting point is also chosen since deeper ResNet with 101 or 152 layers did not improve classification performance for the tiny image cutouts we studied.

\subsubsection{Inception network}
\textbf{Inception} is an architecture that is developed by \cite{2014arXiv1409.4842S} to handle the challenges of effectively capturing multi-scale features and reducing computational complexity in deep neural networks.
The central notion behind the Inception model is to employ parallel convolutional filters of different sizes within a single layer, allowing the network to grasp features at different spatial scales.
These filters are usually composed of $1\times1$, $3\times3$, or $5\times5$ convolutions, along with a $3\times3$ max pooling operation. 
By combining these different-sized filters and pooling operations, Inception enables the network to apprehend both fine-grained details and broader contextual information simultaneously.

The Inception family has undergone several improvements over time, leading to subsequent versions such as InceptionV3, Xception, and InceptionResNetV2 \citep{2015arXiv151200567S,2016arXiv160207261S,2016arXiv161002357C}.
These variants, which we will utilize in this work, incorporated additional design elements, including factorized convolutions, depthwise separable convolutions, batch normalization, or residual connections, to further enhance the gradient flow and training stability.

\subsubsection{Neural architecture search network}
Neural architecture search network, or \textbf{NASNet}, is a category of CNNs that revolutionizes network modeling, developed by \cite{2017arXiv170707012Z}.
Its ability to automatically discover high-performing architectures has remarkably reduced the need for human expertise and computational resources in the design process.
The idea behind NASNet is to utilize a reinforcement learning-based search algorithm to explore a vast search space of potential models. 
NASNet is composed of diverse essential elements. 
The main building block is a ``cell'' structure, represented by a directed acyclic graph, which captures the connectivity pattern of the neural network. 
The search algorithm learns to discover the optimal cell structure, which is then repeated multiple times to construct the complete network.
A notable characteristic of NASNet is the incorporation of skip connections, or residual connections, within its cell structure. 
These connections allow for seamless information flow and facilitate gradient propagation during training, leading to more effective learning.
Additionally, NASNet introduces reduction cells that downsample feature maps spatially, enabling the network to handle larger input images and capture high-level spatial information.
Here, we consider a smaller variant of the NASNet family called NASNetMobile.

\subsubsection{MobileNet}
\textbf{MobileNet} is a class of lightweight CNNs designed for mobile and embedded devices. 
MobileNetV3, developed by \cite{2019arXiv190502244H}, brings several critical improvements over its predecessors -- namely, MobileNetV1 \citep{2017arXiv170404861H} and MobileNetV2 \citep{2018arXiv180104381S} -- to deliver enhanced performance while maintaining efficiency, catering to the limitations of resource-constrained devices.
The architecture of MobileNetV3 incorporates an efficient backbone structure consisting of various lightweight layers, including depthwise separable convolutions, pointwise convolutions, and linear bottleneck layers. 
SE blocks are also employed to recalibrate channel-wise features adaptively, emphasizing their importance and improving the model capacity. 
In addition, the use of hard-swish activation functions introduces non-linearity while keeping the computational cost low.
By leveraging these components, MobileNetV3 reduces complexity while retaining the ability to capture crucial visual features. 
Additionally, the adoption of neural architecture search (NAS) allows it to automatically discover the optimal architecture through an algorithmic exploration of a vast search space. 
MobileNetV3 is available in different versions, and in this case, we employ the MobileNetV3Large, aiming to achieve higher accuracy with a sacrifice of slightly larger model size and calculation requirements compared to its smaller variant.

\subsubsection{EfficientNet}
\textbf{EfficientNet} is a family of CNN architectures presented by \cite{2019arXiv190511946T} that has demonstrated outstanding performance across various computer vision tasks, including image classification, object detection, and semantic segmentation. 
The EfficientNet models are designed to attain cutting-edge performance while being computationally efficient, requiring fewer parameters and computations compared to other architectures.
The pivotal innovation behind EfficientNet is the concept of compound scaling, which uniformly scales the network's depth, width, and resolution, allowing for an optimal trade-off between model size and performance.
The EfficientNet models also employ other techniques to enhance performance, such as the use of mobile inverted bottleneck convolutional (MBConv) layers and a compound coefficient for controlling the number of channels in each layer. 
Further improvement is then proposed by \cite{2021arXiv210400298T} by introducing a new convolutional operation called ``Fused-MBConv'', combining depthwise separable convolutions with inverted bottleneck residual connections.
These approaches additionally improve the efficiency and effectiveness of the models.
Here, we adopt the simplest form of EfficientNet, represented by EfficientNetB0 and EfficientNetV2B0.
Other variants of this family (B1--B7, S--XL, etc.) introduce progressive scaling to increase the model sizes and complexity. 

\subsubsection{Regularized network}
Regularized Network, hereafter \textbf{RegNet}, is a family of CNNs introduced by \cite{2020arXiv200313678R} to address the challenge of network scaling by promoting a design principle that improves both accuracy and efficiency, utilizing adaptive regularization of weights and adjustment of scaling coefficients. 
This approach applies regularization proportional to the magnitude of network weights, which helps prevent overfitting and improves generalization performance \citep{2021arXiv210306877D}.
RegNet architectures also comprise a channel-wise group convolution technique, effectively reducing the computational cost without significantly sacrificing accuracy.

We will implement here two of the smallest RegNet variants.
The first one is RegNetX002, which emphasizes the model depth as the primary scaling factor, has deeper layers, and aims to capture more complex patterns in the data.
The second one is RegNetY002, which prioritizes the model width as the scaling factor, has a wider network and seeks to capture more fine-grained details in the data with increasing feature diversity.

\subsubsection{Vision transformer}
Vision transformer, shortened as \textbf{ViT}, is a state-of-the-art deep learning architecture introduced by \cite{2020arXiv201011929D}. 
It brings the powerful transformer-based architecture, originally designed for natural language processing, to the field of computer vision. 
It also represents a significant departure from traditional CNNs by relying solely on self-attention mechanisms without any convolutional layers.
The principal concept of ViT is to treat an image as a sequence of patches, where each patch is regarded as a token. 
These image patches are flattened and fed into a transformer encoder consisting of multiple stacked self-attention layers and feed-forward neural networks. 
The self-attention mechanism allows the model to capture global dependencies and relationships between different patches in the image, enabling it to learn contextual information and high-level representations.
Since the transformer architecture does not inherently encode spatial information, positional encoding is introduced to provide the model with the relative positions of the image patches. 
This information helps the model understand the spatial structure of the image and retain spatial relationships between different patches during the self-attention process.

However, ViT has shown to be more data-hungry, meaning it typically requires larger amounts of labeled training data to achieve competitive performance compared to CNNs. 
This phenomenon is partly due to its reliance on self-attention mechanisms and the challenges in capturing fine-grained spatial details.
The combination of ViT with Shifted Patch Tokenization (SPT) and Locality Self-Attention (LSA) is an alternative modification that aims to improve the efficiency and effectiveness of ViT models \citep{2021arXiv211213492L}.
SPT shift the image patches by half of their size horizontally and vertically.
This method improves the alignment between the patches and the objects in the image, enhancing the model's ability to capture accurate spatial information. 
On the other hand, LSA restricts the network attention to a local neighborhood of tokens instead of attending to all of them, significantly reducing the computational cost while still capturing relevant contextual information.
Hence, we will use the original ViT model (dubbed as ViT-Vanilla) and ViT with the implementation of SPT and LSA (named as ViT-Lite) for constructing our ensemble network.


\subsection{Training the Network Models}

\begin{figure}[htb!]
	\centering
	\resizebox{\hsize}{!}{\includegraphics{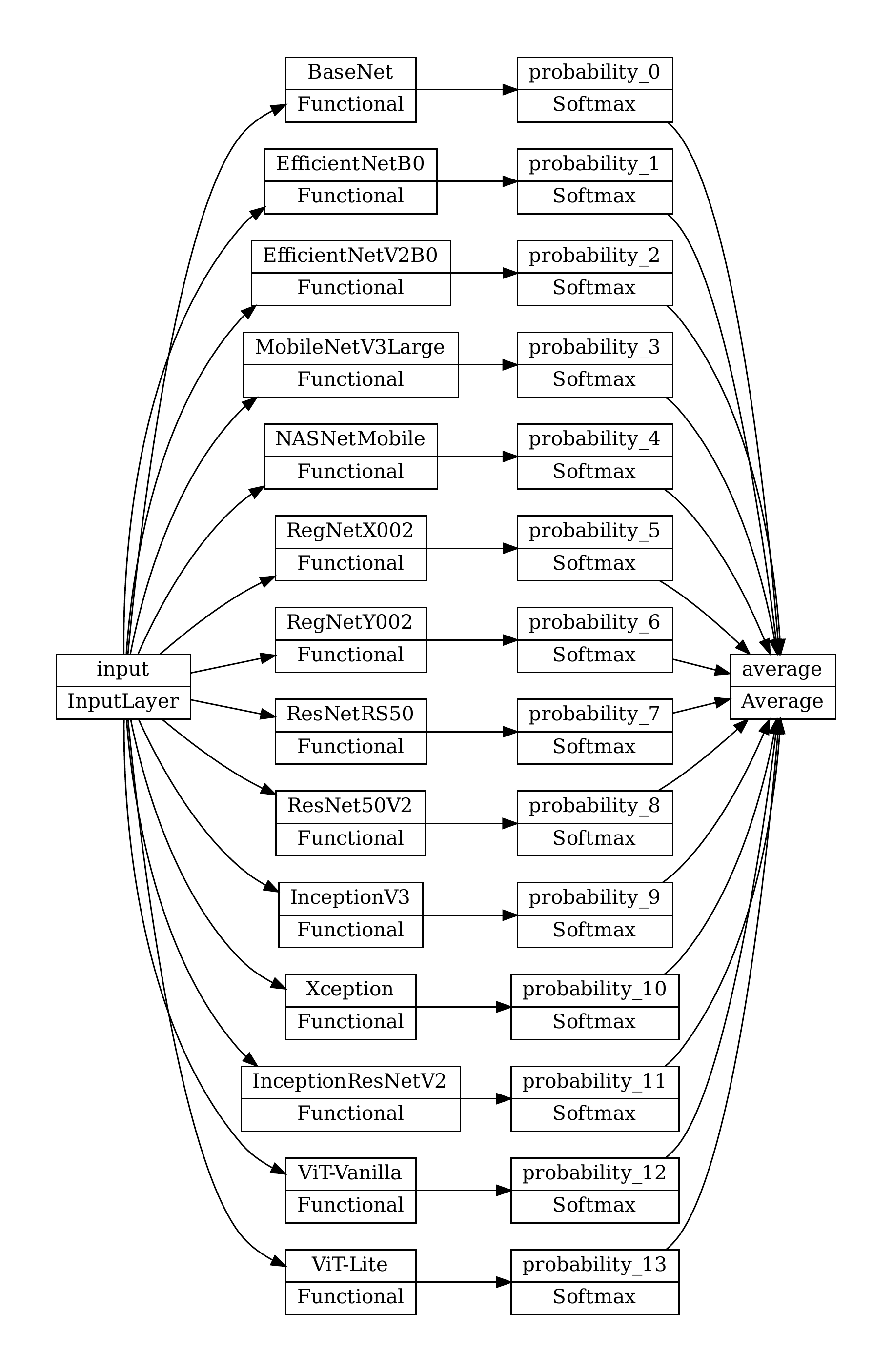}}
	\caption{
		Architecture of our ensemble network, which produces 4 outcomes as the end product -- to be exact, the probability of a candidate being a lensed quasar, a galaxy, an unlensed quasar, and a star -- is shown here.
		The input layer takes the HSC $grizy$-band images and passes them to the functional layers containing 12 CNN and 2 ViT models.
		Each model produces 4 output logits, which are then transformed into a probability distribution that sums to unity using the softmax layer.
		Subsequently, we average the softmax outputs of these 14 networks to produce only 4 final probabilities.
	}
	\label{fig:ensemble_model}
\end{figure}

The CNN and ViT models discussed before need to be separately trained first.
Once the optimum parameters for each classifier are obtained, we will combine them to construct the ensemble networks.
This strategy aims to improve predictive power by averaging the forecasts of multiple instances into a unified and robust decision-making framework.
One shortcoming of this technique is that each model contributes the same proportion to the ensemble forecast, irrespective of how well the network performs. 
A weighted average ensemble is a version of this strategy that weights the role of every ensemble member by its performance on the test dataset. 
This scheme allows high-performing classifiers to contribute more while low-performing models influence less.
This technique may be further generalized by substituting the linear weighted sum model used to integrate the sub-model predictions with any learning algorithm and correspondingly known as layered generalization or stacking. 
A stacked generalization ensemble, as compared to a weighted average ensemble, can utilize the set of forecasts as a context and dynamically select how to weigh the input predictions, possibly leading to higher performance \citep{2021arXiv210402395G}.
However, for simplicity and considering that: (1) manually assigning each of the model contributions in the weighted average ensemble is not straightforward and (2) the stacked generalization ensemble usually needs more independent datasets for its training to prevent overfitting, we will use the model averaging ensemble without the weighing approach instead, as shown in Figure~\ref{fig:ensemble_model}.

Training the network models involves several steps, including data splitting, batch subdivision, and iterative forward and backward propagation.
In this case, the input data is split into training, validation, and test datasets using a ratio of 70:20:10, allowing for proper evaluation and testing of the trained models.
The dataset is further subdivided into smaller batches consisting of 128 samples to facilitate efficient computation by processing a subset of the data at a time.
During the training process, the models undergo iterative forward and backward propagation. 
Forward propagation involves passing the input data through the network and generating predictions. 
These calculations are then compared to the ground-truth labels using the sparse categorical cross-entropy\footnote{\url{https://www.tensorflow.org/api_docs/python/tf/keras/losses/SparseCategoricalCrossentropy}}, a loss function that is often used for multilabel classification tasks. 
The backpropagation step estimates the gradients of the loss function, which are then utilized to update the corresponding weights and biases employing a stochastic gradient descent technique \citep[e.g., Adam optimizer;][]{2014arXiv1412.6980K}. 
This optimization process allows the models to iteratively adjust their parameters, improving their predictions and overall performance.

\begin{figure}[htb!]
	\centering
	\resizebox{\hsize}{!}{\includegraphics{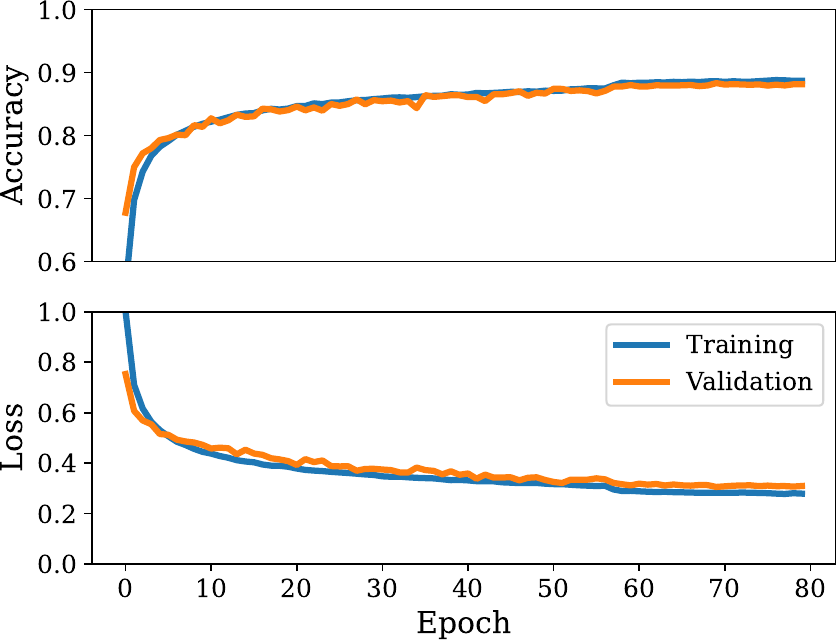}}
	\caption{
		Example of accuracy and loss curves as a function of the training epoch of ViT-Lite, one of the top-performing models, is displayed in the upper and lower panels.
		The other classifiers also show a similar increasing accuracy and decreasing loss trend, indicating that they are able to learn.
		These metrics are inferred by testing the network on the training and validation datasets, which are subsequently depicted as blue and orange lines, respectively.
	}
	\label{fig:accuracy_loss}
\end{figure}

The training and validation losses need to be monitored to check whether the model is able to learn or if overfitting occurs -- in other words, where the network becomes too specialized to the training data and fails to generalize well to new, unseen data.
In an attempt to prevent overfitting and achieve more accurate predictions, we randomly shuffle the training and validation data after each epoch.
At the start, the learning rate is adjusted to $10^{-4}$ while the bias and weight of each neuron are set randomly and subsequently updated during the training.
When learning becomes stagnant, the network models often benefit from reducing the learning rate by a factor of 2--10. 
Due to this reason, we apply a callback function, which lowers the model learning rate by a factor of ten if the loss curve shows a plateau for five consecutive epochs.
After numerous epochs, we stop the training if the lowest average validation loss over multiple runs is reached, or to put it another way, early stopping is applied if the loss difference fails to decrease below $10^{-4}$ over ten consecutive epochs.
Typically, the training cycle could reach 50 to 100 epochs before the optimization is converged, and the classifier performance can not be improved further (see Figure~\ref{fig:accuracy_loss} as reference).
The best model is then stored, which corresponds to the combination of the weights and other parameters that generate the lowest cross-entropy loss tested in the validation dataset.

\subsection{Classifier performance evaluation}

Each of our classifiers will return the probability estimates for an individual tensor that is passed into the networks, indicating whether the respective images contain a lensed quasar, an unlensed galaxy, an ordinary quasar, or a star -- to be exact, $P_\mathrm{lens}$, $P_\mathrm{galaxy}$, $P_\mathrm{quasar}$, and $P_\mathrm{star}$ -- where the sum of this probabilities equals to unity. 
The predicted category is then allocated by selecting the class with the highest likelihood score. 
It is worth noting that $P_\mathrm{lens} = 1$ indicates that there is a strong probability that the categorized images include a lensed quasar.
$P_\mathrm{lens} = 0$, on the other hand, indicates that the cutouts do not comprise a lensed quasar and are more likely to contain contaminating sources. 

As commonly perceived, the excellent accuracy achieved in the training process might have been attributed to overfitting. 
Naturally, we then examine the accuracy-loss learning curves obtained by assessing network predictions on validation and training datasets (see Figure~\ref{fig:accuracy_loss} for an example). 
The training and validation losses are settling and maintaining the same trends after declining for numerous epochs. 
The absence of any overfitting signals -- specifically, the training loss stays to decline while the validation loss starts rising after several epochs -- gives us confidence that our classifier can generalize and learn well.

In order to evaluate the overall performance of the trained models further, another commonly used metric is the receiver operating characteristic (ROC) curve.
The area under the ROC curve (AUROC) provides insights into how effectively a binary classifier distinguishes between two classes as the decision threshold is adjusted.
Therefore, to utilize the ROC curve and calculate the AUROC, we mark lenses as the positive (P) and non-lenses or contaminating sources as the negative (N) cases. 
True positives (TP) are instances in which the model properly predicts the lenses, distinct from true negatives (TN), which are accurate identifications of non-lenses. 
False positives (FP) emerge when the classifier wrongly labels contaminants as lenses. 
Finally, false negatives (FN) are occasions in which the model incorrectly rejects lenses. 
The ROC curve compares the false-positive rate (FPR) to the true-positive rate (TPR) for the unseen test dataset, where:
\begin{equation}
	\mathrm{
		TPR = \frac{TP}{P} = \frac{TP}{TP + FN}
	};\quad
	\mathrm{
		FPR = \frac{FP}{N} = \frac{FP}{FP + TN}
	}.
\end{equation}

\begin{figure}[htb!]
	\centering
	\resizebox{\hsize}{!}{\includegraphics{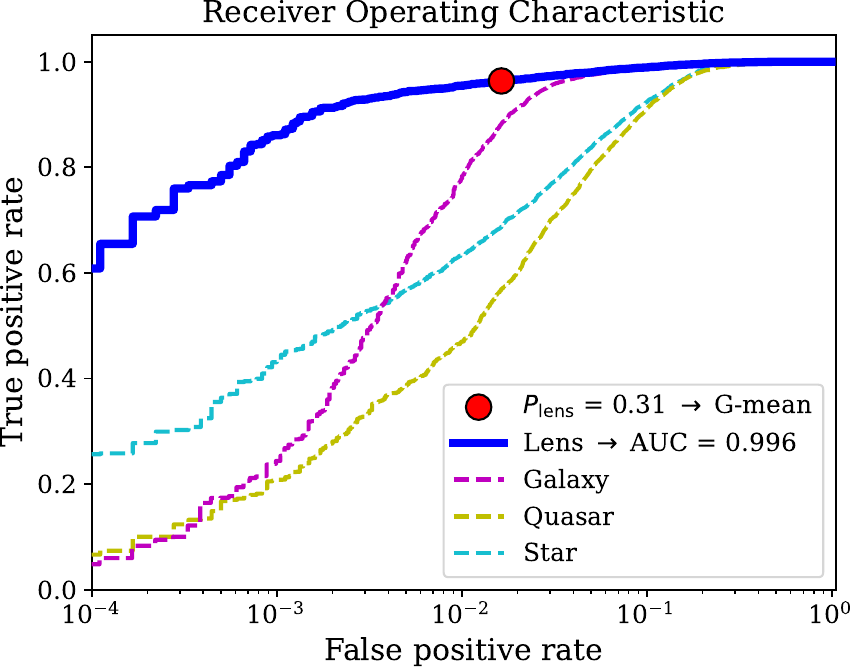}}
	\caption{
		Curves of the receiver operating characteristic and its area under the curve (AUC) calculation are presented.
		A solid blue line indicates the curve for categorizing lensed quasars.
		The curves for classifying galaxies, quasars, and stars, are also illustrated as magenta, yellow, and cyan dashed lines, respectively.
		Finally, a red circle represents the FPR and TPR values for the chosen $P_\mathrm{lens}$ limit.
	}
	\label{fig:roc_curve}
\end{figure}

The ROC curve is then made by gradually raising the probability cutoff from 0 to 1.
This result in AUROC = 1 for a flawless classifier, while AUROC = 0.5 for a classifier that only predicts randomly. 
Since we have four categories for classifying the candidates (i.e., a multilabel classification), we have to binarize each network's prediction using the so-called ``one versus all'' framework. 
As a result, we generate four ROC curves constructed based on the evaluation of the classifier using a previously unseen test dataset and display them in Figure~\ref{fig:roc_curve}, which encompasses the following scenarios:
(1) distinguishing lensed quasars from galaxies and other point-source contaminants, represented by the solid blue line; 
(2) discriminating galaxies from lens systems and other contaminating sources, symbolized by the dashed magenta line;
(3) separating stars from other sources, illustrated by the dashed yellow line;
and (4) classifying quasars with respect to lenses, galaxies, and other sources, expressed by the dashed cyan line. 
Notably, these curves exhibit high AUROC values, indicating the exceptional performance of the classifier in these scenarios.
We then employ the geometric mean or G-mean\footnote{The definition is G-mean~=~$\sqrt{\mathrm{TPR} \times (1-\mathrm{FPR})}$} metric to find a balance of the TPR and FPR ratios based on the ROC curves. 
The highest G-mean score shows an ideal $P_\mathrm{lens}$ threshold for maximizing TPR while reducing FPR. 
In this situation, the reasonable $P_\mathrm{lens}$ and the resulting FPR and TPR for each network model and the combined classifiers are reported in Table~\ref{tab:network_performance}.

It is noteworthy to mention that below the previously mentioned $P_\mathrm{lens}$ limit, the amount of candidates increases rapidly while their quality declines, which means that the visual assessment required at the next stage will be more tedious and less practical. 
Concerning the compromise involving completeness and purity, we have to strike a balance in which the quantity of candidates is reasonable for follow-up observations at the next step.

\section{Results and discussion} \label{sec:result}

\subsection{Final list of lens candidates} \label{sec:final_lens}

After the catalog-level preselection explained before and employing our ensemble network classification, we obtained 3080 surviving targets with $P_\mathrm{lens} > 0.3$.
To further remove the contaminating sources from this list, we consider the relevant astrometric excess noise \citep[AEN;][]{2016A&A...595A...1G} and proper motion significance \citep[PMSIG;][]{2019MNRAS.483.4242L} parameters when the astrometry of these lens candidates is available in the Gaia Data Release 3 catalog \citep{2022arXiv220800211G}.
We note that only about 30\% of our candidates have this astrometric information.
A high value of AEN ($\gtrsim10$~mas) could point to a potential star-forming galaxy, while a significant \mbox{PMSIG} number ($\gtrsim10\sigma$) strongly implies that the system includes a star \citep{2019MNRAS.483.4242L}.
These additional criteria manage to select 2604 sources (1801 unique systems/groups), which are then visually inspected, yielding 210 remaining candidates with high lens probabilities.

We further split these candidates into A and B grades.
Grade A is assigned when the cutout exhibits a definite configuration of strong lensing, even without the assistance of a higher-resolution image. 
This case implies the presence of multiple-imaged sources or the indication of a counter-image, along with the existence of a possible lens galaxy.
On the other hand, grade B means that the candidates display a potentially lensing-like configuration, although visual identification of multiple images is not feasible. 
This category includes scenarios where multiple objects or a single arc-like object are positioned on one side of the central galaxy without a clear counter-image visible on the opposite. 
In addition, we performed a random check on some sources with $P_\mathrm{lens} \leq 0.3$ and found that on some occasions, they show lensing features and are missed by our classifier.
Accordingly, we mark these objects with grade A* or B*, depending on their visual-based quality.
We stress that the list in this category might be incomplete since not all sources with $P_\mathrm{lens} \leq 0.3$ are inspected.


Interestingly, our compilation of candidates encompasses not only lensed quasars but also galaxy-galaxy lenses.
These strongly lensed galaxies are not included explicitly in the training dataset.
However, the lensed point-source lights from the multiply-imaged quasars sometimes could mimic the extended arcs of the galaxy-galaxy lens systems and confuse our ensemble classifier.
Therefore, more information beyond the optical images is required to discriminate between these systems, and we will attempt to check their SEDs later (see Appendix~\ref{sec:sed_fitting}).
In addition, we provide the complete list of our lens candidates in Table~\ref{tab:lens_candidates} of Appendix~\ref{sec:lens_candidates}.

Based on the evaluation applied to the test dataset and reported in Table~\ref{tab:network_performance}, our ensemble classifier appears to have an FPR as low as 1.6\% for recognizing lensed quasars at $z = 1.5$--7.2.
Yet, how many lensed quasars that we expect to discover?
Using the latest estimate from \cite{2022AJ....163..139Y}, assuming an $i$-band 5$\sigma$ depth of $\approx$26~mag, and sky coverage of around $1300\deg^2$, we expect to find about 153 lensed quasars, including 13 quadruply-imaged sources, within the HSC wide-layer footprint.
Approximately 80\% (50\%) of these systems have a separation greater than 0\farcs5 (1\arcsec).
We remark that this number is about two to three times lower compared to the earlier model from \cite{2010MNRAS.405.2579O}, primarily caused by the discrepancy in the details of the simulated quasars.
\citeauthor{2010MNRAS.405.2579O} adopt a steeper faint-end slope of the quasar luminosity function, do not impose an absolute magnitude cut, and disregard redshift evolution of the deflector velocity dispersion function \citep[VDF;][]{2007ApJ...658..884C}.
On the other hand, an improvement from \citeauthor{2022AJ....163..139Y} is made by considering quasars with $i$-band absolute magnitude of $M_i<-20$ and employing a new VDF that decreases with redshift \citep{2019ApJS..245...26K}.

\begin{figure}[htb!]
	\centering
	\centering
	\resizebox{\hsize}{!}{\includegraphics{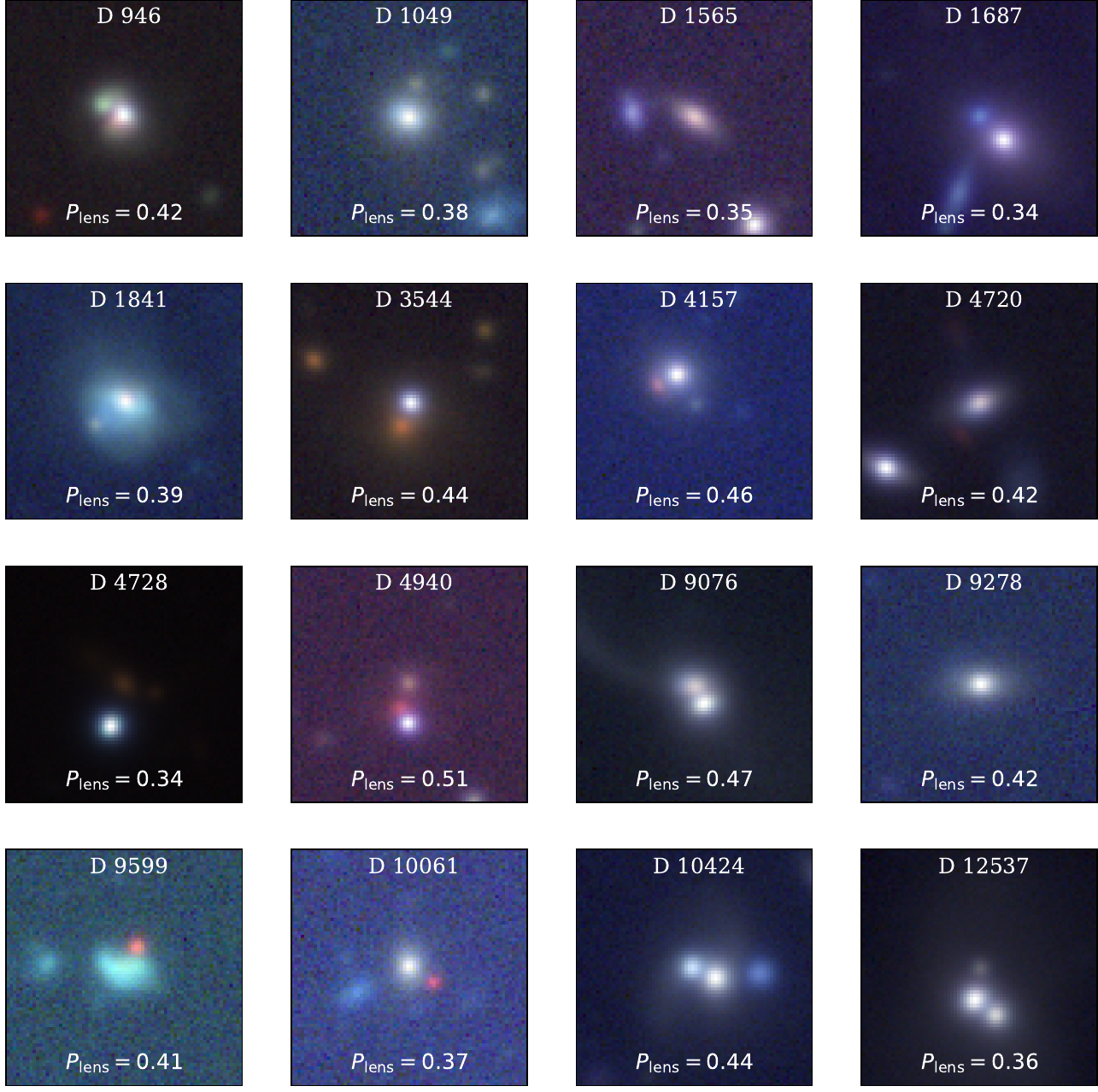}}
	\caption{
		FP sources, or contaminants, identified by our classifier and their lens probabilities are shown here.
		The images are created based on the HSC $grz$-band cutouts with 72 pixels ($\approx$12\arcsec) on a side, colorized, and square-root stretched.
	}
	\label{fig:bad_target}
\end{figure}

\begin{table*}[htb!]
	\small
	\caption{
		Performances of the individual and ensemble network models evaluated on the test dataset are reported here. 
		We train the models using computing nodes that contain Intel Xeon E5-2680 2.4 GHz, 28 cores, and 252 GiB RAM each.
	}
	\label{tab:network_performance}
	\centering
	\begin{tabular}{ccccccccc}
		\hline\hline
		Architecture & Parameters & CPU Hours & TPR & FPR & AUROC & Candidates & Lens & Model Reference \\
		\hline
		BaseNet & 581,828 & 104 & 0.906 & 0.097 & 0.973 & 62,326 & 13 & \cite{2023ApJ...943..150A} \\
		RegNetX002 & 2,338,692 & 1202 & 0.924 & 0.072 & 0.983 & 160,852 & 17 & \cite{2020arXiv200313678R} \\
		RegNetY002 & 2,816,896 & 1317 & 0.899 & 0.078 & 0.977 & 192,797 & 13 & \cite{2020arXiv200313678R} \\
		MobileNetV3Large & 3,000,484 & 148 & 0.938 & 0.046 & 0.989 & 190,808 & 16 & \cite{2019arXiv190502244H} \\
		EfficientNetB0 & 4,055,275 & 314 & 0.945 & 0.042 & 0.991 & 147,705 & 19 & \cite{2019arXiv190511946T} \\
		NASNetMobile & 4,274,520 & 434 & 0.948 & 0.048 & 0.991 & 169,309 & 17 & \cite{2017arXiv170707012Z} \\
		EfficientNetV2B0 & 5,925,012 & 243 & 0.958 & 0.030 & 0.994 & 151,319 & 15 & \cite{2021arXiv210400298T} \\
		ViT-Vanilla & 9,208,772 & 1185 & 0.949 & 0.050 & 0.990 & 24,642 & 16 & \cite{2020arXiv201011929D} \\
		ViT-Lite & 9,230,060 & 3499 & 0.960 & 0.029 & 0.994 & 12,872 & 16 & \cite{2021arXiv211213492L} \\
		Xception & 20,870,252 & 1041 & 0.970 & 0.017 & 0.997 & 216,620 & 17 & \cite{2016arXiv160207261S} \\
		InceptionV3 & 21,811,556 & 549 & 0.967 & 0.020 & 0.996 & 186,190 & 17 & \cite{2015arXiv151200567S} \\
		ResNet50V2 & 23,579,268 & 1094 & 0.967 & 0.021 & 0.996 & 164,722 & 18 & \cite{2016arXiv160305027H} \\
		ResNetRS50 & 33,705,060 & 1245 & 0.966 & 0.021 & 0.996 & 155,262 & 16 & \cite{2021arXiv210307579B} \\
		InceptionResNetV2 & 54,343,460 & 1630 & 0.966 & 0.014 & 0.996 & 150,435 & 18 & \cite{2016arXiv161002357C} \\
		\hline
		Ensemble & 195,741,135 & \nodata & 0.963 & 0.016 & 0.996 & 3080 & 16 & This work \\
		\hline
	\end{tabular}
	\tablefoot{
		Columns, from left to right, correspond to the architecture name, the number of parameters in the model, the total CPU hours required to train the network, the resulting TPR, the FPR, the AUROC, the number of selected lens candidates, the number of recovered known lensed quasars, and the literature explaining the architecture design.
	}
\end{table*}

While our current strategy appears to yield reasonable candidate samples, we believe there is an opportunity for optimization. 
The occurrence of FP-classified sources is, at present, not zero. 
Fortunately, trained astronomers can swiftly rule out the spurious sources in this candidate list, as presented in Figure~\ref{fig:bad_target}.
Based on our visual inspection, they are highly improbable to be lenses or have no apparent indicators of strong lensing characteristics.
Moreover, they have a variety of visible forms, such as irregular galaxies, spiral arms, or groups of multiple sources that imitate lensing arcs. 
Aside from that, sources with unusual morphologies that do not belong to either category in the data used for the training can get unexpected network-based classification scores. 
Due to these reasons, our network models have been trained iteratively by including the sample of identified FPs to keep improving the classifier performances.

\subsection{Selection of high-redshift lensed quasars}
Lensed quasars at $z\gtrsim6$ is another intriguing case we want to explore because so far, only 2 lenses have been found at this distance among $\approx$300 known quasars throughout the whole sky \citep{2019ApJ...870L..11F,2023AJ....165..191Y}.
Subsequently, we expect to discover at least 2 lensed quasars in the current dataset if we consider the high-$z$ lensed quasar fraction of $\approx$1\% \citep{2022ApJ...925..169Y} among $\sim$150 quasars at $z\gtrsim6$ that have been found within the HSC footprint \citep{2022ApJS..259...18M}.
However, this estimate could be much larger, depending on the chosen model.
For example, \cite{2019ApJ...870L..12P} suggest a lens fraction of $>$4\%, which results in the expected number of lenses of more than 6.
This tension might be the result of unaccounted-for biases that have not been thoroughly examined, in particular, the disparity between the adopted quasar luminosity function and deflector VDF used by \citeauthor{2022ApJ...925..169Y} and \citeauthor{2019ApJ...870L..12P}.

In retrospect, the reasons why many lenses are left undiscovered might be very evident. 
Most methods to select quasar candidates have included extra magnitude cuts or full ``dropout'' criteria at all bandpasses bluer than the Ly$\alpha$ emission \citep[e.g.,][and reference therein]{2022ApJS..259...18M,2022arXiv221206907F}.
This approach makes sense because the light emitted by $z\gtrsim6$ quasars at wavelengths shorter of Ly$\alpha$ is severely absorbed by the foreground IGM, forming a prominent break in the spectrum and thus a primary marker for the quasar preselection. 
To put it another way, we are not anticipating any substantial flux emanating from the $g$ or $r$ bands for quasars at these redshifts. 
This characteristic, however, is not true if the lens galaxies exist at $0.1 \lesssim z \lesssim 1.5$, which might produce substantial emission at the observed wavelengths of $\lambda_\mathrm{obs} \lesssim 8000$.

Testing our methodology against prior quasar selection approaches validates the expanded selection space. 
If we imposed an extra cut of S/N($g$, $r) < 5$, or the dropout criterion, none of the mock lenses generated in Section~\ref{sec:lens_simulation} would survive. 
In simpler terms, we would overlook all lens systems featuring luminous galaxies as deflectors.
Hence, to include prospective lens systems, we have eliminated such dropout requirements in our photometric preselection.
Instead, we exploit the whole spatial and color information by using the associated multiband images and process them with our deep learning classifier.

The intersection with earlier search methods occurs when the deflectors are fainter and less massive, resulting in compact lenses with small $\theta_\mathrm{E}$ and image separations.
These systems are expected to have quasar-dominated light and may exhibit a slightly extended shape in ground-based images. 
However, when their lensing mass is below a certain threshold, conventional quasar search techniques can still detect them, but demonstrating their true lensing nature becomes challenging. 
Employing higher-resolution, space-based imaging would greatly widen the range of parameters accessible for this diagnostic.

Accordingly, as an effort to: 
(1) detect compact, faint lenses with the light dominated by bright $z\gtrsim6$ quasars,
(2) identify quasars in the cases that they are well separated ($\gtrsim0\farcs5$) from the associated foreground deflectors,
and (3) recognize unwanted sources (e.g., binary stars and quasar-star pairs) solely using their catalog-level photometric information, 
we also implement a standard object classification using the SED fitting method.
Our approach involves pinpointing candidates of quasars, galaxies, and stars based on their multiwavelength data, estimating photometric redshifts, and reassessing our final list of targets (see Appendix~\ref{sec:sed_fitting} for more details).
However, we did not discover any new $z\gtrsim6$ quasars using the current dataset and selection method.
Instead, all of our lens candidates are located at lower redshifts. 
In future work, we aim to relax the current photometric preselection to uncover high-$z$ lensed quasars. 

\subsection{Quasar selection completeness}

\begin{figure}[htb!]
	\centering
	\centering
	\resizebox{\hsize}{!}{\includegraphics{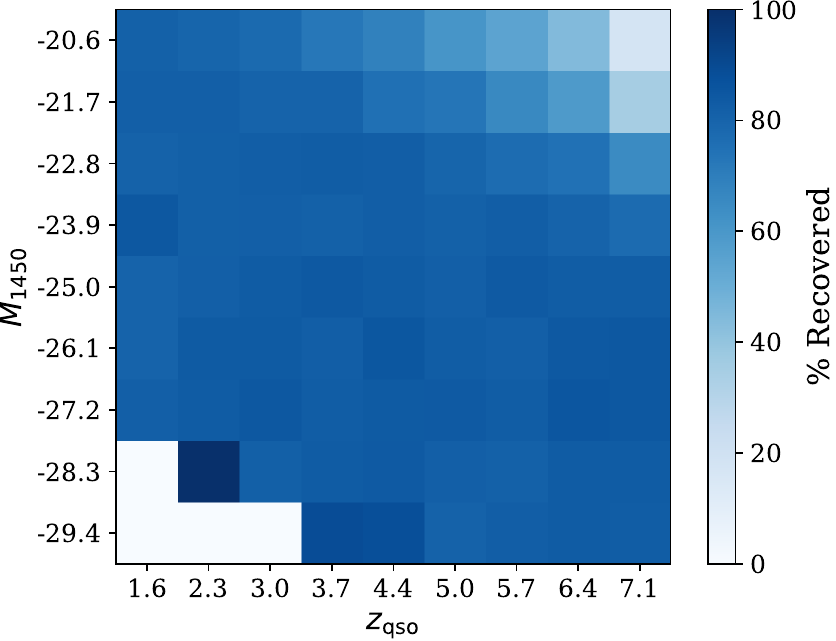}}
	\caption{
		Model of the quasar selection function that we make use of in this work. 
		The percentage of recovery rate corresponds to the number of mock quasars in each ($M_{1450}$, $z$) bin that is successfully recovered by our selection method.
	}
	\label{fig:completeness}
\end{figure}

As previously stated, the parent population of quasars in our simulation follows a uniform distribution within the redshifts of $1.5 \leq z \leq 7.2$ and absolute magnitudes of $-30 \leq M_{1450} \leq -20$.
Without the contribution of strong gravitational lensing, our ensemble network classifier can only discover quasars with $M_{1450}\lesssim-22$ at $z\gtrsim6$. 
Fortunately, the lensing event could shift this boundary towards a lower luminosity territory, depending on the factor of magnification values.
To examine this in more depth, we initially define our selection function (or completeness) as the proportion of simulated quasars with specific $M_{1450}$, $z$, and intrinsic SEDs that are successfully identified by our selection criteria. 
The outcomes are displayed in Figure~\ref{fig:completeness}. 
Quasars with inherently low brightness can only be detected if they experience a substantial magnification boost from gravitational lensing, and such occasions are relatively rare. 
Therefore, as the quasars possess inherently lower luminosity, our completeness rate diminishes at a given redshift.
As an extra note, since in Section~\ref{sec:lens_simulation} we exclude sources with $y$-band magnitude $>$15~mag to discard unusually bright objects or saturated images, we missed all intrinsically luminous quasars with $M_{1450} \lesssim -27$ at low redshifts ($z\lesssim3$), which might not even exist in the real universe.

Then, looking at Figure~\ref{fig:lensgal_dist}, our classifier could recover most of the lensing configurations without a significant bias.
A slight decrease in recovery rate is apparent in the cases of systems with deflector redshifts of $z_\mathrm{gal} \gtrsim1$ and magnitudes of $i\gtrsim21$, which might be caused by farther lens galaxies having fainter apparent fluxes and more challenging to be detected.
Though these sources could still be identified if the lensed quasars are well separated, these phenomena can be exacerbated for the lens systems that are too compact due to small lens masses.
For example, we notice a slight drop in lens recovery rate in Figure~\ref{fig:lensgal_dist} for systems with $\sigma_v \lesssim 100$~km~s$^{-1}$ and $\theta_\mathrm{E} \lesssim 0\farcs5$.

\subsection{Evaluation with independent datasets} \label{sec:known_lenses_eval}

\begin{figure*}[htb!]
	\centering
	\centering
	\resizebox{\hsize}{!}{\includegraphics{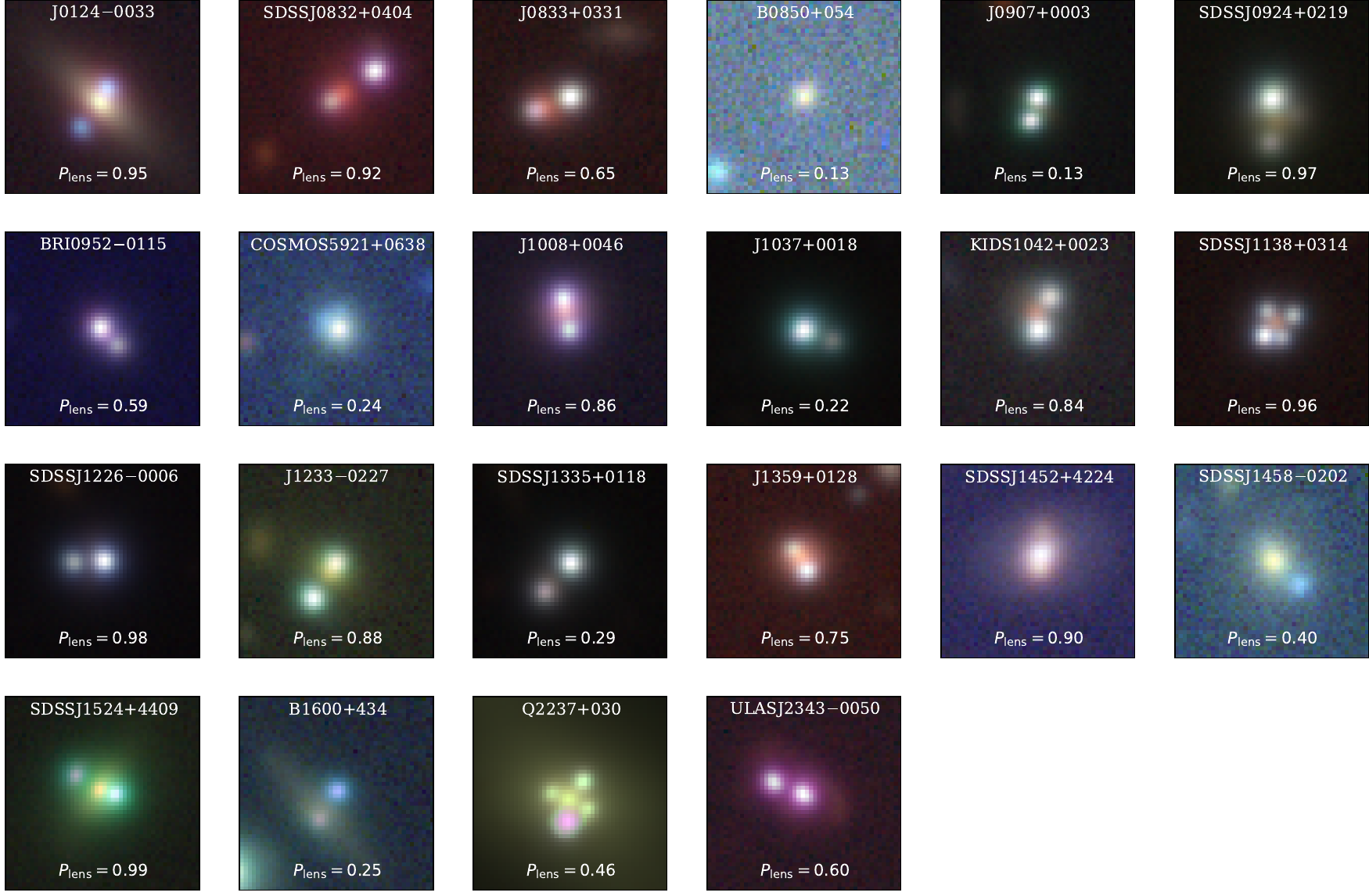}}
	\caption{
		Example of discovered lenses compiled from the literature (see main text). 
		The images are generated using $48\times48$ pixels ($\approx$8$\arcsec$ on a side) of HSC $grz$ cutouts, colorized, and square-root stretched. The lens probabilities, assigned by our ensemble classifier, of each system are also shown in the figure.
	}
	\label{fig:CLemon_test}
\end{figure*}

We also conduct a supplementary test with an alternative dataset assembled using the list of known quasars compiled in the GLQD \citep{1992Gemin..36....1M,1995MNRAS.274L..25J,2003AJ....126..666I,2004PASJ...56..399O,2007AJ....133..214M,2008AJ....135..520O,2008AJ....135..496I,2008MNRAS.387..741J,2012AJ....143..119I,2015MNRAS.448.1446A,2016MNRAS.456.1595M,2018MNRAS.475.2086A,2018MNRAS.477L..70W,2018MNRAS.480.1163S,2018MNRAS.479.5060L,2019MNRAS.483.4242L,2019arXiv191208977K,2020A&A...636A..87C,2020MNRAS.494.3491L,2021MNRAS.502.1487J,2023MNRAS.520.3305L}.
In this case, HSC images are available for about 22 of the 220 lenses in the database (see Figure~\ref{fig:CLemon_test}).
To evaluate the completeness and purity of our ensemble classifier, we first combine these known lenses with a sample of contaminants compiled in Section~\ref{sec:training_input}, which includes galaxies, stars, and quasars, with the non-lensed sources expected to outweigh the lens population by a factor of a few thousand.
As an outcome, our model correctly identifies 16 known lensed quasars, resulting in a TPR (or completeness) of 72.7\% and an FPR of 1.6\%.
In addition, we discover that the classifier has a purity\footnote{
	Assuming that the fraction of lenses among all sources in the universe is in the order of $\rm S=10^{-3}$, or corresponds to 1 lens per 1000 objects, we then define the purity as $\rm AP = TPR \times S / (TPR \times S + FPR \times (1-S))$.
} of 5.6\% in identifying the lens candidates.

Other individual model explained in Section~\ref{sec:network_architecture} demonstrates impressive performance when assessed against the test dataset, with an AUROC exceeding 97.3\% and a median false positive rate as low as 3.6\%.
However, they still encounter difficulties in generalizing to real-world data. 
This thing is evident from the presence of numerous spurious sources identified by each classifier. 
For example, InceptionResNetV2, ResNet50V2, and EfficientNetB0 are three best-performing classifiers, recovering 18 to 19 known lensed quasars -- that is, a completeness of 81.8--86.4\%.
Yet, these networks give high scores to more than a hundred thousand sources, making the required visual inspection for the resulting lens candidates at the later stage extremely time-consuming.
Better selection purity is achieved by ViT-Vanilla and ViT-Lite, which have a completeness of 72.7\% while keeping the number of lens candidates in the order of ten to twenty thousand.
Then, a much better improvement is uncovered by ensembling the predictions of all of the CNNs and ViTs via model averaging, leading to a reduction in impurities by up to a factor of $\approx50$.
This ensemble network manages to reduce the number of candidates to just a few thousand while maintaining the completeness of 72.7\%.


When evaluated against real lens systems, the model seems to perform worse than when tested against simulated lenses, which is indicated by the declining TPR from 96.3\% to 72.7\% (see Table~\ref{tab:network_performance}).
This lower performance is somewhat expected and might be attributed to the uniqueness of some of the lens systems that are not taken into account by our simulation. 
Some missed lenses might contain arcs or counter images that are too dim to be recognized, contamination of the bright deflector lights, compact lenses, saturated images, or other factors. 
Still, our classifier can generalize and obtain a high enough accuracy for our objectives, where the purpose of our network-based classification is to reduce the false candidates as much as possible before proceeding with the visual inspection and compiling the final list of targets.

In addition, to assess the flexibility of our ensemble model when applied to the next-generation ground-based survey data, we perform one more test using a mock dataset that closely resembles the LSST photometry \citep{2019ApJ...873..111I}.
This dataset, kindly provided by \cite{2022AJ....163..139Y}, encompasses 3628 strongly-lensed quasars at $z\gtrsim5$. 
Employing our ensemble model directly, we successfully identified 3095 systems of the parent sample, yielding a completeness of 85.3\%. 
We expect that applying transfer learning by retraining our classifier on a subset of this dataset could increase the network performance further \citep[e.g.,][]{2022A&A...664A...4T,2022MNRAS.515.5121B}.
This outcome illustrates the adaptability and robustness of our ensemble model, showcasing its ability to excel in the upcoming LSST data.

\section{Summary and conclusion} \label{sec:conclusion}

In this paper, we conduct a systematic hunt for lensed quasars at $1.5 \leq z \leq 7.2$ by exploiting the HSC, UKIRT, VISTA, unWISE, and Gaia data.
Our approach is divided into two key stages. 
First, we use catalog-level information to preselect the candidates based on their photometric color, decreasing the number of sources from $\sim$60~million to only 892,609. 
Second, we use an ensemble of CNN and ViT classifiers to assess the relative likelihood of each source being a lens or contaminant, yielding 3080 prevailing candidates.
It is worth noting that the training input is created by overlaying deflected point-source lights on the images of real HSC galaxies. 
This strategy allows us to generate realistic strong-lens simulations and concentrate on identifying systems within the Einstein radii of $\theta_\mathrm{E}<5\arcsec$. 
We then obtain 210 newly found lens candidates after inspecting their astrometric data when available and visually evaluating the objects with the lens probability of $P_\mathrm{lens} > 0.3$.
These findings indicate that automated neural network-based classifiers, with minimal human involvement, are promising for spotting lensed quasars in big datasets.

The technique presented in this paper is readily applicable to seeking out galaxy-quasar lenses across a wide range of redshifts.
It also appears to be suitable for next-generation surveys such as Euclid \citep{2011arXiv1110.3193L,2022A&A...662A.112E}, which will provide high-resolution NIR imagery over a large portion of the extragalactic sky, and Rubin Observatory Legacy Survey of Space and Time \citep{2019ApJ...873..111I}, which will have extensive optical multiband data. 
In this case, modifications to the bandpass profiles, seeing values, and image scale will be essential to obtain optimal results. 
In addition, adopting more complex galaxy mass profiles beyond the SIE model might also help to enhance the classifier performance. 
To fully exploit the scientific potential of our catalog of lenses, it is essential to conduct spectroscopic observations to confirm the redshifts of the deflectors and sources, along with high-resolution imaging to perform accurate lens modeling.

\begin{acknowledgements}
	
We thank Alessandro Sonnenfeld, the referee of this paper, for his constructive and insightful comments.
We are also grateful to James Chan and Simona Vegetti for their valuable contributions and fruitful discussions, which have significantly improved the quality of this manuscript.
This research is supported in part by the Excellence Cluster ORIGINS, which is funded by the Deutsche Forschungsgemeinschaft (DFG, German Research Foundation) under Germany's Excellence Strategy -- EXC-2094 -- 390783311.
SHS thanks the Max Planck Society for support through the Max Planck Fellowship. 
SS acknowledges financial support through grants PRIN-MIUR 2017WSCC32 and 2020SKSTHZ.
YS acknowledges the support from the China Manned Spaced Project (No. CMS-CSST-2021-A07 and CMS-CSST-2021-A12).
ATJ is supported by the Program Riset ITB 2023.

This work is based on data collected at the Subaru Telescope and retrieved from the HSC data archive system, operated by the Subaru Telescope and Astronomy Data Center at the National Astronomical Observatory of Japan.
The Hyper Suprime-Cam (HSC) collaboration includes the astronomical communities of Japan, Taiwan, and Princeton University. 
The HSC instrumentation and software were developed by the National Astronomical Observatory of Japan (NAOJ), the Kavli Institute for the Physics and Mathematics of the Universe (Kavli IPMU), the University of Tokyo, the High Energy Accelerator Research Organization (KEK), the Academia Sinica Institute for Astronomy and Astrophysics in Taiwan (ASIAA), and Princeton University. 
Funding was contributed by the FIRST program from the Japanese Cabinet Office, the Ministry of Education, Culture, Sports, Science and Technology (MEXT), the Japan Society for the Promotion of Science (JSPS), the Japan Science and Technology Agency (JST), the Toray Science Foundation, NAOJ, Kavli IPMU, KEK, ASIAA, and Princeton University. 
This paper makes use of software developed for the Large Synoptic Survey Telescope. 
We thank the LSST Project for making their code available as free software at \url{http://dm.lsst.org}.

This project has included data from the Sloan Digital Sky Survey (SDSS).
Funding for SDSS-IV has been provided by the Alfred P. Sloan Foundation, the U.S. Department of Energy Office of Science, and the Participating Institutions. 
SDSS-IV acknowledges support and resources from the Center for High-Performance Computing at the University of Utah. 
The SDSS website is \url{https://www.sdss.org/}.
SDSS-IV is managed by the Astrophysical Research Consortium for the Participating Institutions of the SDSS Collaboration.

The unWISE catalog utilized in this paper is based on data products from the Wide-field Infrared Survey Explorer, which is a joint project of the University of California, Los Angeles, and NEOWISE, which is a project of the Jet Propulsion Laboratory/California Institute of Technology. 
WISE and NEOWISE are funded by the National Aeronautics and Space Administration.

We acknowledge the use of the VHS, VIKING, UKIDSS, and UHS data.

This work has made use of data from the European Space Agency (ESA) mission Gaia (\url{https://www.cosmos.esa.int/gaia}), processed by the Gaia Data Processing and Analysis Consortium (DPAC, \url{https://www.cosmos.esa.int/web/gaia/dpac/consortium}). 
Funding for the DPAC has been provided by national institutions, in particular, the institutions participating in the Gaia Multilateral Agreement.

\end{acknowledgements}

\tiny{
	\noindent
	\textit{Facilities.} ESO:VISTA (VIRCAM), Gaia, Sloan (eBOSS/BOSS), Subaru (HSC), UKIRT (WFCAM), WISE.
}

\vspace{1mm}

\tiny{
	\noindent
	\textit{Software.}
	Astropy \citep{2013A&A...558A..33A,2018AJ....156..123A},
	EAZY \citep{2008ApJ...686.1503B},	
	Matplotlib \citep{2021zndo....592536C},
	NumPy \citep{2020Natur.585..357H},
	Pandas \citep{2022zndo...3509134R},
	PyAutoLens \citep{2021JOSS....6.2825N},	
	SIMQSO \citep{2013ApJ...768..105M},
	TensorFlow \citep{2016arXiv160508695A,2022zndo...4724125D}.
}

%
   \bibliographystyle{aa} 
   \bibliography{biblio} 
%

\onecolumn
\begin{appendix}
\section{Eliminating spurious sources with spectral modeling} \label{sec:sed_fitting}
We implement SED fitting as an additional practice to discover unresolved lens systems with the light dominated by the background quasar and to remove spurious sources in our lens search. 
In principle, we want to separate the candidates of quasars, galaxies, and stars based on their multiwavelength data and estimate the associated photometric redshifts.
Therefore, the \texttt{eazy-py}\footnote{\url{https://github.com/gbrammer/eazy-py}} module, a Pythonic photometric redshift tool based on \texttt{EAZY} \citep{2008ApJ...686.1503B}, will be used to implement the SED modeling. 
The way it works is by going across a grid of spectral templates, matching them to the photometry of the targets, and trying to discover the best model. 
Here, we select the best models with the lowest reduced chi-square ($\chi^2_\mathrm{red}$) as solutions.
The candidates with a high likelihood of being quasars are chosen based on the derived $\chi^2_\mathrm{red}$ of the quasar ($\chi^2_\mathrm{red, q}$), galaxy ($\chi^2_\mathrm{red, g}$), brown dwarf ($\chi^2_\mathrm{red, d}$), and star ($\chi^2_\mathrm{red, s}$) templates along with their associated ratios.
Hence, the sources which are best fitted with models of stars or brown dwarfs will be removed from our list of lens candidates.

To establish the templates, we first compile the brown dwarf spectra from the SpeX Prism Library\footnote{\url{http://pono.ucsd.edu/~adam/browndwarfs/spexprism/library.html}} \citep{2014ASInC..11....7B}.
This database contains 360 spectra of M5--M9, L0--L9, and T0--T8 stars with wavelength spanning from 0.625\,$\mu$m to 2.55\,$\mu$m.
Following the prescription of \cite{2020ApJ...903...34A}, we then extend these templates into the wavelength covered by unWISE bands -- namely, W1~(3.4\,$\mu$m) and W2~(4.6\,$\mu$m).
We also add the stellar models\footnote{\url{http://www.eso.org/sci/facilities/paranal/decommissioned/isaac/tools/lib.html}} provided by \cite{1998PASP..110..863P}, which contains 131 spectra in the range of 1150--25,000~\AA\ to include the SED of main sequence stars.

After that, we utilize the latest version of the XMM-COSMOS galaxy and active galactic nucleus (AGN) SEDs \citep{2007ApJ...663...81P,2009ApJ...690.1250S,2011ApJ...742...61S}.
These templates are discussed in detail by \cite{2017ApJ...850...66A} as part of their work on estimating the redshift of X-ray AGNs.
Similar to what has been done by \cite{2021A&A...648A...4D}, we then apply the dust reddening by employing the attenuation levels of $0 \leq A_V \leq 2$, following the \cite{2000ApJ...533..682C} extinction law.
While the original template list covers a wide range of galaxy spectral types, we only pick the SEDs of luminous quasars dominated by the continuum and broad-line emissions for our purpose.

Next, to ascertain that our targets do not resemble unlensed galaxies, we fit them with the set of SEDs established from the Flexible Stellar Population Synthesis code \citep[FSPS;][]{2009ApJ...699..486C,2010ApJ...708...58C,2010ApJ...712..833C,2010ascl.soft10043C}. 
These templates are composed of a combination of stellar lights, nebular lines, and MIR dust-reprocessed emissions. 
They also encompass ultraviolet to infrared wavelengths and contain information about stellar population properties, such as ages, metallicities, and initial mass functions.

\begin{figure*}[htb!]
	\centering
	\resizebox{\hsize}{!}{\includegraphics{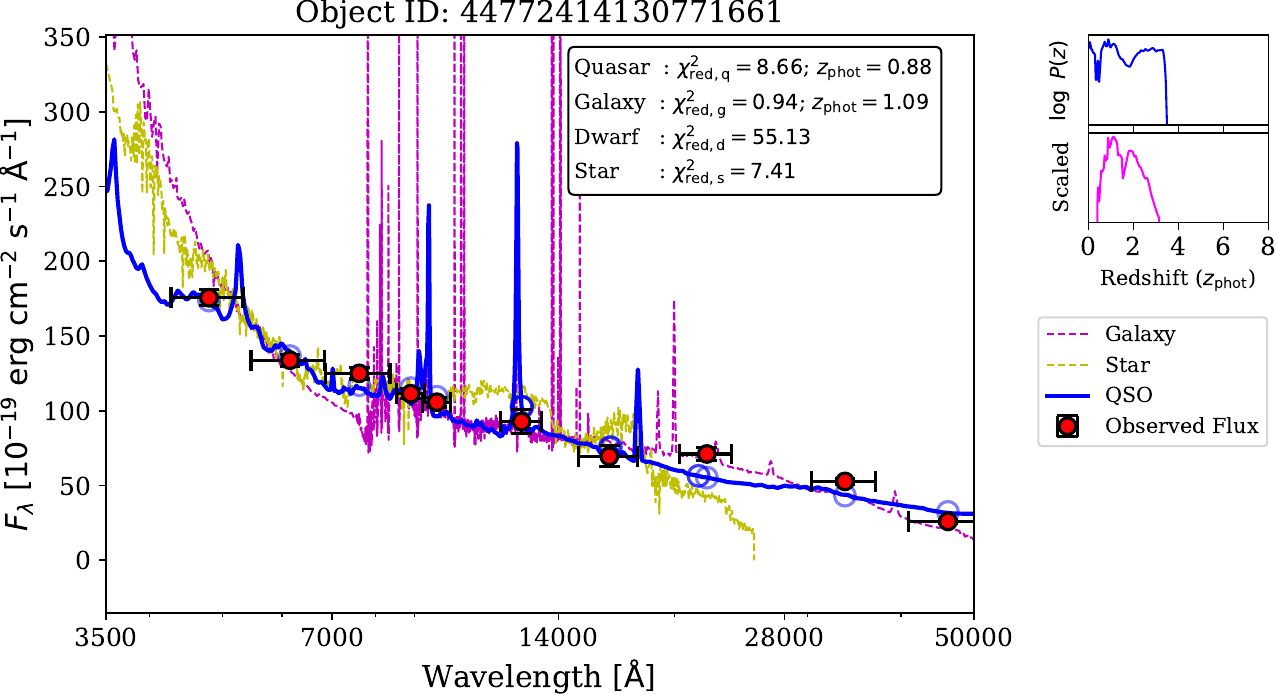}}
	\caption{
		SED fitting result for a lensed quasar candidate is presented. 
		In the main panel, the observed photometry of the source is depicted as red circles with error bars modeled with three distinct templates. 
		The best-fit quasar spectral template is represented by a blue line, while the synthesized photometry is illustrated with blue circles. 
		The yellow and magenta colors indicate the best-fit models using non-lensed star and galaxy templates, respectively. 
		The right panel showcases the photometric redshift probability density functions obtained by fitting the data to templates of unlensed quasars (cyan line) and galaxies (magenta line).
	}
	\label{fig:sed_fit}
\end{figure*}

A grid of SED sets is then assembled by distributing the quasar and galaxy templates in the redshifts of $0 \leq z \leq 8$ with a step size of $\Delta z = 0.005$. 
We have to note that in these SED models, we incorporate the attenuation produced by the H\,\textsc{i} in the IGM using the analytical approximation from \cite{2014MNRAS.442.1805I}.
While the star, brown dwarf, and quasar templates are fitted utilizing the single template mode in \texttt{eazy-py}, the galaxy models are matched to each source's photometry with non-negative linear combinations, allowing each component to contribute to the fit.
An example of our SED fitting result is displayed in Figure~\ref{fig:sed_fit}.

\section{Complete list of lens candidates} \label{sec:lens_candidates}
\begin{figure*}[htb!]
	\centering
	\resizebox{\hsize}{!}{\includegraphics{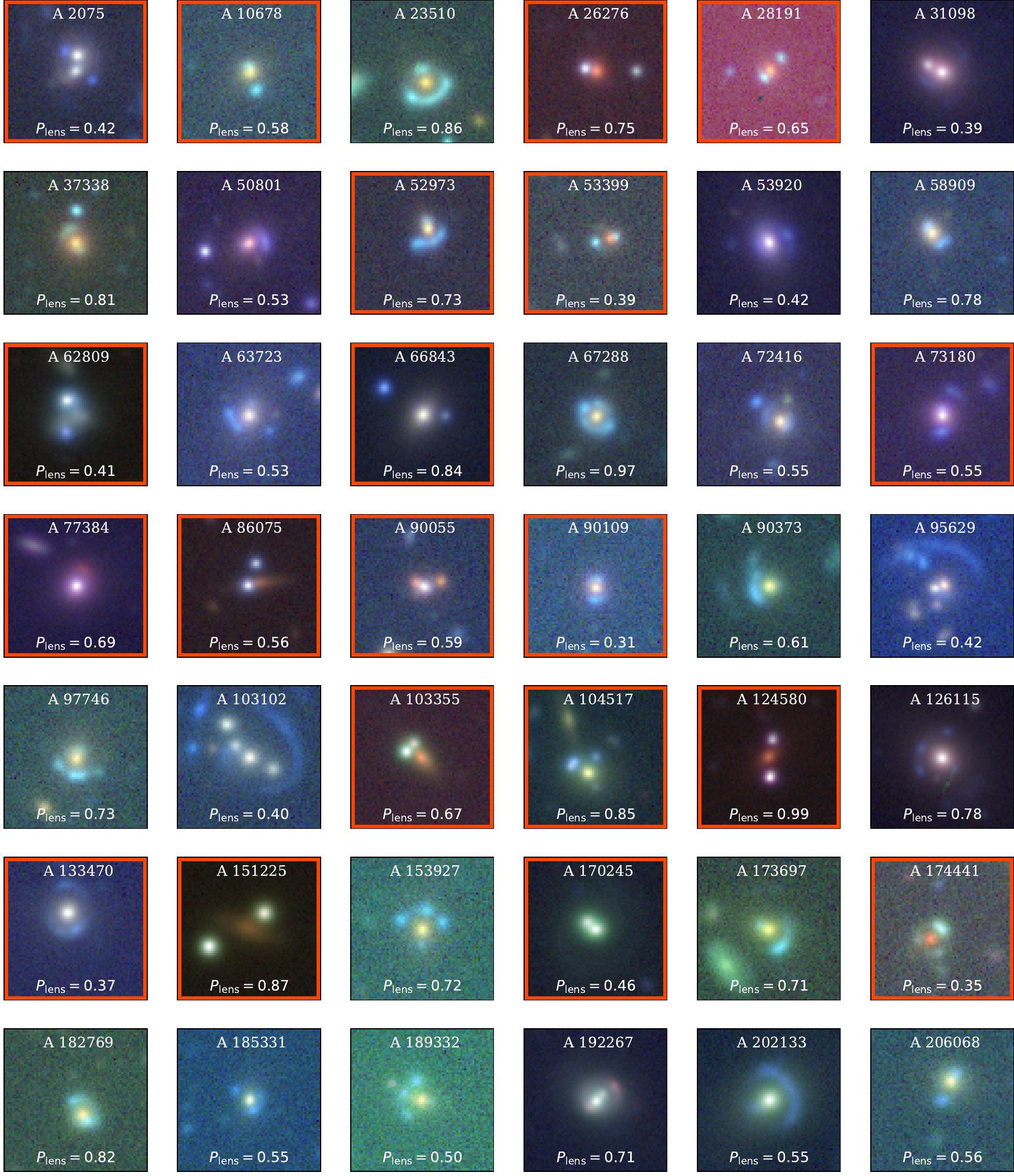}}
	\caption{
		The $12\arcsec\times12\arcsec$ $grz$-color images of our lens candidates are shown.
		To enhance the visual contrast of each HSC cutout, we apply a square-root stretch to the fluxes.
		We also display the grade determined from our visual inspection and the identification number of each target on top of the panel.
		At the bottom side, we report the lens probability inferred by our automated classifier.
		Newly discovered sources in this work, along with some candidates that are also independently published by \cite{2023arXiv230405425C} that are not listed in the MLD and GLQD, are marked with red rectangles.
	}
	\label{fig:lens_candidates}
\end{figure*}

We present here the complete list of our lens candidates selected in the main text. 
The photometry measured within a 2\arcsec\ aperture diameter of each system is reported in Table~\ref{tab:lens_candidates}, while the associated color images are displayed in Figure~\ref{fig:lens_candidates}.

\begin{center}
\centering
\scriptsize
\begin{longtable}{cccccccccccc}
	\caption{List of rediscovered strong lenses and newly found lensed quasar candidates.} \\
	\label{tab:lens_candidates} \\
	\hline\hline
	ID & Name & $g$ & $r$ & $i$ & $z$ & $y$ & $J$ & W1 & $P_\mathrm{lens}$ & Grade & Reference \\
	\hline
	\endfirsthead
	\caption{continued.}\\
	\hline\hline
	ID & Name & $g$ & $r$ & $i$ & $z$ & $y$ & $J$ & W1 & $P_\mathrm{lens}$ & Grade & Reference \\
	\hline
	\endhead
	\hline
	\endfoot
	2075 & J000.67141+02.81442 & $22.89 \pm 0.01$ & $21.13 \pm 0.01$ & $20.20 \pm 0.00$ & $19.85 \pm 0.01$ & $19.60 \pm 0.01$ & $18.95 \pm 0.05$ & $16.68 \pm 0.01$ & 0.42 & A & \nodata \\
	10678 & J003.99032$-$00.17125 & $23.73 \pm 0.02$ & $22.39 \pm 0.01$ & $21.07 \pm 0.01$ & $20.53 \pm 0.00$ & $20.28 \pm 0.01$ & $19.61 \pm 0.11$ & $18.15 \pm 0.01$ & 0.58 & A & \nodata \\
	23510 & J015.65975+01.98275 & $23.69 \pm 0.02$ & $23.32 \pm 0.02$ & $22.68 \pm 0.02$ & $22.24 \pm 0.02$ & $22.01 \pm 0.04$ &  \nodata  &  \nodata  & 0.86 & A & [1], A \\
	26276 & J016.54800+00.00320 & $23.77 \pm 0.03$ & $22.35 \pm 0.01$ & $21.15 \pm 0.00$ & $20.86 \pm 0.01$ & $20.48 \pm 0.01$ & $19.80 \pm 0.10$ & $18.94 \pm 0.03$ & 0.75 & A & \nodata \\
	28191 & J017.13796+00.54755 & $23.62 \pm 0.02$ & $22.40 \pm 0.01$ & $21.49 \pm 0.01$ & $20.99 \pm 0.01$ & $20.69 \pm 0.01$ & $20.09 \pm 0.11$ & $18.76 \pm 0.03$ & 0.65 & A & \nodata \\
	31098 & J018.11680+01.28198 & $22.35 \pm 0.01$ & $20.64 \pm 0.00$ & $19.78 \pm 0.00$ & $19.54 \pm 0.00$ & $19.24 \pm 0.00$ & $18.55 \pm 0.03$ & $17.44 \pm 0.01$ & 0.39 & A & [2], B \\
	37338 & J020.07557+00.19049 & $23.30 \pm 0.02$ & $21.87 \pm 0.01$ & $20.76 \pm 0.00$ & $20.51 \pm 0.00$ & $20.17 \pm 0.01$ & $19.85 \pm 0.14$ & $18.46 \pm 0.02$ & 0.81 & A & [3], A \\
	50801 & J029.07556$-$01.12976 & $23.02 \pm 0.01$ & $21.38 \pm 0.01$ & $20.40 \pm 0.00$ & $20.01 \pm 0.00$ & $19.74 \pm 0.01$ & $19.19 \pm 0.07$ & $18.05 \pm 0.01$ & 0.53 & A & [3], B \\
	52973 & J029.38138$-$03.51638 & $22.31 \pm 0.01$ & $22.05 \pm 0.01$ & $22.01 \pm 0.01$ & $21.94 \pm 0.02$ & $21.87 \pm 0.03$ &  \nodata  &  \nodata  & 0.73 & A & \nodata \\
	53399 & J029.44110$-$05.46330 & $23.25 \pm 0.02$ & $22.98 \pm 0.02$ & $22.88 \pm 0.02$ & $22.63 \pm 0.03$ & $22.90 \pm 0.08$ &  \nodata  &  \nodata  & 0.39 & A & \nodata \\
	53920 & J029.51855$-$00.48850 & $21.87 \pm 0.00$ & $20.20 \pm 0.00$ & $19.54 \pm 0.00$ & $19.27 \pm 0.00$ & $19.03 \pm 0.00$ & $18.58 \pm 0.04$ & $17.96 \pm 0.01$ & 0.42 & A & [3], C \\
	58909 & J030.19785$-$03.73786 & $23.62 \pm 0.02$ & $23.14 \pm 0.02$ & $22.54 \pm 0.01$ & $22.21 \pm 0.02$ & $21.91 \pm 0.03$ &  \nodata  &  \nodata  & 0.78 & A & [1], A \\
	62809 & J030.71468+01.72214 & $20.24 \pm 0.00$ & $20.35 \pm 0.00$ & $20.05 \pm 0.00$ & $20.18 \pm 0.00$ & $20.13 \pm 0.01$ &  \nodata  &  \nodata  & 0.41 & A & \nodata \\
	63723 & J030.83909$-$01.27991 & $24.15 \pm 0.03$ & $21.92 \pm 0.01$ & $20.70 \pm 0.00$ & $20.23 \pm 0.00$ & $20.02 \pm 0.01$ & $19.27 \pm 0.03$ & $18.20 \pm 0.02$ & 0.53 & A & [3], B \\
	66843 & J031.26944$-$01.05439 & $21.44 \pm 0.00$ & $19.86 \pm 0.00$ & $19.21 \pm 0.00$ & $18.94 \pm 0.00$ & $18.69 \pm 0.00$ & $18.30 \pm 0.03$ & $17.93 \pm 0.01$ & 0.84 & A & \nodata \\
	67288 & J031.32085$-$01.38902 & $22.49 \pm 0.01$ & $21.53 \pm 0.00$ & $20.51 \pm 0.00$ & $20.10 \pm 0.00$ & $19.89 \pm 0.00$ & $19.20 \pm 0.03$ & $18.20 \pm 0.02$ & 0.97 & A & [4], B \\
	72416 & J032.04487$-$02.33828 & $24.00 \pm 0.03$ & $23.28 \pm 0.02$ & $22.72 \pm 0.02$ & $22.36 \pm 0.02$ & $22.03 \pm 0.03$ &  \nodata  &  \nodata  & 0.55 & A & [1], B \\
	73180 & J032.14510$-$00.29402 & $22.87 \pm 0.01$ & $21.33 \pm 0.00$ & $20.39 \pm 0.00$ & $20.10 \pm 0.00$ & $19.76 \pm 0.00$ & $19.20 \pm 0.05$ & $18.25 \pm 0.02$ & 0.55 & A & \nodata \\
	77384 & J032.70043+00.09788 & $22.15 \pm 0.01$ & $20.37 \pm 0.00$ & $19.72 \pm 0.00$ & $19.48 \pm 0.00$ & $19.13 \pm 0.00$ & $18.48 \pm 0.03$ & $17.42 \pm 0.01$ & 0.69 & A & \nodata \\
	86075 & J033.83087+01.46423 & $21.35 \pm 0.00$ & $20.84 \pm 0.00$ & $20.54 \pm 0.00$ & $20.37 \pm 0.00$ & $20.28 \pm 0.01$ & $19.81 \pm 0.10$ & $19.01 \pm 0.03$ & 0.56 & A & \nodata \\
	90055 & J034.35612$-$01.98059 & $23.15 \pm 0.02$ & $22.06 \pm 0.01$ & $20.88 \pm 0.00$ & $20.42 \pm 0.00$ & $20.20 \pm 0.01$ & $19.55 \pm 0.08$ & $18.04 \pm 0.01$ & 0.59 & A & \nodata \\
	90109 & J034.36326$-$00.77953 & $23.92 \pm 0.02$ & $22.70 \pm 0.01$ & $21.58 \pm 0.01$ & $21.02 \pm 0.01$ & $20.71 \pm 0.01$ & $20.01 \pm 0.14$ & $18.74 \pm 0.02$ & 0.31 & A & \nodata \\
	90373 & J034.40482$-$05.22486 & $23.36 \pm 0.01$ & $21.92 \pm 0.01$ & $20.67 \pm 0.00$ & $20.18 \pm 0.00$ & $19.99 \pm 0.00$ & $19.35 \pm 0.03$ & $18.19 \pm 0.01$ & 0.61 & A & [3], B \\
	95629 & J035.12880+00.73936 & $24.38 \pm 0.04$ & $22.12 \pm 0.01$ & $20.83 \pm 0.00$ & $20.08 \pm 0.00$ & $19.91 \pm 0.00$ & $19.19 \pm 0.07$ & $17.42 \pm 0.01$ & 0.42 & A & [5], B \\
	97746 & J035.41724$-$02.17223 & $23.72 \pm 0.02$ & $22.30 \pm 0.01$ & $21.10 \pm 0.00$ & $20.63 \pm 0.00$ & $20.37 \pm 0.01$ & $19.46 \pm 0.06$ & $18.25 \pm 0.02$ & 0.73 & A & [1], B \\
	103102 & J036.04321$-$03.60151 & $24.21 \pm 0.03$ & $22.32 \pm 0.01$ & $20.90 \pm 0.00$ & $20.40 \pm 0.00$ & $20.11 \pm 0.01$ & $19.15 \pm 0.02$ &  \nodata  & 0.40 & A & [3], A \\
	103355 & J036.07595+02.78420 & $23.42 \pm 0.03$ & $21.71 \pm 0.01$ & $20.64 \pm 0.01$ & $20.28 \pm 0.01$ & $19.97 \pm 0.01$ & $18.87 \pm 0.05$ & $17.86 \pm 0.01$ & 0.67 & A & [6], A \\
	104517 & J036.23320+01.45236 & $23.22 \pm 0.02$ & $22.85 \pm 0.02$ & $22.25 \pm 0.01$ & $22.07 \pm 0.01$ & $21.66 \pm 0.03$ &  \nodata  &  \nodata  & 0.85 & A & \nodata \\
	124580 & J038.94324$-$02.32565 & $22.88 \pm 0.01$ & $21.43 \pm 0.01$ & $20.58 \pm 0.00$ & $20.25 \pm 0.00$ & $20.16 \pm 0.01$ & $19.51 \pm 0.05$ & $18.81 \pm 0.02$ & 0.99 & A & \nodata \\
	126115 & J039.15545$-$03.53893 & $21.09 \pm 0.00$ & $19.60 \pm 0.00$ & $19.10 \pm 0.00$ & $18.82 \pm 0.00$ & $18.56 \pm 0.00$ & $17.87 \pm 0.02$ & $17.58 \pm 0.01$ & 0.78 & A & [1], A \\
	133470 & J128.29814+02.75330 & $23.36 \pm 0.02$ & $22.40 \pm 0.01$ & $21.77 \pm 0.01$ & $21.49 \pm 0.01$ & $21.31 \pm 0.01$ &  \nodata  &  \nodata  & 0.37 & A & \nodata \\
	151225 & J132.11535+00.80992 & $22.82 \pm 0.01$ & $21.63 \pm 0.01$ & $20.85 \pm 0.00$ & $20.55 \pm 0.00$ & $20.28 \pm 0.01$ & $19.66 \pm 0.03$ & $18.56 \pm 0.02$ & 0.87 & A & \nodata \\
	153927 & J132.69420+00.65150 & $25.53 \pm 0.10$ & $23.11 \pm 0.02$ & $21.89 \pm 0.01$ & $20.97 \pm 0.01$ & $20.73 \pm 0.01$ & $19.95 \pm 0.07$ & $18.52 \pm 0.02$ & 0.72 & A & [3], A \\
	170245 & J135.64663+03.75000 & $22.07 \pm 0.00$ & $20.75 \pm 0.00$ & $19.93 \pm 0.00$ & $19.73 \pm 0.00$ & $19.65 \pm 0.00$ & $18.60 \pm 0.04$ & $18.41 \pm 0.02$ & 0.46 & A & \nodata \\
	173697 & J136.03305$-$00.99807 & $23.42 \pm 0.01$ & $22.14 \pm 0.01$ & $20.96 \pm 0.00$ & $20.55 \pm 0.00$ & $20.27 \pm 0.01$ & $19.73 \pm 0.11$ & $18.54 \pm 0.02$ & 0.71 & A & [1], A \\
	174441 & J136.12370$-$01.04093 & $23.38 \pm 0.02$ & $22.78 \pm 0.01$ & $22.57 \pm 0.01$ & $22.50 \pm 0.02$ & $22.39 \pm 0.04$ & $20.47 \pm 0.06$ &  \nodata  & 0.35 & A & \nodata \\
	182769 & J137.09317$-$01.13083 & $23.37 \pm 0.02$ & $23.29 \pm 0.02$ & $22.95 \pm 0.01$ & $22.83 \pm 0.03$ & $22.60 \pm 0.04$ &  \nodata  &  \nodata  & 0.82 & A & [1], A \\
	185331 & J137.41080+00.47847 & $23.68 \pm 0.02$ & $22.39 \pm 0.01$ & $21.11 \pm 0.00$ & $20.49 \pm 0.00$ & $20.31 \pm 0.01$ & $19.94 \pm 0.07$ & $18.56 \pm 0.02$ & 0.55 & A & [3], A \\
	189332 & J137.95382+04.31466 & $24.63 \pm 0.06$ & $22.87 \pm 0.02$ & $21.54 \pm 0.01$ & $20.85 \pm 0.01$ & $20.78 \pm 0.01$ & $20.04 \pm 0.10$ & $18.26 \pm 0.02$ & 0.50 & A & [3], A \\
	192267 & J138.37957+00.65171 & $21.47 \pm 0.00$ & $19.94 \pm 0.00$ & $19.12 \pm 0.00$ & $18.69 \pm 0.00$ & $18.55 \pm 0.00$ & $18.08 \pm 0.01$ & $16.91 \pm 0.01$ & 0.71 & A & [3], B \\
	202133 & J139.76917+03.61072 & $22.03 \pm 0.01$ & $20.53 \pm 0.00$ & $19.57 \pm 0.00$ & $19.25 \pm 0.00$ & $19.13 \pm 0.00$ & $18.56 \pm 0.04$ & $17.62 \pm 0.01$ & 0.55 & A & [1], A \\
	206068 & J140.33668+04.74140 & $23.08 \pm 0.02$ & $23.34 \pm 0.04$ & $22.49 \pm 0.02$ & $22.31 \pm 0.03$ & $22.76 \pm 0.08$ &  \nodata  &  \nodata  & 0.56 & A & [1], B \\
	212853 & J141.43748+00.28413 & $24.96 \pm 0.06$ & $22.81 \pm 0.01$ & $21.47 \pm 0.00$ & $20.87 \pm 0.00$ & $20.55 \pm 0.01$ & $19.85 \pm 0.06$ & $18.30 \pm 0.02$ & 0.33 & A & [3], A \\
	225350 & J144.24382+03.97925 & $22.69 \pm 0.01$ & $21.03 \pm 0.01$ & $20.00 \pm 0.00$ & $19.65 \pm 0.00$ & $19.46 \pm 0.00$ & $18.83 \pm 0.03$ & $17.95 \pm 0.01$ & 0.37 & A & [7], B \\
	231613 & J145.88780$-$00.72746 & $23.24 \pm 0.02$ & $23.10 \pm 0.02$ & $22.58 \pm 0.01$ & $22.51 \pm 0.02$ & $22.24 \pm 0.04$ & $19.87 \pm 0.11$ & $18.71 \pm 0.02$ & 0.49 & A & \nodata \\
	231935 & J145.95031+00.99094 & $21.44 \pm 0.00$ & $21.21 \pm 0.00$ & $20.64 \pm 0.00$ & $20.59 \pm 0.00$ & $20.38 \pm 0.01$ & $19.73 \pm 0.06$ & $18.21 \pm 0.02$ & 0.48 & A & [1], A \\
	244751 & J148.68976+03.39569 & $22.76 \pm 0.01$ & $21.61 \pm 0.00$ & $20.71 \pm 0.00$ & $20.33 \pm 0.00$ & $20.08 \pm 0.01$ & $19.52 \pm 0.05$ & $18.29 \pm 0.02$ & 0.93 & A & \nodata \\
	256582 & J151.21551$-$00.52884 & $23.21 \pm 0.02$ & $22.43 \pm 0.01$ & $21.64 \pm 0.00$ & $21.19 \pm 0.01$ & $20.89 \pm 0.01$ & $19.87 \pm 0.06$ &  \nodata  & 0.70 & A & \nodata \\
	271265 & J154.30296+00.04452 & $21.91 \pm 0.01$ & $21.27 \pm 0.00$ & $20.85 \pm 0.00$ & $20.52 \pm 0.00$ & $20.32 \pm 0.01$ & $18.06 \pm 0.03$ & $18.57 \pm 0.02$ & 0.45 & A & \nodata \\
	289037 & J158.06272$-$00.61727 & $23.91 \pm 0.03$ & $22.86 \pm 0.02$ & $21.49 \pm 0.00$ & $20.82 \pm 0.01$ & $20.67 \pm 0.01$ & $19.90 \pm 0.07$ & $18.26 \pm 0.02$ & 0.39 & A & \nodata \\
	302023 & J160.00216+00.76367 & $23.36 \pm 0.01$ & $21.97 \pm 0.01$ & $20.93 \pm 0.00$ & $20.46 \pm 0.00$ & $20.29 \pm 0.01$ & $19.64 \pm 0.07$ & $18.07 \pm 0.01$ & 0.39 & A & \nodata \\
	306069 & J160.59736+00.25589 & $22.88 \pm 0.01$ & $21.43 \pm 0.00$ & $20.38 \pm 0.00$ & $19.96 \pm 0.00$ & $19.95 \pm 0.00$ & $19.21 \pm 0.06$ & $18.09 \pm 0.01$ & 0.92 & A & [8], B \\
	327044 & J163.72886+03.74769 & $23.04 \pm 0.02$ & $22.61 \pm 0.01$ & $22.43 \pm 0.01$ & $22.21 \pm 0.02$ & $22.14 \pm 0.04$ &  \nodata  &  \nodata  & 0.46 & A & \nodata \\
	327503 & J163.80007+00.11517 & $24.07 \pm 0.04$ & $24.02 \pm 0.05$ & $23.39 \pm 0.02$ & $23.12 \pm 0.03$ & $22.78 \pm 0.07$ &  \nodata  &  \nodata  & 0.38 & A & \nodata \\
	379069 & J171.99042+04.40394 & $21.86 \pm 0.01$ & $21.64 \pm 0.00$ & $21.47 \pm 0.00$ & $21.30 \pm 0.01$ & $20.95 \pm 0.01$ &  \nodata  &  \nodata  & 0.71 & A & [6], A \\
	380704 & J172.25016$-$01.70394 & $22.61 \pm 0.01$ & $21.15 \pm 0.00$ & $19.94 \pm 0.00$ & $19.58 \pm 0.00$ & $19.40 \pm 0.00$ & $18.81 \pm 0.04$ & $17.50 \pm 0.01$ & 0.52 & A & [9], B \\
	386421 & J173.11350+03.06973 & $24.60 \pm 0.05$ & $23.09 \pm 0.02$ & $21.80 \pm 0.01$ & $21.02 \pm 0.01$ & $20.80 \pm 0.01$ & $20.19 \pm 0.12$ & $18.68 \pm 0.02$ & 0.38 & A & \nodata \\
	395957 & J174.35873+01.93274 & $22.29 \pm 0.01$ & $22.09 \pm 0.01$ & $21.98 \pm 0.01$ & $21.81 \pm 0.02$ & $21.77 \pm 0.03$ & $20.00 \pm 0.04$ &  \nodata  & 0.37 & A & \nodata \\
	398137 & J174.62120+03.96756 & $22.33 \pm 0.01$ & $20.64 \pm 0.00$ & $19.83 \pm 0.00$ & $19.46 \pm 0.00$ & $19.24 \pm 0.00$ & $18.68 \pm 0.03$ & $17.69 \pm 0.01$ & 0.82 & A & \nodata \\
	419901 & J177.23536+02.21434 & $24.11 \pm 0.03$ & $23.80 \pm 0.03$ & $23.23 \pm 0.02$ & $22.92 \pm 0.04$ & $22.69 \pm 0.06$ &  \nodata  &  \nodata  & 0.90 & A & \nodata \\
	420962 & J177.36317+02.40689 & $24.00 \pm 0.02$ & $22.30 \pm 0.01$ & $21.03 \pm 0.00$ & $20.33 \pm 0.00$ & $20.11 \pm 0.01$ & $19.63 \pm 0.06$ & $18.51 \pm 0.02$ & 0.36 & A & \nodata \\
	445898 & J180.32619$-$00.21206 & $24.97 \pm 0.07$ & $22.75 \pm 0.01$ & $21.32 \pm 0.00$ & $20.78 \pm 0.00$ & $20.51 \pm 0.01$ & $19.94 \pm 0.12$ & $18.61 \pm 0.02$ & 0.56 & A & [10], C \\
	447019 & J180.45179+01.97132 & $22.75 \pm 0.01$ & $21.04 \pm 0.00$ & $20.27 \pm 0.00$ & $20.02 \pm 0.00$ & $19.77 \pm 0.00$ & $19.25 \pm 0.05$ & $18.67 \pm 0.02$ & 0.86 & A & \nodata \\
	449138 & J180.73692+00.65845 & $23.54 \pm 0.02$ & $21.66 \pm 0.00$ & $20.38 \pm 0.00$ & $19.90 \pm 0.00$ & $19.62 \pm 0.00$ & $18.80 \pm 0.03$ & $17.43 \pm 0.01$ & 0.47 & A & [11], A \\
	456623 & J181.74491$-$00.11098 & $22.69 \pm 0.01$ & $20.91 \pm 0.00$ & $20.02 \pm 0.00$ & $19.65 \pm 0.00$ & $19.48 \pm 0.00$ & $18.14 \pm 0.01$ &  \nodata  & 0.37 & A & \nodata \\
	467002 & J183.12973+04.01508 & $24.19 \pm 0.06$ & $23.12 \pm 0.01$ & $22.03 \pm 0.01$ & $21.44 \pm 0.01$ & $21.15 \pm 0.02$ & $20.28 \pm 0.12$ & $19.04 \pm 0.03$ & 0.42 & A & \nodata \\
	479739 & J185.05108+04.11398 & $22.12 \pm 0.01$ & $20.56 \pm 0.00$ & $19.78 \pm 0.00$ & $19.60 \pm 0.00$ & $19.34 \pm 0.00$ & $18.84 \pm 0.03$ & $18.02 \pm 0.01$ & 0.68 & A & [5], C \\
	479947 & J185.07904+01.21536 & $21.84 \pm 0.00$ & $21.69 \pm 0.00$ & $21.47 \pm 0.00$ & $21.33 \pm 0.01$ & $21.39 \pm 0.01$ &  \nodata  &  \nodata  & 0.95 & A & [6], A \\
	481124 & J185.25917+00.31505 & $24.66 \pm 0.04$ & $24.10 \pm 0.04$ & $23.34 \pm 0.02$ & $22.83 \pm 0.04$ & $22.32 \pm 0.05$ &  \nodata  &  \nodata  & 0.70 & A & [1], A \\
	486410 & J186.05248+01.28670 & $22.79 \pm 0.01$ & $21.92 \pm 0.01$ & $20.75 \pm 0.00$ & $20.28 \pm 0.00$ & $20.11 \pm 0.00$ & $19.55 \pm 0.06$ & $18.05 \pm 0.01$ & 0.96 & A & [2], A \\
	495027 & J187.45272+01.61460 & $23.45 \pm 0.02$ & $21.55 \pm 0.00$ & $20.17 \pm 0.00$ & $19.67 \pm 0.00$ & $19.49 \pm 0.00$ & $18.90 \pm 0.04$ &  \nodata  & 0.80 & A & [2], B \\
	500215 & J188.20587$-$01.51725 & $23.75 \pm 0.03$ & $23.47 \pm 0.03$ & $22.99 \pm 0.02$ & $22.74 \pm 0.02$ & $22.61 \pm 0.04$ &  \nodata  &  \nodata  & 0.37 & A & \nodata \\
	516365 & J190.34810+00.26748 & $23.85 \pm 0.03$ & $23.56 \pm 0.03$ & $22.86 \pm 0.03$ & $22.46 \pm 0.03$ & $22.23 \pm 0.04$ & $19.57 \pm 0.11$ &  \nodata  & 0.65 & A & [2], B \\
	553965 & J198.39826+00.57921 & $26.73 \pm nan$ & $23.90 \pm 0.03$ & $22.60 \pm 0.01$ & $22.16 \pm 0.02$ & $21.94 \pm 0.03$ &  \nodata  &  \nodata  & 0.35 & A & \nodata \\
	559372 & J199.90890$-$00.69446 & $23.22 \pm 0.02$ & $22.18 \pm 0.01$ & $21.12 \pm 0.00$ & $20.63 \pm 0.00$ & $20.37 \pm 0.01$ &  \nodata  &  \nodata  & 0.73 & A & \nodata \\
	582855 & J205.96651+01.13814 & $22.86 \pm 0.01$ & $21.52 \pm 0.00$ & $20.41 \pm 0.00$ & $20.02 \pm 0.00$ & $19.84 \pm 0.00$ & $19.14 \pm 0.04$ & $18.12 \pm 0.01$ & 0.36 & A & [3], B \\
	598833 & J209.25645+01.07771 & $24.91 \pm 0.08$ & $22.80 \pm 0.01$ & $21.57 \pm 0.01$ & $20.74 \pm 0.01$ & $20.43 \pm 0.01$ & $19.88 \pm 0.08$ & $18.14 \pm 0.01$ & 0.56 & A & \nodata \\
	612727 & J212.06575+01.17503 & $22.23 \pm 0.01$ & $21.31 \pm 0.00$ & $20.83 \pm 0.00$ & $20.39 \pm 0.00$ & $20.34 \pm 0.01$ & $19.79 \pm 0.12$ & $18.66 \pm 0.02$ & 0.38 & A & \nodata \\
	617342 & J212.90213$-$01.03770 & $24.02 \pm 0.04$ & $23.01 \pm 0.02$ & $22.01 \pm 0.01$ & $21.22 \pm 0.01$ & $20.95 \pm 0.01$ & $20.20 \pm 0.05$ & $18.90 \pm 0.02$ & 0.60 & A & [10], A \\
	619815 & J213.25027$-$01.43561 & $23.37 \pm 0.02$ & $22.13 \pm 0.01$ & $20.86 \pm 0.00$ & $20.47 \pm 0.00$ & $20.10 \pm 0.01$ & $19.71 \pm 0.09$ & $18.37 \pm 0.02$ & 0.78 & A & [3], A \\
	631397 & J214.87668+43.69162 & $23.25 \pm 0.02$ & $21.76 \pm 0.01$ & $20.70 \pm 0.00$ & $20.36 \pm 0.00$ & $20.17 \pm 0.01$ & $19.78 \pm 0.11$ & $18.66 \pm 0.02$ & 0.61 & A & [3], B \\
	633575 & J215.20170+00.12608 & $23.81 \pm 0.02$ & $22.18 \pm 0.01$ & $21.15 \pm 0.00$ & $20.75 \pm 0.00$ & $20.62 \pm 0.01$ & $19.35 \pm 0.09$ & $18.20 \pm 0.02$ & 0.53 & A & [3], B \\
	633707 & J215.22372+00.93918 & $26.61 \pm nan$ & $26.22 \pm nan$ & $23.89 \pm 0.03$ & $23.45 \pm 0.05$ & $23.14 \pm 0.07$ & $19.50 \pm 0.03$ &  \nodata  & 0.50 & A & [12], B \\
	635073 & J215.41219+44.43825 & $22.02 \pm 0.01$ & $20.77 \pm 0.00$ & $20.11 \pm 0.00$ & $19.88 \pm 0.00$ & $19.66 \pm 0.01$ & $19.20 \pm 0.07$ & $18.23 \pm 0.01$ & 0.67 & A & [13], B \\
	637832 & J215.82575$-$00.20738 & $26.24 \pm 0.20$ & $26.28 \pm 0.32$ & $24.13 \pm 0.04$ & $23.81 \pm 0.06$ & $23.54 \pm 0.09$ &  \nodata  &  \nodata  & 0.35 & A & \nodata \\
	641575 & J216.36693$-$01.25125 & $22.88 \pm 0.01$ & $21.20 \pm 0.00$ & $20.22 \pm 0.00$ & $19.86 \pm 0.00$ & $19.63 \pm 0.00$ & $19.11 \pm 0.04$ & $18.19 \pm 0.01$ & 0.92 & A & [3], B \\
	665591 & J219.70144+00.37881 & $23.40 \pm 0.02$ & $23.21 \pm 0.02$ & $22.88 \pm 0.02$ & $22.70 \pm 0.03$ & $22.49 \pm 0.04$ & $20.22 \pm 0.16$ &  \nodata  & 0.34 & A & \nodata \\
	666055 & J219.75576+00.85495 & $23.66 \pm 0.02$ & $22.43 \pm 0.01$ & $21.13 \pm 0.00$ & $20.55 \pm 0.00$ & $20.18 \pm 0.01$ & $19.63 \pm 0.07$ & $18.06 \pm 0.01$ & 0.32 & A & [3], B \\
	670584 & J220.25645+01.56248 & $23.57 \pm 0.02$ & $23.36 \pm 0.03$ & $22.78 \pm 0.01$ & $22.44 \pm 0.03$ & $22.21 \pm 0.04$ &  \nodata  &  \nodata  & 0.56 & A & [9], B \\
	673740 & J220.62905$-$00.39813 & $22.30 \pm 0.01$ & $20.53 \pm 0.00$ & $19.71 \pm 0.00$ & $19.30 \pm 0.00$ & $19.14 \pm 0.00$ & $18.43 \pm 0.03$ & $17.63 \pm 0.01$ & 0.42 & A & [3], B \\
	676872 & J220.97918$-$00.12522 & $22.20 \pm 0.00$ & $21.62 \pm 0.00$ & $21.02 \pm 0.00$ & $20.65 \pm 0.00$ & $20.52 \pm 0.01$ & $20.10 \pm 0.14$ & $18.89 \pm 0.03$ & 0.87 & A & [11], B \\
	693983 & J223.15304$-$00.36196 & $22.73 \pm 0.01$ & $22.61 \pm 0.01$ & $22.41 \pm 0.01$ & $22.18 \pm 0.02$ & $21.92 \pm 0.03$ &  \nodata  & $19.00 \pm 0.03$ & 0.62 & A & [1], A \\
	696749 & J223.51032+43.76905 & $23.29 \pm 0.02$ & $22.09 \pm 0.01$ & $20.96 \pm 0.00$ & $20.64 \pm 0.01$ & $20.30 \pm 0.01$ & $19.70 \pm 0.09$ & $18.38 \pm 0.01$ & 0.77 & A & [1], B \\
	703052 & J224.38570$-$01.98819 & $22.10 \pm 0.01$ & $21.03 \pm 0.00$ & $20.17 \pm 0.00$ & $19.79 \pm 0.00$ & $19.64 \pm 0.00$ & $19.11 \pm 0.07$ & $18.10 \pm 0.01$ & 0.79 & A & [1], A \\
	734136 & J238.82391+41.86073 & $22.38 \pm 0.02$ & $21.18 \pm 0.01$ & $20.15 \pm 0.00$ & $19.90 \pm 0.00$ & $19.52 \pm 0.01$ & $18.94 \pm 0.06$ & $17.85 \pm 0.01$ & 0.94 & A & [3], A \\
	734908 & J239.61106+43.47521 & $21.93 \pm 0.00$ & $20.33 \pm 0.00$ & $19.45 \pm 0.00$ & $19.11 \pm 0.00$ & $18.94 \pm 0.00$ & $18.46 \pm 0.03$ & $17.88 \pm 0.01$ & 0.55 & A & [1], A \\
	736419 & J240.78050+43.23936 & $22.64 \pm 0.01$ & $21.07 \pm 0.00$ & $19.91 \pm 0.00$ & $19.46 \pm 0.00$ & $19.28 \pm 0.00$ & $18.84 \pm 0.05$ & $17.61 \pm 0.01$ & 0.40 & A & [3], B \\
	747538 & J247.92922+43.66667 & $22.97 \pm 0.01$ & $21.47 \pm 0.00$ & $20.64 \pm 0.00$ & $20.30 \pm 0.00$ & $20.20 \pm 0.01$ & $19.37 \pm 0.09$ & $18.57 \pm 0.01$ & 0.41 & A & \nodata \\
	771737 & J333.82346+04.74438 & $22.29 \pm 0.01$ & $20.75 \pm 0.00$ & $20.05 \pm 0.00$ & $19.79 \pm 0.00$ & $19.62 \pm 0.00$ & $19.19 \pm 0.07$ & $18.79 \pm 0.02$ & 0.35 & A & \nodata \\
	786779 & J336.48464+04.16030 & $20.98 \pm 0.00$ & $20.82 \pm 0.00$ & $20.41 \pm 0.00$ & $20.21 \pm 0.00$ & $20.11 \pm 0.01$ & $19.60 \pm 0.09$ & $18.27 \pm 0.02$ & 0.59 & A & [6], A \\
	793292 & J337.67363$-$00.01736 & $24.13 \pm 0.03$ & $22.58 \pm 0.01$ & $21.65 \pm 0.01$ & $21.27 \pm 0.01$ & $21.07 \pm 0.02$ &  \nodata  &  \nodata  & 0.36 & A & \nodata \\
	799335 & J338.82285+02.37503 & $22.30 \pm 0.01$ & $20.67 \pm 0.00$ & $20.01 \pm 0.00$ & $19.46 \pm 0.00$ & $19.21 \pm 0.00$ & $18.66 \pm 0.03$ & $17.81 \pm 0.01$ & 0.77 & A & \nodata \\
	802722 & J339.38975+00.83767 & $22.82 \pm 0.01$ & $21.29 \pm 0.00$ & $20.20 \pm 0.00$ & $19.73 \pm 0.00$ & $19.59 \pm 0.00$ & $19.01 \pm 0.05$ & $17.98 \pm 0.01$ & 0.35 & A & [3], A \\
	805701 & J339.89447+02.58520 & $25.38 \pm 0.12$ & $23.62 \pm 0.03$ & $22.28 \pm 0.01$ & $21.34 \pm 0.01$ & $20.86 \pm 0.02$ & $19.86 \pm 0.12$ & $18.03 \pm 0.01$ & 0.39 & A & [3], A \\
	809323 & J340.47762+00.05874 & $22.19 \pm 0.01$ & $20.97 \pm 0.00$ & $20.19 \pm 0.00$ & $19.78 \pm 0.00$ & $19.71 \pm 0.00$ & $19.09 \pm 0.06$ & $18.32 \pm 0.02$ & 0.97 & A & [1], B \\
	810052 & J340.59043+00.19585 & $23.39 \pm 0.02$ & $22.53 \pm 0.01$ & $22.05 \pm 0.01$ & $21.66 \pm 0.01$ & $21.48 \pm 0.02$ &  \nodata  &  \nodata  & 0.33 & A & [12], A \\
	814445 & J341.24882+01.86662 & $21.63 \pm 0.00$ & $20.90 \pm 0.00$ & $20.72 \pm 0.00$ & $20.70 \pm 0.01$ & $20.73 \pm 0.01$ & $19.81 \pm 0.08$ & $19.37 \pm 0.04$ & 0.35 & A & \nodata \\
	837433 & J346.34029$-$00.03659 & $22.59 \pm 0.01$ & $20.80 \pm 0.00$ & $19.85 \pm 0.00$ & $19.49 \pm 0.00$ & $19.42 \pm 0.00$ & $18.87 \pm 0.03$ & $17.87 \pm 0.01$ & 1.00 & A & [1], A \\
	855465 & J350.50896+00.66983 & $23.71 \pm 0.03$ & $22.31 \pm 0.01$ & $21.51 \pm 0.01$ & $21.24 \pm 0.01$ & $20.97 \pm 0.01$ & $20.39 \pm 0.11$ & $19.41 \pm 0.04$ & 0.37 & A & \nodata \\
	857588 & J350.94195$-$00.51052 & $23.73 \pm 0.03$ & $22.62 \pm 0.01$ & $21.49 \pm 0.01$ & $20.92 \pm 0.01$ & $20.72 \pm 0.01$ & $19.78 \pm 0.11$ & $18.50 \pm 0.02$ & 0.87 & A & [11], B \\
	859449 & J351.28928+00.85322 & $22.74 \pm 0.01$ & $21.30 \pm 0.00$ & $20.41 \pm 0.00$ & $20.13 \pm 0.00$ & $19.99 \pm 0.00$ & $19.39 \pm 0.09$ & $18.62 \pm 0.02$ & 0.65 & A & [3], B \\
	860199 & J351.44998+00.62763 & $21.89 \pm 0.00$ & $20.35 \pm 0.00$ & $19.48 \pm 0.00$ & $19.26 \pm 0.00$ & $19.07 \pm 0.00$ & $18.43 \pm 0.03$ & $17.66 \pm 0.01$ & 0.84 & A & [3], B \\
	867306 & J352.87700+00.62595 & $23.38 \pm 0.02$ & $21.92 \pm 0.01$ & $20.81 \pm 0.00$ & $20.54 \pm 0.00$ & $20.36 \pm 0.00$ & $19.31 \pm 0.08$ & $17.91 \pm 0.01$ & 0.98 & A & [3], A \\
	867677 & J352.94338+01.64599 & $22.49 \pm 0.01$ & $20.80 \pm 0.00$ & $19.90 \pm 0.00$ & $19.66 \pm 0.00$ & $19.44 \pm 0.00$ & $18.98 \pm 0.04$ & $17.90 \pm 0.01$ & 0.70 & A & [3], A \\
	877180 & J354.92148+00.01208 & $21.12 \pm 0.00$ & $20.82 \pm 0.00$ & $20.58 \pm 0.00$ & $20.49 \pm 0.00$ & $20.19 \pm 0.01$ & $19.44 \pm 0.06$ & $18.48 \pm 0.02$ & 0.99 & A & [5], A \\
	889892 & J359.05923+02.52107 & $21.72 \pm 0.00$ & $20.11 \pm 0.00$ & $19.39 \pm 0.00$ & $19.10 \pm 0.00$ & $18.95 \pm 0.00$ & $18.51 \pm 0.03$ & $17.90 \pm 0.01$ & 0.69 & A & \nodata \\
	891812 & J359.72171+01.40182 & $23.12 \pm 0.01$ & $21.51 \pm 0.00$ & $20.49 \pm 0.00$ & $20.12 \pm 0.00$ & $19.73 \pm 0.00$ & $19.30 \pm 0.06$ & $18.12 \pm 0.01$ & 0.57 & A & [3], B \\
	51372 & J029.16338$-$02.04496 & $23.24 \pm 0.02$ & $21.53 \pm 0.01$ & $20.73 \pm 0.00$ & $20.39 \pm 0.00$ & $20.17 \pm 0.01$ & $19.50 \pm 0.07$ & $18.96 \pm 0.03$ & 0.36 & B & \nodata \\
	66605 & J031.23188$-$02.14772 & $20.39 \pm 0.00$ & $19.84 \pm 0.00$ & $19.40 \pm 0.00$ & $19.30 \pm 0.00$ & $19.17 \pm 0.00$ &  \nodata  &  \nodata  & 0.78 & B & \nodata \\
	68390 & J031.48690$-$02.14298 & $23.12 \pm 0.01$ & $21.42 \pm 0.00$ & $20.20 \pm 0.00$ & $19.76 \pm 0.00$ & $19.53 \pm 0.00$ & $18.87 \pm 0.03$ & $17.69 \pm 0.01$ & 0.64 & B & [3], C \\
	75191 & J032.43238+00.62463 & $21.19 \pm 0.00$ & $20.10 \pm 0.00$ & $19.67 \pm 0.00$ & $19.41 \pm 0.00$ & $19.31 \pm 0.00$ & $18.79 \pm 0.05$ & $18.12 \pm 0.01$ & 0.51 & B & \nodata \\
	77117 & J032.66791+02.19009 & $20.54 \pm 0.00$ & $19.86 \pm 0.00$ & $19.24 \pm 0.00$ & $18.96 \pm 0.00$ & $18.88 \pm 0.00$ & $18.00 \pm 0.02$ & $17.70 \pm 0.01$ & 0.37 & B & \nodata \\
	84320 & J033.59293$-$02.91591 & $23.89 \pm 0.03$ & $23.76 \pm 0.04$ & $23.44 \pm 0.03$ & $23.30 \pm 0.05$ & $22.47 \pm 0.04$ &  \nodata  &  \nodata  & 0.51 & B & \nodata \\
	93880 & J034.90323$-$00.91156 & $23.29 \pm 0.01$ & $21.63 \pm 0.01$ & $20.51 \pm 0.00$ & $20.11 \pm 0.00$ & $19.86 \pm 0.00$ & $19.45 \pm 0.09$ & $18.24 \pm 0.02$ & 0.50 & B & \nodata \\
	126604 & J039.22622$-$03.93275 & $23.34 \pm 0.01$ & $21.82 \pm 0.01$ & $20.77 \pm 0.00$ & $20.37 \pm 0.00$ & $20.13 \pm 0.01$ & $19.48 \pm 0.07$ & $18.45 \pm 0.02$ & 0.40 & B & \nodata \\
	147439 & J131.32696+02.18556 & $26.23 \pm 0.18$ & $25.17 \pm 0.11$ & $23.54 \pm 0.03$ & $23.20 \pm 0.03$ & $22.65 \pm 0.04$ &  \nodata  &  \nodata  & 0.34 & B & \nodata \\
	148498 & J131.54947+01.44483 & $24.09 \pm 0.02$ & $23.81 \pm 0.03$ & $23.76 \pm 0.03$ & $23.34 \pm 0.04$ & $23.39 \pm 0.08$ &  \nodata  &  \nodata  & 0.31 & B & \nodata \\
	157312 & J133.42600+01.38805 & $21.88 \pm 0.00$ & $20.43 \pm 0.00$ & $19.78 \pm 0.00$ & $19.41 \pm 0.00$ & $19.26 \pm 0.00$ & $18.82 \pm 0.05$ & $18.35 \pm 0.02$ & 0.73 & B & \nodata \\
	163604 & J134.67324+03.89196 & $22.40 \pm 0.01$ & $20.88 \pm 0.00$ & $20.15 \pm 0.00$ & $19.80 \pm 0.00$ & $19.74 \pm 0.00$ & $19.10 \pm 0.04$ &  \nodata  & 0.60 & B & \nodata \\
	168262 & J135.38983+04.60396 & $24.16 \pm 0.04$ & $22.68 \pm 0.02$ & $22.57 \pm 0.03$ & $21.86 \pm 0.02$ & $21.74 \pm 0.03$ &  \nodata  &  \nodata  & 0.58 & B & \nodata \\
	170439 & J135.66838+04.82183 & $20.83 \pm 0.00$ & $19.70 \pm 0.00$ & $18.76 \pm 0.00$ & $18.28 \pm 0.00$ & $18.20 \pm 0.00$ &  \nodata  &  \nodata  & 0.56 & B & \nodata \\
	170930 & J135.73538+04.23009 & $24.57 \pm 0.04$ & $22.87 \pm 0.01$ & $21.64 \pm 0.01$ & $20.93 \pm 0.01$ & $20.69 \pm 0.01$ &  \nodata  &  \nodata  & 0.47 & B & \nodata \\
	174479 & J136.12723+04.44683 & $26.83 \pm nan$ & $22.85 \pm 0.01$ & $22.16 \pm 0.02$ & $21.65 \pm 0.01$ & $21.41 \pm 0.02$ &  \nodata  &  \nodata  & 0.53 & B & \nodata \\
	183185 & J137.14126+00.13506 & $22.71 \pm 0.01$ & $21.69 \pm 0.00$ & $20.58 \pm 0.00$ & $20.03 \pm 0.00$ & $19.80 \pm 0.00$ & $19.23 \pm 0.04$ & $17.79 \pm 0.01$ & 0.46 & B & [1], B \\
	184295 & J137.27788+03.39111 & $22.29 \pm 0.01$ & $21.20 \pm 0.00$ & $20.75 \pm 0.00$ & $20.49 \pm 0.00$ & $20.29 \pm 0.01$ &  \nodata  &  \nodata  & 0.70 & B & \nodata \\
	187379 & J137.67904$-$00.14358 & $23.73 \pm 0.02$ & $22.42 \pm 0.01$ & $21.78 \pm 0.01$ & $21.05 \pm 0.01$ & $20.81 \pm 0.01$ &  \nodata  &  \nodata  & 0.33 & B & \nodata \\
	188765 & J137.86759+03.86639 & $22.14 \pm 0.00$ & $21.71 \pm 0.00$ & $21.01 \pm 0.00$ & $21.00 \pm 0.01$ & $20.93 \pm 0.01$ & $19.96 \pm 0.08$ & $18.93 \pm 0.03$ & 0.42 & B & \nodata \\
	189415 & J137.96492$-$00.73540 & $21.56 \pm 0.00$ & $21.29 \pm 0.00$ & $20.71 \pm 0.00$ & $20.31 \pm 0.00$ & $20.04 \pm 0.00$ & $19.28 \pm 0.07$ & $18.25 \pm 0.02$ & 0.52 & B & \nodata \\
	209801 & J140.92640$-$00.92415 & $22.57 \pm 0.01$ & $21.68 \pm 0.01$ & $21.14 \pm 0.00$ & $21.01 \pm 0.01$ & $20.73 \pm 0.01$ & $19.40 \pm 0.03$ &  \nodata  & 0.40 & B & \nodata \\
	216863 & J142.17544+01.32812 & $22.44 \pm 0.01$ & $21.02 \pm 0.00$ & $20.05 \pm 0.00$ & $19.74 \pm 0.00$ & $19.41 \pm 0.00$ & $18.99 \pm 0.03$ & $18.12 \pm 0.01$ & 0.73 & B & [3], B \\
	231509 & J145.86521$-$01.91481 & $22.28 \pm 0.01$ & $20.57 \pm 0.00$ & $19.57 \pm 0.00$ & $19.32 \pm 0.00$ & $19.09 \pm 0.00$ & $18.54 \pm 0.03$ & $17.66 \pm 0.01$ & 0.34 & B & [1], B \\
	235815 & J146.80482+02.79552 & $19.65 \pm 0.00$ & $19.61 \pm 0.00$ & $19.40 \pm 0.00$ & $19.46 \pm 0.00$ & $19.42 \pm 0.00$ & $19.53 \pm 0.07$ & $17.46 \pm 0.01$ & 0.40 & B & \nodata \\
	238123 & J147.29565+00.14604 & $21.85 \pm 0.01$ & $20.39 \pm 0.00$ & $19.54 \pm 0.00$ & $19.27 \pm 0.00$ & $19.14 \pm 0.00$ &  \nodata  &  \nodata  & 0.37 & B & \nodata \\
	238737 & J147.43061+00.09343 & $18.84 \pm 0.00$ & $18.75 \pm 0.00$ & $18.65 \pm 0.00$ & $18.56 \pm 0.00$ & $18.45 \pm 0.00$ & $18.18 \pm 0.02$ & $16.97 \pm 0.01$ & 0.56 & B & \nodata \\
	248368 & J149.45906+01.75185 & $23.56 \pm 0.02$ & $23.04 \pm 0.01$ & $22.59 \pm 0.01$ & $22.34 \pm 0.02$ & $22.16 \pm 0.03$ &  \nodata  &  \nodata  & 0.35 & B & \nodata \\
	254401 & J150.70590+00.51154 & $22.70 \pm 0.01$ & $21.35 \pm 0.00$ & $20.27 \pm 0.00$ & $20.00 \pm 0.00$ & $19.78 \pm 0.00$ & $19.02 \pm 0.03$ & $17.68 \pm 0.01$ & 0.64 & B & \nodata \\
	268088 & J153.63551$-$00.99589 & $23.78 \pm 0.02$ & $22.12 \pm 0.01$ & $21.08 \pm 0.00$ & $20.76 \pm 0.00$ & $20.76 \pm 0.02$ & $19.43 \pm 0.05$ & $18.61 \pm 0.02$ & 0.32 & B & \nodata \\
	297613 & J159.35616+02.52577 & $22.65 \pm 0.01$ & $21.79 \pm 0.01$ & $21.06 \pm 0.00$ & $20.87 \pm 0.01$ & $20.61 \pm 0.01$ & $20.23 \pm 0.12$ & $19.14 \pm 0.03$ & 0.32 & B & \nodata \\
	335597 & J165.10725+00.33480 & $23.06 \pm 0.01$ & $21.43 \pm 0.00$ & $20.53 \pm 0.00$ & $20.11 \pm 0.00$ & $20.02 \pm 0.01$ & $19.58 \pm 0.08$ & $18.84 \pm 0.03$ & 0.76 & B & \nodata \\
	353496 & J168.05397+02.45084 & $22.56 \pm 0.01$ & $20.87 \pm 0.00$ & $20.05 \pm 0.00$ & $19.71 \pm 0.00$ & $19.53 \pm 0.00$ & $19.06 \pm 0.04$ & $18.28 \pm 0.02$ & 0.34 & B & [9], B \\
	376283 & J171.57689+04.10215 & $21.35 \pm 0.01$ & $20.17 \pm 0.00$ & $19.59 \pm 0.00$ & $19.13 \pm 0.00$ & $19.10 \pm 0.00$ & $18.64 \pm 0.03$ & $17.41 \pm 0.01$ & 0.34 & B & \nodata \\
	396577 & J174.43839$-$00.97846 & $22.60 \pm 0.01$ & $21.98 \pm 0.01$ & $21.09 \pm 0.00$ & $20.79 \pm 0.01$ & $20.59 \pm 0.01$ & $20.10 \pm 0.10$ & $19.20 \pm 0.04$ & 0.69 & B & \nodata \\
	398473 & J174.66094$-$00.90362 & $23.32 \pm 0.01$ & $22.30 \pm 0.01$ & $21.17 \pm 0.00$ & $20.78 \pm 0.00$ & $20.60 \pm 0.01$ & $19.92 \pm 0.08$ & $19.08 \pm 0.03$ & 0.56 & B & [1], B \\
	402941 & J175.17507$-$00.72703 & $22.95 \pm 0.01$ & $22.88 \pm 0.02$ & $22.60 \pm 0.01$ & $22.48 \pm 0.02$ & $22.20 \pm 0.03$ &  \nodata  &  \nodata  & 0.54 & B & [14], C \\
	403630 & J175.25814+00.71887 & $23.45 \pm 0.02$ & $21.93 \pm 0.01$ & $21.06 \pm 0.00$ & $20.70 \pm 0.00$ & $20.54 \pm 0.01$ & $20.02 \pm 0.09$ & $19.11 \pm 0.03$ & 0.40 & B & \nodata \\
	408028 & J175.79711$-$01.65959 & $22.87 \pm 0.01$ & $22.32 \pm 0.01$ & $21.66 \pm 0.01$ & $21.15 \pm 0.01$ & $20.96 \pm 0.01$ &  \nodata  &  \nodata  & 0.93 & B & [1], A \\
	421588 & J177.44370$-$00.36609 & $22.83 \pm 0.01$ & $21.64 \pm 0.01$ & $20.75 \pm 0.00$ & $20.40 \pm 0.00$ & $20.26 \pm 0.01$ & $19.74 \pm 0.11$ & $18.84 \pm 0.03$ & 0.45 & B & [3], B \\
	424090 & J177.75274+02.48334 & $20.51 \pm 0.00$ & $20.20 \pm 0.00$ & $20.00 \pm 0.00$ & $19.87 \pm 0.00$ & $19.93 \pm 0.00$ & $19.56 \pm 0.07$ & $18.39 \pm 0.02$ & 0.32 & B & [6], C \\
	428220 & J178.26416+03.81270 & $23.20 \pm 0.01$ & $21.81 \pm 0.00$ & $20.51 \pm 0.00$ & $20.10 \pm 0.00$ & $19.82 \pm 0.00$ & $19.22 \pm 0.02$ & $18.03 \pm 0.01$ & 0.71 & B & \nodata \\
	429921 & J178.46614$-$00.88752 & $23.32 \pm 0.02$ & $21.63 \pm 0.00$ & $20.43 \pm 0.00$ & $19.99 \pm 0.00$ & $19.80 \pm 0.00$ & $19.14 \pm 0.04$ & $17.91 \pm 0.01$ & 0.42 & B & [3], C \\
	439418 & J179.59017+01.17014 & $23.64 \pm 0.02$ & $22.04 \pm 0.01$ & $20.69 \pm 0.00$ & $20.28 \pm 0.00$ & $19.97 \pm 0.00$ & $19.34 \pm 0.04$ & $18.09 \pm 0.01$ & 0.51 & B & \nodata \\
	447096 & J180.46200+04.89641 & $22.62 \pm 0.01$ & $21.08 \pm 0.00$ & $20.24 \pm 0.00$ & $19.94 \pm 0.00$ & $19.67 \pm 0.00$ & $19.18 \pm 0.04$ & $18.54 \pm 0.02$ & 0.61 & B & \nodata \\
	452511 & J181.17000$-$00.04080 & $24.94 \pm 0.06$ & $24.63 \pm 0.06$ & $23.98 \pm 0.03$ & $23.41 \pm 0.04$ & $23.10 \pm 0.08$ & $20.23 \pm 0.05$ &  \nodata  & 0.59 & B & \nodata \\
	457430 & J181.85322$-$00.56724 & $22.20 \pm 0.01$ & $20.74 \pm 0.00$ & $20.14 \pm 0.00$ & $19.86 \pm 0.00$ & $19.67 \pm 0.00$ & $19.20 \pm 0.05$ & $18.39 \pm 0.02$ & 0.33 & B & \nodata \\
	463617 & J182.66988+03.56547 & $22.94 \pm 0.01$ & $22.67 \pm 0.01$ & $21.99 \pm 0.01$ & $21.64 \pm 0.01$ & $21.41 \pm 0.02$ &  \nodata  &  \nodata  & 0.35 & B & \nodata \\
	472144 & J183.88950$-$00.97856 & $22.18 \pm 0.01$ & $20.38 \pm 0.00$ & $19.48 \pm 0.00$ & $19.11 \pm 0.00$ & $19.07 \pm 0.00$ & $18.26 \pm 0.02$ &  \nodata  & 0.55 & B & \nodata \\
	485479 & J185.90854+00.81986 & $21.34 \pm 0.00$ & $20.28 \pm 0.00$ & $19.85 \pm 0.00$ & $19.47 \pm 0.00$ & $19.61 \pm 0.00$ & $18.89 \pm 0.03$ & $18.14 \pm 0.01$ & 0.36 & B & \nodata \\
	510783 & J189.64253$-$00.72405 & $22.25 \pm 0.01$ & $20.51 \pm 0.00$ & $19.58 \pm 0.00$ & $19.27 \pm 0.00$ & $19.14 \pm 0.00$ & $18.53 \pm 0.02$ & $17.64 \pm 0.01$ & 0.52 & B & [9], B \\
	564583 & J201.38247$-$01.35862 & $20.48 \pm 0.00$ & $20.02 \pm 0.00$ & $19.58 \pm 0.00$ & $19.43 \pm 0.00$ & $19.31 \pm 0.00$ &  \nodata  &  \nodata  & 0.78 & B & \nodata \\
	566973 & J202.04765+00.26065 & $20.04 \pm 0.00$ & $19.88 \pm 0.00$ & $19.75 \pm 0.00$ & $19.65 \pm 0.00$ & $19.67 \pm 0.00$ & $18.13 \pm 0.02$ &  \nodata  & 0.38 & B & \nodata \\
	574483 & J203.97402+00.94847 & $23.16 \pm 0.02$ & $22.23 \pm 0.01$ & $21.15 \pm 0.00$ & $20.69 \pm 0.00$ & $20.38 \pm 0.01$ & $19.80 \pm 0.08$ & $18.67 \pm 0.02$ & 0.74 & B & \nodata \\
	577157 & J204.68695$-$01.15195 & $22.54 \pm 0.01$ & $21.61 \pm 0.01$ & $20.91 \pm 0.00$ & $20.56 \pm 0.01$ & $20.31 \pm 0.01$ & $19.32 \pm 0.02$ & $18.20 \pm 0.02$ & 0.32 & B & \nodata \\
	600996 & J209.72304$-$02.25715 & $23.30 \pm 0.03$ & $21.95 \pm 0.01$ & $20.56 \pm 0.00$ & $20.19 \pm 0.01$ & $19.90 \pm 0.01$ & $19.31 \pm 0.05$ & $18.15 \pm 0.01$ & 0.38 & B & [1], B \\
	605855 & J210.71856$-$00.29353 & $26.53 \pm 0.33$ & $23.10 \pm 0.02$ & $22.44 \pm 0.01$ & $22.18 \pm 0.02$ & $21.94 \pm 0.03$ & $17.90 \pm 0.01$ &  \nodata  & 0.34 & B & \nodata \\
	616231 & J212.70374$-$01.16278 & $22.64 \pm 0.01$ & $22.30 \pm 0.01$ & $22.13 \pm 0.01$ & $22.06 \pm 0.01$ & $21.88 \pm 0.02$ &  \nodata  &  \nodata  & 0.63 & B & [11], B \\
	628110 & J214.43249$-$00.12044 & $24.31 \pm 0.03$ & $22.89 \pm 0.01$ & $21.69 \pm 0.00$ & $20.83 \pm 0.00$ & $20.51 \pm 0.01$ & $19.80 \pm 0.09$ & $18.28 \pm 0.02$ & 0.31 & B & [10], C \\
	639109 & J216.01461+00.13081 & $23.32 \pm 0.01$ & $21.68 \pm 0.01$ & $20.73 \pm 0.00$ & $20.34 \pm 0.00$ & $20.22 \pm 0.00$ & $19.92 \pm 0.10$ & $18.45 \pm 0.02$ & 0.37 & B & \nodata \\
	640138 & J216.18209$-$00.97517 & $21.53 \pm 0.00$ & $20.75 \pm 0.00$ & $20.51 \pm 0.00$ & $20.28 \pm 0.00$ & $20.31 \pm 0.01$ & $18.84 \pm 0.02$ & $19.16 \pm 0.03$ & 0.87 & B & \nodata \\
	640692 & J216.24943+00.91866 & $24.44 \pm 0.04$ & $23.61 \pm 0.02$ & $23.17 \pm 0.02$ & $22.86 \pm 0.03$ & $22.70 \pm 0.05$ & $19.25 \pm 0.02$ &  \nodata  & 0.51 & B & [2], B \\
	644701 & J216.81354$-$00.06799 & $21.67 \pm 0.00$ & $20.87 \pm 0.00$ & $20.61 \pm 0.00$ & $20.50 \pm 0.00$ & $19.93 \pm 0.00$ & $19.70 \pm 0.10$ &  \nodata  & 0.69 & B & \nodata \\
	646669 & J217.10934+01.25957 & $21.00 \pm 0.00$ & $20.27 \pm 0.00$ & $19.96 \pm 0.00$ & $19.82 \pm 0.00$ & $19.37 \pm 0.00$ & $19.27 \pm 0.05$ & $18.79 \pm 0.02$ & 0.63 & B & \nodata \\
	647015 & J217.16807+00.33684 & $21.56 \pm 0.00$ & $20.67 \pm 0.00$ & $20.23 \pm 0.00$ & $20.15 \pm 0.00$ & $19.60 \pm 0.00$ & $19.59 \pm 0.09$ & $18.28 \pm 0.02$ & 0.31 & B & \nodata \\
	652509 & J218.04209$-$00.43369 & $22.66 \pm 0.01$ & $21.27 \pm 0.00$ & $20.35 \pm 0.00$ & $20.01 \pm 0.00$ & $19.87 \pm 0.00$ & $19.03 \pm 0.04$ &  \nodata  & 0.33 & B & \nodata \\
	653043 & J218.12022$-$00.20166 & $21.39 \pm 0.00$ & $20.86 \pm 0.00$ & $20.72 \pm 0.00$ & $20.16 \pm 0.00$ & $20.46 \pm 0.01$ & $19.83 \pm 0.10$ & $19.25 \pm 0.03$ & 0.49 & B & \nodata \\
	655414 & J218.44478+43.81510 & $23.41 \pm 0.02$ & $21.80 \pm 0.01$ & $20.58 \pm 0.00$ & $20.29 \pm 0.00$ & $20.26 \pm 0.01$ & $19.13 \pm 0.06$ & $18.26 \pm 0.01$ & 0.51 & B & \nodata \\
	665084 & J219.63533$-$00.39081 & $25.24 \pm 0.09$ & $22.82 \pm 0.01$ & $21.28 \pm 0.00$ & $20.45 \pm 0.00$ & $20.16 \pm 0.00$ & $19.51 \pm 0.09$ & $17.73 \pm 0.01$ & 0.48 & B & [1], B \\
	666651 & J219.82191$-$00.65496 & $24.18 \pm 0.03$ & $22.22 \pm 0.01$ & $20.80 \pm 0.00$ & $20.35 \pm 0.00$ & $20.08 \pm 0.00$ & $19.56 \pm 0.07$ & $18.31 \pm 0.02$ & 0.51 & B & [10], B \\
	691834 & J222.88068$-$01.67836 & $21.83 \pm 0.01$ & $20.47 \pm 0.00$ & $19.75 \pm 0.00$ & $19.47 \pm 0.00$ & $19.35 \pm 0.00$ & $18.81 \pm 0.04$ & $18.54 \pm 0.02$ & 0.39 & B & \nodata \\
	694866 & J223.26168$-$00.70981 & $20.53 \pm 0.00$ & $20.12 \pm 0.00$ & $19.76 \pm 0.00$ & $19.68 \pm 0.00$ & $19.41 \pm 0.00$ & $19.21 \pm 0.06$ & $18.32 \pm 0.02$ & 0.36 & B & \nodata \\
	703099 & J224.39263+43.67893 & $22.71 \pm 0.01$ & $21.14 \pm 0.00$ & $20.23 \pm 0.00$ & $19.97 \pm 0.00$ & $19.76 \pm 0.00$ & $19.16 \pm 0.05$ & $18.14 \pm 0.01$ & 0.85 & B & [3], B \\
	720082 & J229.29540+44.12778 & $23.98 \pm 0.02$ & $22.74 \pm 0.01$ & $21.67 \pm 0.01$ & $21.16 \pm 0.01$ & $20.91 \pm 0.01$ & $20.43 \pm 0.13$ & $19.05 \pm 0.02$ & 0.70 & B & [3], B \\
	725696 & J233.42927+42.83937 & $23.87 \pm 0.02$ & $23.23 \pm 0.02$ & $22.44 \pm 0.01$ & $22.07 \pm 0.02$ & $21.69 \pm 0.02$ &  \nodata  &  \nodata  & 0.61 & B & \nodata \\
	736755 & J241.05542+43.37910 & $22.70 \pm 0.01$ & $21.90 \pm 0.00$ & $21.21 \pm 0.00$ & $20.93 \pm 0.01$ & $20.81 \pm 0.01$ & $20.40 \pm 0.20$ & $19.67 \pm 0.04$ & 0.53 & B & \nodata \\
	751225 & J249.97872+43.62234 & $21.46 \pm 0.00$ & $20.12 \pm 0.00$ & $19.54 \pm 0.00$ & $19.20 \pm 0.00$ & $19.27 \pm 0.00$ & $18.69 \pm 0.04$ & $18.22 \pm 0.01$ & 0.50 & B & \nodata \\
	764918 & J332.57071+02.12085 & $22.17 \pm 0.01$ & $20.47 \pm 0.00$ & $19.58 \pm 0.00$ & $19.17 \pm 0.00$ & $19.03 \pm 0.00$ & $18.03 \pm 0.02$ & $17.07 \pm 0.01$ & 0.37 & B & \nodata \\
	768999 & J333.27886$-$00.51026 & $22.71 \pm 0.01$ & $21.52 \pm 0.01$ & $20.29 \pm 0.00$ & $19.87 \pm 0.00$ & $19.66 \pm 0.01$ & $19.05 \pm 0.06$ & $17.20 \pm 0.01$ & 0.61 & B & [3], B \\
	787110 & J336.53490+01.15634 & $22.00 \pm 0.01$ & $21.43 \pm 0.00$ & $21.28 \pm 0.00$ & $21.13 \pm 0.01$ & $21.13 \pm 0.01$ &  \nodata  & $19.21 \pm 0.03$ & 0.32 & B & \nodata \\
	788642 & J336.82352+00.45240 & $22.82 \pm 0.01$ & $21.65 \pm 0.00$ & $20.73 \pm 0.00$ & $20.39 \pm 0.00$ & $20.11 \pm 0.01$ & $19.59 \pm 0.10$ & $18.63 \pm 0.02$ & 0.43 & B & \nodata \\
	793076 & J337.63423$-$00.12474 & $21.31 \pm 0.00$ & $19.96 \pm 0.00$ & $19.31 \pm 0.00$ & $19.05 \pm 0.00$ & $18.78 \pm 0.00$ & $18.26 \pm 0.02$ & $17.87 \pm 0.01$ & 0.37 & B & \nodata \\
	795264 & J338.04846+01.99870 & $23.30 \pm 0.02$ & $21.65 \pm 0.00$ & $20.50 \pm 0.00$ & $20.03 \pm 0.00$ & $19.83 \pm 0.00$ & $19.35 \pm 0.06$ & $17.82 \pm 0.01$ & 0.48 & B & [3], B \\
	806087 & J339.96430+02.52779 & $21.29 \pm 0.00$ & $19.80 \pm 0.00$ & $19.23 \pm 0.00$ & $18.89 \pm 0.00$ & $18.77 \pm 0.00$ & $17.82 \pm 0.02$ & $17.60 \pm 0.01$ & 0.39 & B & \nodata \\
	811241 & J340.78956+01.87792 & $21.52 \pm 0.00$ & $19.88 \pm 0.00$ & $19.24 \pm 0.00$ & $18.93 \pm 0.00$ & $18.85 \pm 0.00$ & $18.12 \pm 0.02$ & $17.68 \pm 0.01$ & 0.33 & B & \nodata \\
	816392 & J341.56034+05.97432 & $25.44 \pm 0.30$ & $24.63 \pm 0.24$ & $24.04 \pm 0.18$ & $22.09 \pm 0.02$ & $21.88 \pm 0.05$ &  \nodata  &  \nodata  & 0.81 & B & \nodata \\
	816796 & J341.62357+00.45897 & $22.09 \pm 0.01$ & $21.28 \pm 0.01$ & $20.78 \pm 0.00$ & $20.58 \pm 0.00$ & $20.67 \pm 0.01$ & $20.05 \pm 0.14$ & $19.56 \pm 0.05$ & 0.48 & B & \nodata \\
	819437 & J342.07629+01.12530 & $21.99 \pm 0.01$ & $20.86 \pm 0.00$ & $20.21 \pm 0.00$ & $19.79 \pm 0.00$ & $19.73 \pm 0.00$ & $17.82 \pm 0.02$ & $17.75 \pm 0.01$ & 0.91 & B & \nodata \\
	856994 & J350.82679+00.83850 & $22.59 \pm 0.01$ & $21.12 \pm 0.00$ & $20.22 \pm 0.00$ & $19.90 \pm 0.00$ & $19.77 \pm 0.00$ & $19.11 \pm 0.07$ & $18.22 \pm 0.02$ & 0.31 & B & \nodata \\
	860355 & J351.48909$-$00.87412 & $24.88 \pm 0.09$ & $22.43 \pm 0.02$ & $21.28 \pm 0.01$ & $20.84 \pm 0.01$ & $20.54 \pm 0.01$ & $19.82 \pm 0.12$ & $17.91 \pm 0.01$ & 0.59 & B & [3], A \\
	862256 & J351.88254+00.14394 & $21.72 \pm 0.00$ & $20.45 \pm 0.00$ & $19.88 \pm 0.00$ & $19.72 \pm 0.00$ & $19.63 \pm 0.00$ & $18.68 \pm 0.04$ & $18.48 \pm 0.02$ & 0.36 & B & \nodata \\
	872424 & J353.80073+02.02693 & $21.60 \pm 0.01$ & $20.57 \pm 0.00$ & $20.23 \pm 0.00$ & $20.15 \pm 0.00$ & $19.33 \pm 0.00$ & $19.45 \pm 0.11$ & $18.35 \pm 0.02$ & 0.87 & B & \nodata \\
	874771 & J354.30706+00.93668 & $20.67 \pm 0.00$ & $20.38 \pm 0.00$ & $19.92 \pm 0.00$ & $20.12 \pm 0.00$ & $20.11 \pm 0.00$ &  \nodata  &  \nodata  & 0.48 & B & \nodata \\
	1520 & J000.48800+00.56474 & $23.71 \pm 0.02$ & $22.22 \pm 0.01$ & $21.48 \pm 0.01$ & $21.14 \pm 0.01$ & $20.99 \pm 0.01$ & $20.12 \pm 0.08$ & $19.43 \pm 0.04$ & 0.10 & A* & \nodata \\
	3978 & J001.34828$-$00.04974 & $24.07 \pm 0.03$ & $22.44 \pm 0.01$ & $21.19 \pm 0.00$ & $20.74 \pm 0.01$ & $20.51 \pm 0.01$ & $19.39 \pm 0.06$ & $17.86 \pm 0.01$ & 0.19 & B* & \nodata \\
	11816 & J004.61623+00.18388 & $22.88 \pm 0.01$ & $22.18 \pm 0.01$ & $21.58 \pm 0.01$ & $21.46 \pm 0.01$ & $21.46 \pm 0.02$ & $20.88 \pm 0.24$ & $19.53 \pm 0.05$ & 0.08 & B* & \nodata \\
	14736 & J006.25319$-$00.45575 & $23.36 \pm 0.02$ & $22.54 \pm 0.01$ & $21.73 \pm 0.01$ & $21.42 \pm 0.01$ & $21.35 \pm 0.02$ &  \nodata  &  \nodata  & 0.25 & A* & [6], A \\
	56047 & J029.81149$-$00.42173 & $22.17 \pm 0.01$ & $20.49 \pm 0.00$ & $19.82 \pm 0.00$ & $19.50 \pm 0.00$ & $19.26 \pm 0.00$ & $18.80 \pm 0.05$ & $17.87 \pm 0.01$ & 0.25 & B* & [3], C \\
	212090 & J141.32130+02.08565 & $23.49 \pm 0.02$ & $22.59 \pm 0.01$ & $22.03 \pm 0.01$ & $21.85 \pm 0.01$ & $21.56 \pm 0.02$ &  \nodata  &  \nodata  & 0.06 & A* & \nodata \\
	257799 & J151.47453$-$00.11954 & $23.65 \pm 0.02$ & $23.21 \pm 0.02$ & $22.71 \pm 0.01$ & $22.76 \pm 0.02$ & $22.57 \pm 0.05$ &  \nodata  &  \nodata  & 0.14 & B* & \nodata \\
	287412 & J157.83342+03.23086 & $22.54 \pm 0.01$ & $21.07 \pm 0.00$ & $20.09 \pm 0.00$ & $19.79 \pm 0.00$ & $19.55 \pm 0.00$ & $18.86 \pm 0.04$ & $17.61 \pm 0.01$ & 0.27 & A* & \nodata \\
	315399 & J162.00962$-$01.53456 & $21.67 \pm 0.01$ & $21.00 \pm 0.00$ & $20.63 \pm 0.00$ & $20.45 \pm 0.00$ & $20.38 \pm 0.01$ & $19.89 \pm 0.08$ & $19.10 \pm 0.03$ & 0.13 & A* & \nodata \\
	337878 & J165.48269+03.49853 & $22.45 \pm 0.01$ & $20.80 \pm 0.00$ & $20.03 \pm 0.00$ & $19.68 \pm 0.00$ & $19.55 \pm 0.00$ & $18.94 \pm 0.04$ & $18.28 \pm 0.02$ & 0.25 & A* & [7], C \\
	350705 & J167.62605+03.45060 & $22.69 \pm 0.01$ & $22.00 \pm 0.01$ & $21.52 \pm 0.00$ & $21.23 \pm 0.01$ & $21.09 \pm 0.02$ & $20.54 \pm 0.06$ & $18.87 \pm 0.03$ & 0.16 & A* & \nodata \\
	371839 & J170.92342$-$00.64420 & $24.75 \pm 0.05$ & $23.29 \pm 0.02$ & $21.66 \pm 0.00$ & $21.17 \pm 0.01$ & $20.73 \pm 0.01$ & $20.23 \pm 0.12$ & $18.89 \pm 0.03$ & 0.29 & A* & \nodata \\
	387152 & J173.20414+02.24483 & $23.19 \pm 0.02$ & $21.69 \pm 0.01$ & $20.27 \pm 0.00$ & $19.83 \pm 0.00$ & $19.56 \pm 0.00$ & $19.07 \pm 0.04$ & $17.64 \pm 0.01$ & 0.24 & B* & \nodata \\
	418565 & J177.06399$-$01.15220 & $22.08 \pm 0.01$ & $20.60 \pm 0.00$ & $20.03 \pm 0.00$ & $19.45 \pm 0.00$ & $19.48 \pm 0.01$ & $18.85 \pm 0.03$ & $18.06 \pm 0.01$ & 0.28 & B* & \nodata \\
	426556 & J178.05915+00.52403 & $21.66 \pm 0.00$ & $20.47 \pm 0.00$ & $19.73 \pm 0.00$ & $19.41 \pm 0.00$ & $19.17 \pm 0.00$ & $18.61 \pm 0.03$ & $17.54 \pm 0.01$ & 0.30 & B* & [1], B \\
	429855 & J178.45593+02.52472 & $24.88 \pm 0.06$ & $22.43 \pm 0.01$ & $20.97 \pm 0.00$ & $20.53 \pm 0.00$ & $20.16 \pm 0.01$ & $19.26 \pm 0.06$ & $18.33 \pm 0.02$ & 0.15 & A* & [7], B \\
	430680 & J178.55360$-$00.81515 & $24.08 \pm 0.03$ & $23.21 \pm 0.02$ & $22.21 \pm 0.01$ & $21.59 \pm 0.01$ & $21.38 \pm 0.02$ & $20.63 \pm 0.10$ & $19.40 \pm 0.04$ & 0.10 & A* & \nodata \\
	497701 & J187.85952+03.51182 & $24.21 \pm 0.06$ & $22.88 \pm 0.01$ & $21.81 \pm 0.01$ & $20.99 \pm 0.01$ & $20.82 \pm 0.01$ & $20.49 \pm 0.17$ & $18.79 \pm 0.02$ & 0.13 & A* & \nodata \\
	532151 & J192.65234$-$01.15956 & $23.26 \pm 0.01$ & $22.84 \pm 0.01$ & $21.94 \pm 0.01$ & $21.40 \pm 0.01$ & $21.17 \pm 0.01$ &  \nodata  &  \nodata  & 0.09 & A* & \nodata \\
	588195 & J206.95203+01.03627 & $23.02 \pm 0.01$ & $22.66 \pm 0.01$ & $22.34 \pm 0.01$ & $22.21 \pm 0.02$ & $21.92 \pm 0.03$ &  \nodata  &  \nodata  & 0.05 & B* & \nodata \\
	588206 & J206.95377$-$00.44817 & $22.87 \pm 0.01$ & $22.26 \pm 0.01$ & $21.10 \pm 0.00$ & $20.59 \pm 0.00$ & $20.66 \pm 0.01$ & $19.72 \pm 0.06$ & $18.00 \pm 0.01$ & 0.03 & B* & \nodata \\
	594485 & J208.21891+00.92516 & $23.66 \pm 0.02$ & $23.31 \pm 0.02$ & $22.87 \pm 0.02$ & $22.51 \pm 0.02$ & $22.20 \pm 0.05$ & $20.06 \pm 0.09$ &  \nodata  & 0.20 & B* & \nodata \\
	598787 & J209.24615$-$00.89224 & $22.28 \pm 0.01$ & $21.25 \pm 0.00$ & $20.76 \pm 0.00$ & $20.46 \pm 0.00$ & $20.19 \pm 0.01$ & $19.31 \pm 0.04$ &  \nodata  & 0.28 & B* & \nodata \\
	603402 & J210.21613$-$00.66769 & $23.18 \pm 0.01$ & $21.64 \pm 0.00$ & $20.95 \pm 0.00$ & $20.62 \pm 0.00$ & $20.50 \pm 0.01$ & $19.63 \pm 0.04$ &  \nodata  & 0.30 & B* & [2], B \\
	675053 & J220.77984$-$00.20460 & $21.67 \pm 0.00$ & $20.31 \pm 0.00$ & $19.01 \pm 0.00$ & $18.47 \pm 0.00$ & $18.32 \pm 0.00$ & $17.44 \pm 0.02$ &  \nodata  & 0.09 & B* & \nodata \\
	681390 & J221.52312+02.19926 & $22.29 \pm 0.01$ & $21.69 \pm 0.00$ & $21.46 \pm 0.00$ & $21.38 \pm 0.01$ & $21.15 \pm 0.02$ & $20.90 \pm 0.20$ &  \nodata  & 0.05 & B* & \nodata \\
	703241 & J224.41869+01.64938 & $23.32 \pm 0.01$ & $21.70 \pm 0.00$ & $20.50 \pm 0.00$ & $20.13 \pm 0.00$ & $19.85 \pm 0.00$ & $19.36 \pm 0.05$ & $18.26 \pm 0.02$ & 0.31 & B* & \nodata \\
	728141 & J235.16864+43.59177 & $26.26 \pm 0.15$ & $24.69 \pm 0.06$ & $22.82 \pm 0.02$ & $21.95 \pm 0.02$ & $21.38 \pm 0.02$ & $20.46 \pm 0.18$ & $18.54 \pm 0.01$ & 0.16 & A* & \nodata \\
	731856 & J237.09818+42.72398 & $21.25 \pm 0.00$ & $20.92 \pm 0.00$ & $20.87 \pm 0.00$ & $20.64 \pm 0.00$ & $20.85 \pm 0.02$ & $20.22 \pm 0.13$ & $20.29 \pm 0.06$ & 0.17 & B* & \nodata \\
	786435 & J336.42475+01.08681 & $23.10 \pm 0.01$ & $21.78 \pm 0.01$ & $20.92 \pm 0.00$ & $20.63 \pm 0.00$ & $20.42 \pm 0.01$ & $19.89 \pm 0.09$ &  \nodata  & 0.19 & B* & \nodata \\
	808168 & J340.30123$-$00.04728 & $22.46 \pm 0.01$ & $21.82 \pm 0.00$ & $21.08 \pm 0.00$ & $20.74 \pm 0.01$ & $20.55 \pm 0.01$ & $19.73 \pm 0.12$ & $18.12 \pm 0.01$ & 0.13 & B* & \nodata \\
	866341 & J352.70165$-$00.94598 & $21.80 \pm 0.01$ & $20.92 \pm 0.00$ & $20.62 \pm 0.00$ & $20.46 \pm 0.00$ & $20.46 \pm 0.01$ & $20.36 \pm 0.18$ & $19.48 \pm 0.05$ & 0.14 & B* & \nodata \\
\end{longtable}
\tablefoot{
	Column (1): identification number for each candidate. 
	Column (2): name of the source. 
	Columns (3)--(7): HSC $grizy$-band magnitudes and the corresponding 1$\sigma$ errors.
	Column (8): NIR $J$-band magnitude.
	Column (9): unWISE W1-band magnitude.
	Column (10): lens probability based on the ensemble network classification.
	Column (11): grade after visual inspection.
	Column (12): references for the list of lenses or candidates published by earlier works, along with their reported grade.
	The magnitudes are in AB values, corrected for Galactic extinction based on the dust map of \cite{2019ApJS..240...30S} and considering the \cite{1999PASP..111...63F} reddening equation.
	Based on the visual inspection grades, best and good lens candidates with $P_\mathrm {lens} > 0.3$ are marked with A and B, respectively.
	On the other hand, sources with $P_\mathrm{lens} \leq 0.3$ but show lensing features and might be missed by our classifier are marked with grade A* or B*, depending on their quality.
	To name the candidates, we follow the ``JRRR.rrrrr+DD.ddddd'' convention, where RRR.rrrrr and +DD.ddddd are, respectively, the R.A. and decl. in decimal degrees (J2000).
}
\tablebib{
	[1] \cite{2021AandA...653L...6C}, 
	[2] \cite{2021AandA...653L...6C}, 
	[3] \cite{2022AandA...662A...4S}, 
	[4] \cite{2019ApJS..243...17J}, 
	[5] \cite{2022arXiv220602764S}, 
	[6] \cite{2023arXiv230405425C}, 
	[7] \cite{2022PASJ...74.1209W}, 
	[8] \cite{2020ApJ...899...30L}, 
	[9] \cite{2019MNRAS.484.3879P}, 
	[10] \cite{2020AandA...642A.148S}, 
	[11] \cite{2020MNRAS.495.1291J}, 
	[12] \cite{2018PASJ...70S..29S}, 
	[13] \cite{2022ApJ...932..107S}, 
	[14] \cite{2020ApJ...894...78H}
}
\end{center}

\let\thefigureoriginal\thefigure
\renewcommand{\thefigure}{\ref*{fig:lens_candidates}. continued}

\begin{figure*}[htb!]
	\centering
	\resizebox{\hsize}{!}{\includegraphics{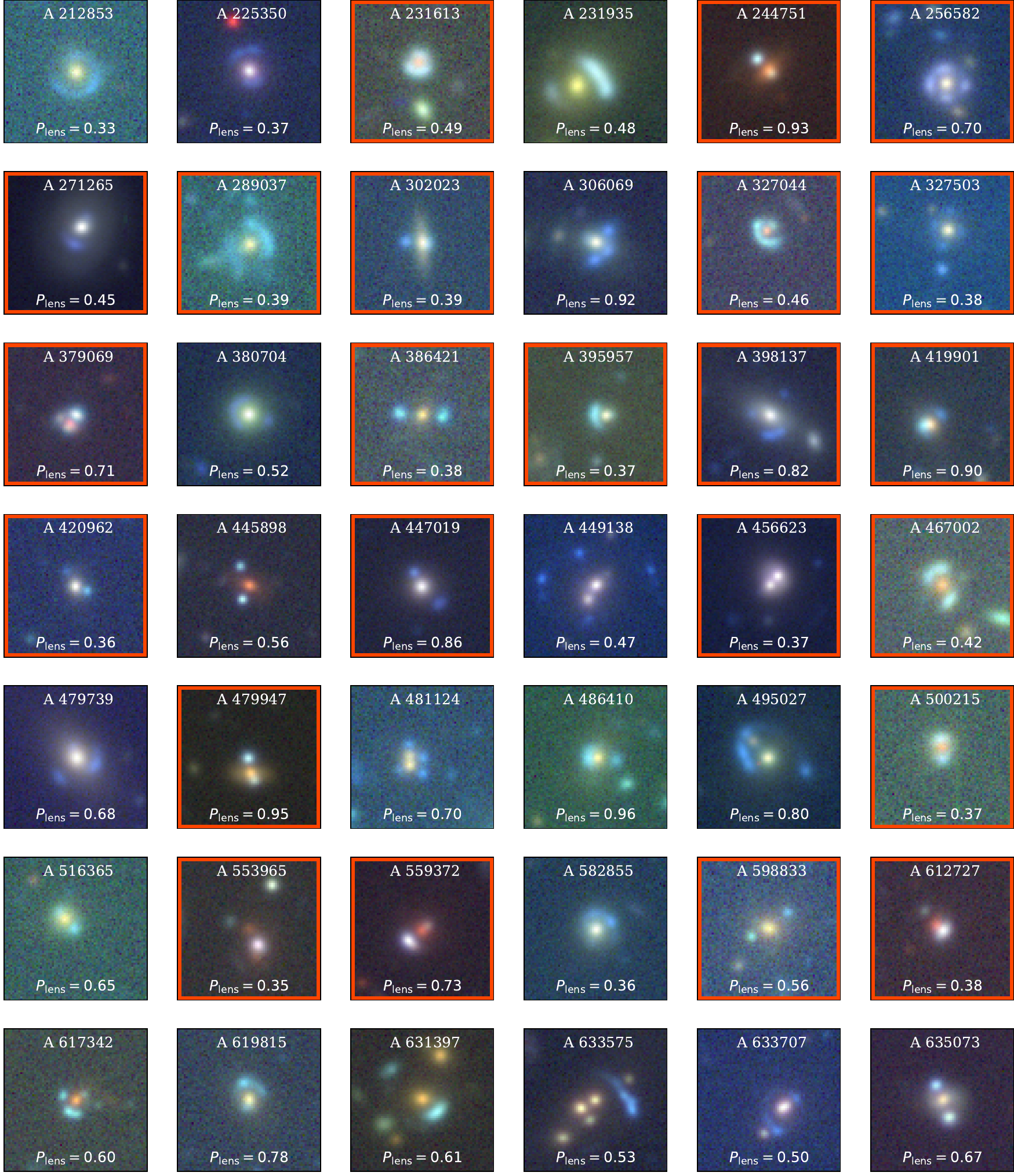}}
	\caption{}
\end{figure*}
\begin{figure*}[htb!]
	\centering
	\resizebox{\hsize}{!}{\includegraphics{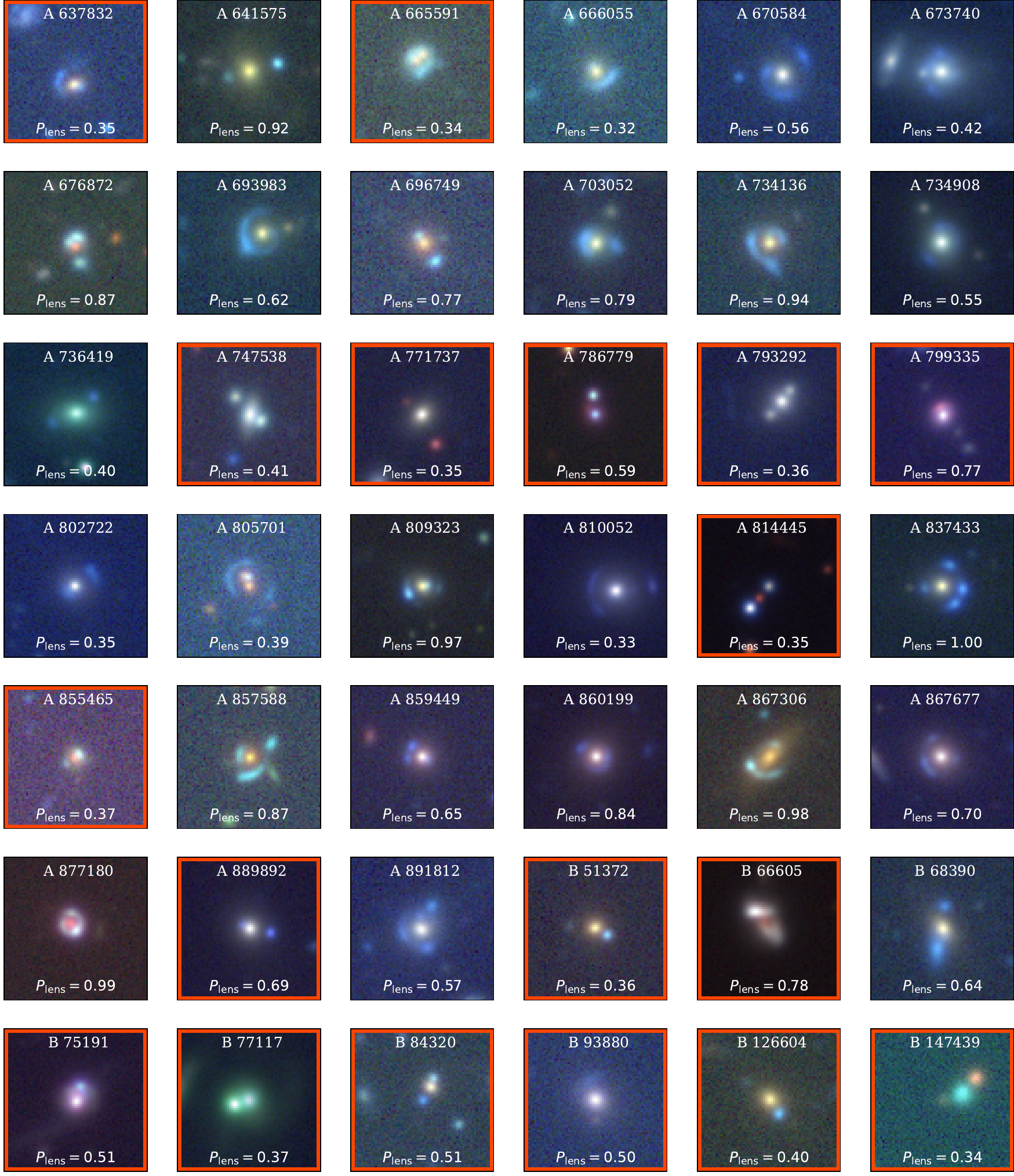}}
	\caption{}
\end{figure*}
\begin{figure*}[htb!]
	\centering
	\resizebox{\hsize}{!}{\includegraphics{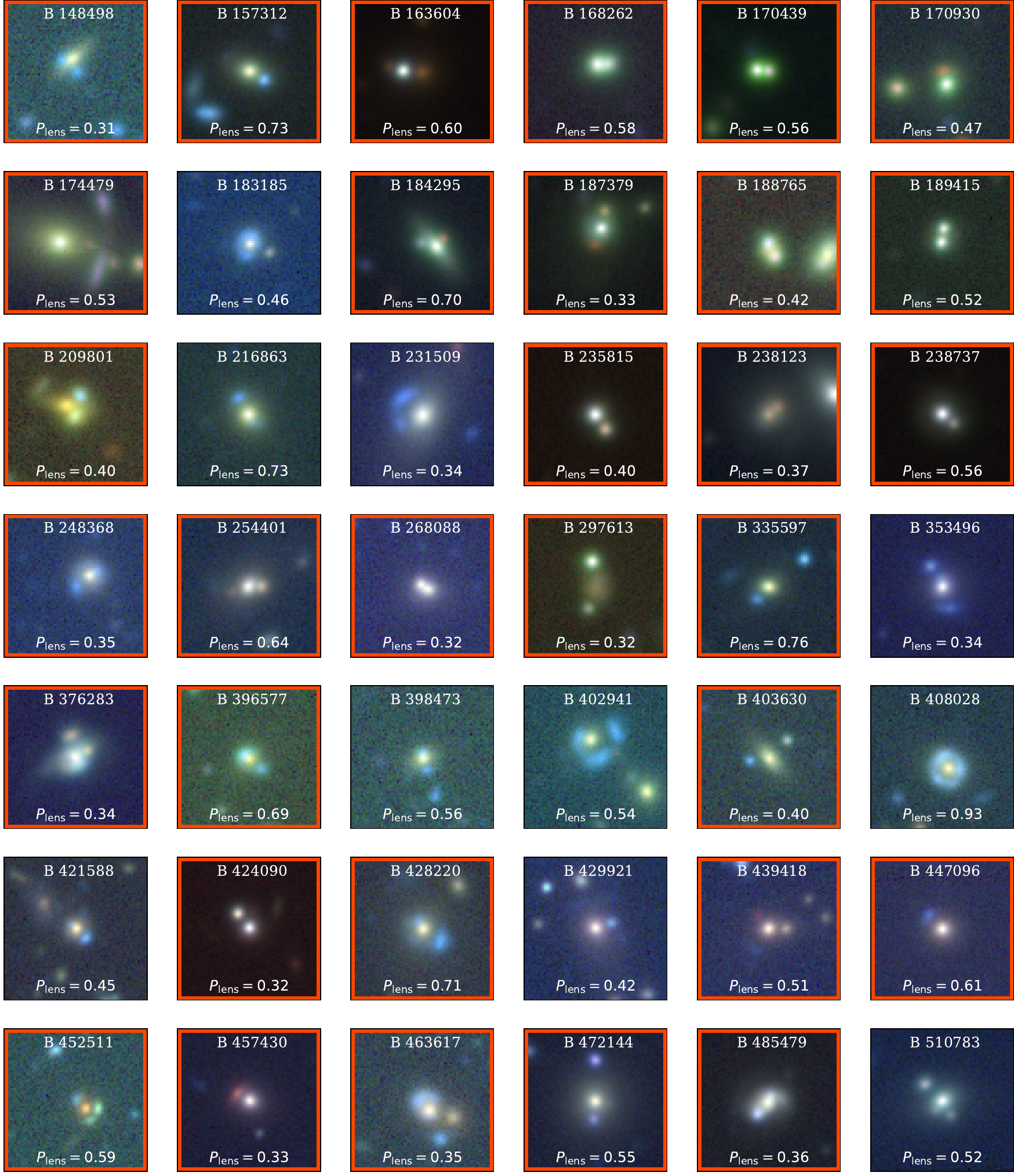}}
	\caption{}
\end{figure*}
\begin{figure*}[htb!]
	\centering
	\resizebox{\hsize}{!}{\includegraphics{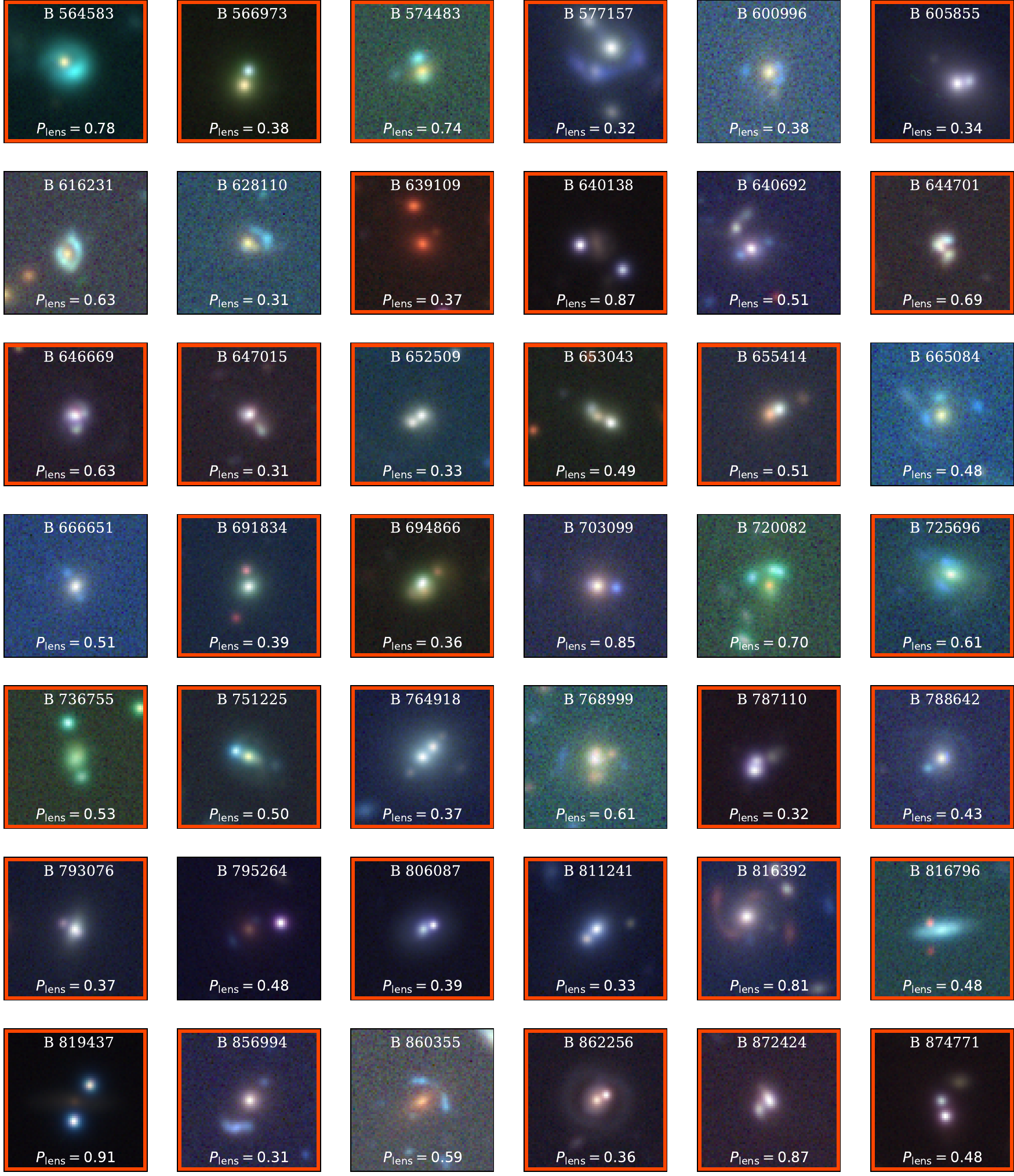}}
	\caption{}
\end{figure*}
\begin{figure*}[htb!]
	\centering
	\resizebox{\hsize}{!}{\includegraphics{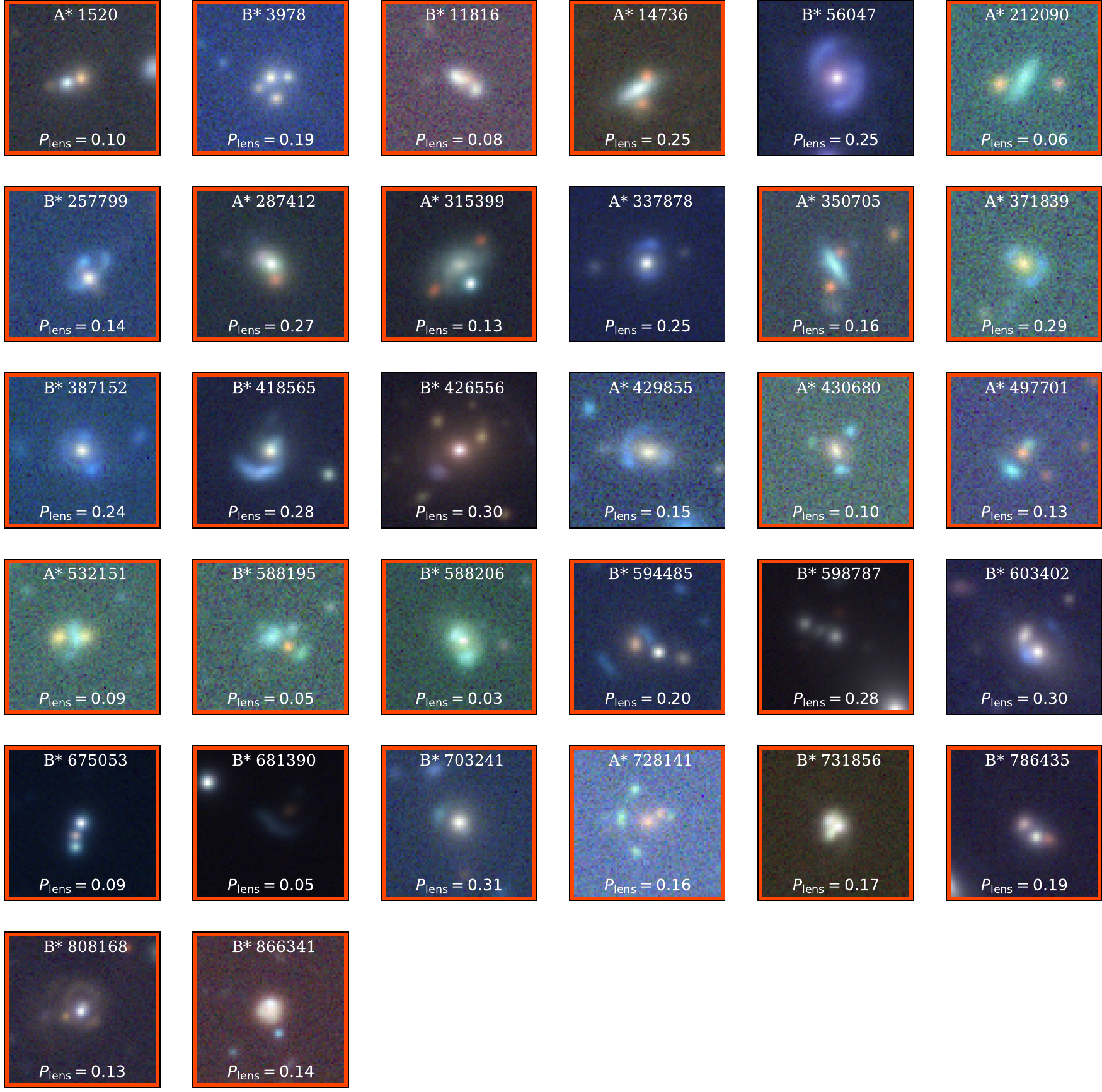}}
	\caption{}
\end{figure*}

\let\thefigure\thefigureoriginal
\setcounter{figure}{2}

\end{appendix}

\end{document}